\documentclass[12pt]{article}
\usepackage{styles/mystyle} 
\usepackage{dsfont} 

\begin{document}

\title{Disentangling the sources of cyber risk premia}
\date{\today}

\author{Loïc Maréchal\thanks{HEC Lausanne, University of Lausanne. \href{mailto:loic.marechal@unil.ch}{loic.marechal@unil.ch}} $ $ and Nathan Monnet\thanks{Swiss Finance Institute, École Polytechnique Fédérale de Lausanne - Cyber-Defence Campus, armasuisse S+T. \href{mailto:nathan.monnet@armasuisse.ch}{nathan.monnet@armasuisse.ch}\newline\newline This document results from a research project funded by the Cyber-Defence Campus, armasuisse Science and Technology, and was initially written as Nathan Monnet's Master thesis. We appreciate helpful comments from seminar participants at the Cyber Alp Retreat 2024. We also thank Julien Hugonnier and Michel Dubois for their invaluable comments. Corresponding author: Loïc Maréchal e-mail: \href{mailto:loic.marechal@unil.ch}{loic.marechal@unil.ch}}}

\renewcommand{\thefootnote}{\fnsymbol{footnote}}
\singlespacing
\maketitle
\vspace{-.2in}

\begin{abstract}
We use a methodology based on a machine learning algorithm to quantify firms' cyber risks based on their disclosures and a dedicated cyber corpus. The model can identify paragraphs related to determined cyber-threat types and accordingly attribute several related cyber scores to the firm. The cyber scores are unrelated to other firms' characteristics. Stocks with high cyber scores significantly outperform other stocks. The long-short cyber risk factors have positive risk premia, are robust to all factors’ benchmarks, and help price returns. Furthermore, we suggest the market does not distinguish between different types of cyber risks but instead views them as a single, aggregate cyber risk.

\end{abstract}

\medskip


\thispagestyle{empty}
\clearpage

\onehalfspacing
\setcounter{footnote}{0}
\renewcommand{\thefootnote}{\arabic{footnote}}
\setcounter{page}{1}

\section{Introduction}

Cyber-insurance contracts and cybersecurity solutions have become crucial for private and public organizations in the widespread and costly context of cyber incidents. These countermeasures, however, have costs that are challenging to estimate. This paper aims to overcome this challenge using natural language processing, clustering methods, and state-of-the-art asset pricing techniques to disentangle and quantify the risk premia of various cyber threats.

The main focus of this paper is to disentangle the different cyber risks faced by firms and the effects on expected returns through risk premia channels. To do this, we collect financial fillings, monthly returns, and other firm characteristics for over 7000 firms listed on US stock markets between January 2007 and December 2023. We use a neural network called \quotes{Paragraph Vector} in combination with the MITRE ATT\&CK cybersecurity knowledgebase and clustering techniques to score each firm's filing based on its various types of cyber risk.

We identify four types of cyber attacks that emerge from the textual cluster structures of MITRE ATT\&CK. We establish scores to quantify the similarity between the annual statements of firms, the 10-Ks, and the identified types of cyber attacks from the knowledgebase MITRE ATT\&CK. We find that the four cyber scores present no correlation with standard firms' characteristics known to help price stock returns and weak correlations with textual non-semantic variables of the annual statement (the highest, $0.36$, correlates with the length of section 1.A., in the 10-Ks). As previously observed in \cite{CelenyMaréchal2023}, who use the same neural network, the resulting aggregation of the various cyber scores shows increasing trends, with scores increasing by $0.04$ from 2007 to 2023. Also, specific industries from the Fama-French 12-industries classification display higher cyber scores, with Business Equipment and Telephone and Television Transmission being the highest.

Sorting firms into cyber scores-based portfolios, we observe monotonically increasing average excess returns along the sorts. All average excess returns of all portfolios are statistically significant at the 1\% level, and investing in a portfolio that enters a long (short) position in the top (bottom) cyber scores firm is statistically significant at the 5\% level. After controlling for common risk factors, the aforementioned results remain valid at the 5\% and 10\% levels in the top portfolios. 

The risk premia associated with the different types of cyber risk are also manifest at the 5\% level in the cross-section with \cite{FamaMacBeth1973} regressions. Using additional pricing factors related to cyber-based portfolios improves pricing ability. We demonstrate that joint alphas of various assets tend to decrease in \cite{GibbonsRossShanken1989} tests. Using the Bayesian approach of \cite{BarillasShanken2018}, we also show that the optimal subset of factors pricing stock returns invariably includes the cyber-based factors.

Additional tests reveal that, although various types of cyber risk exist, the market does not differentiate between them and perceives them as a single aggregate cyber risk. Finally, we conduct an event study to evaluate the performance of a cyber-based portfolio during the massive SolarWinds cyber attack in December 2020. Contrary to previous studies, in particular, \cite{FlorackisLoucaMichaelyWeber2023}, no significant conclusions can be drawn from this event regarding the performance of our cyber-based portfolios around cyberattacks.

The remainder of this work proceeds as follows. Section 2 introduces the related literature and develops hypotheses. Section 3 details the data and the methodology, Section 5 the results, and Section 6 concludes.

\section{Literature review}

\subsection{Sentiment analysis and text classification}

This study connects to several strands of the literature on sentiment analysis and its application to financial markets. First, it relates to the growing body of research that uses textual data to extract economic insights from corporate filings and other financial disclosures (\textit{e.g.}, \citealp{AntweilerFrank2004,Garcia2013,Arslan-AyaydinBoudtThewissen2016}). Early studies, such as \cite{FeldmanGovindarajLivnatSegal2010}, examine the Management Discussion and Analysis (MD\&A) sections of 10-Q and 10-K filings and demonstrated that changes in the tone of non-financial information—measured by the frequency of positive and negative words—correlate with both short-term market returns and longer-term excess returns. This study highlighted the importance of textual sentiment in SEC filings, laying the groundwork for more nuanced measures of market sentiment. Similarly, \cite{JegadeeshWu2013} introduce a return-based term weighting scheme for measuring document tone in 10-K filings. They demonstrate that the tone of these documents, especially when weighted using positive and negative dictionaries, effectively predicts market reactions around filing dates. Their method also generalizes well to other financial documents, such as IPO prospectuses, further expanding the applicability of sentiment analysis in financial contexts.

Building on these foundational studies, \cite{AntweilerFrank2004} apply sentiment analysis to online forums, showing how sentiment derived from financial discussions could predict market returns and trading volumes. \cite{Garcia2013} extend the scope of sentiment analysis to financial news, revealing that the predictive power of sentiment on stock returns is particularly pronounced during recessions. These studies collectively show that textual sentiment, whether derived from financial disclosures or public discourse, is critical in anticipating market behavior. Furthermore, \cite{BodnarukLoughranMcDonald2015} advance textual analysis by developing a linguistic measure of firm-level financial constraints based on 10-K disclosures. Their approach, which identifies constraining words such as \quotes{required} or \quotes{obligations}, proved more effective than traditional financial constraint indexes in predicting liquidity events, offering an innovative way to assess financial health through text.

This study contributes to the literature on textual analysis to assess specific risks. \cite{HassanHollandervLentTahoun2019} and \cite{SautnerLentVilkovZhang2023} apply sentiment analysis to earnings conference calls to measure firm-level political risk and climate risk, respectively. \cite{HassanHollandervLentTahoun2019} use political language to measure political risk, showing that political discussions during earnings calls significantly influence firm behavior and stock volatility. In contrast, \cite{SautnerLentVilkovZhang2023} focus on climate-related language, using a keyword discovery algorithm to assess the extent of firms’ climate risk exposure. These studies demonstrate how sentiment analysis can be tailored to identify specific types of risks beyond traditional financial metrics, thus expanding the scope of risk analysis to include socio-political and environmental factors.

Further extending the application of sentiment analysis, \cite{CalomirisMamaysky2019} incorporate advanced text-processing techniques, such as word flow measures, to predict market returns and risks across both developed and emerging markets. Analyzing sentiment, frequency, and entropy, they demonstrate that textual data could capture latent market risks more effectively than traditional methods. This approach complements the earlier work of \cite{JegadeeshWu2013} and \cite{BodnarukLoughranMcDonald2015} by showing how textual measures can be fine-tuned to assess not just sentiment but also financial constraints and specific risk factors, such as market volatility or liquidity challenges.

This study ties into the broader literature on how textual sentiment influences firm and market behavior. \cite{AntweilerFrank2004} show that sentiment derived from public discussions could affect stock prices and trading volumes. \cite{Arslan-AyaydinBoudtThewissen2016} demonstrate how managerial incentives shape the tone of earnings press releases, influencing market reactions.

\subsubsection{Vector representation of paragraphs and topics clustering}

Our study also connects to several key advancements in the application of vector representations and topic clustering for analyzing textual data. First, it relates to developing distributed representations of paragraphs, most notably the Paragraph Vector (doc2vec) model introduced by \cite{LeMikolov2014}. doc2vec extends the word vector framework to sentences, paragraphs, and entire documents by capturing semantic meanings in fixed-length vectors through neural network training. This technique allows for a richer, more context-aware text representation, outperforming traditional bag-of-words models. However, its performance is highly sensitive to various hyperparameters and data configurations, as shown by \cite{LauBaldwin2016}. Their empirical evaluation of doc2vec provides critical insights into optimizing the model for effective real-world use, emphasizing the importance of careful tuning when applying this method to different datasets.

\cite{AdosoglouLombardoPardalos2021} use doc2vec to analyze a vast corpus of 10-K filings from 1998 to 2018. They compare doc2vec with traditional dictionary-based approaches and found that vector representations captured subtle semantic shifts in financial disclosures that were predictive of future abnormal returns. Specifically, they develop the Semantic Similarity Portfolio (SSP) strategy, which identified firms with minimal year-on-year semantic changes in their filings—particularly those with high cosine similarity scores ($>0.95$) as strong candidates for achieving significant future risk-adjusted returns, up to 10\% annually. While their approach showcased the potential of doc2vec in financial text analysis, it also underscored limitations, such as the computational cost of training models and the need to account for executive turnover that might influence document language.

Regarding topic clustering, recent studies have extended the utility of vector-based representations. \cite{CalomirisMamaysky2019} apply unsupervised clustering techniques, such as the Louvain method, to group documents by topics and assess their relevance in predicting market outcomes. They create a network of document similarities by constructing vectors of word occurrences for each document and calculating similarity scores. The Louvain method was then employed to detect sub-networks or clusters, which they defined as topics. This clustering approach allowed them to identify significant thematic structures in large corpora of news articles. Such clustered topics could provide valuable insights into future market behavior, especially in capturing collective sentiment shifts.

Further reinforcing the benefits of vector-based methods, \cite{CuriskisDrakeOsbornKennedy2020} compare various document clustering and topic modeling techniques using social media text data. Their findings underscored the effectiveness of document and word embeddings, particularly doc2vec, for document clustering tasks. By outperforming traditional \quotes{tf-idf} based approaches and other topic modeling techniques, doc2vec embeddings, combined with k-means clustering, yielded superior results across different datasets. They observe that doc2vec embeddings performed consistently well regardless of document length, although optimal training epoch requirements varied with document size. These results suggest that doc2vec is a robust and adaptable tool for clustering tasks across various text lengths and types.

\subsection{Cyber risk and expected stock returns}

Finally, our study builds on a growing body of literature that examines the intersection of cybersecurity risk and asset pricing. \cite{JamilovReyTahoun2023W} develop a dictionary-based measure of cyber risk exposure using quarterly earnings calls from over 13,000 firms in 85 countries from 2002 to 2021. Their dictionary of cyber-related terms was validated by demonstrating predictability for future cyberattacks and correlations with stock market outcomes and realized volatility. Importantly, the study shows that cyber risk measures can forecast cyberattacks in subsequent quarters and document geographical patterns in cyber risk exposure. The findings are significant for global investors, as U.S. equity holdings in foreign countries predict the destination countries' cyber exposure. Additionally, the authors explore the pricing of cyber risk in the options market, showing that firms with higher cyber risk exposure face higher costs for market-based protection against price and variance risks. These results suggest that cyber risk has real economic implications and is increasingly integrated into stock market expectations.

\cite{FlorackisLoucaMichaelyWeber2023} present a cybersecurity risk measure based on textual analysis of the \quotes{Item 1. A. Risk Factors} section of 10-K filings from 2007 to 2018. By comparing these risk factors across firms using cosine similarity measures, they effectively identify companies with significant cybersecurity risk exposure. The measure correlates with various firm characteristics, such as size, growth opportunities, and R\&D expenditures, and predicts future cyberattacks. Furthermore, the authors show that portfolios composed of firms with high cybersecurity risk scores earn a significant return premium of up to 8.3\% annually. This risk premium is robust across different specifications, confirming that investors require compensation for holding stocks exposed to cybersecurity risks. The study also demonstrates that the cybersecurity-based portfolio underperforms during periods of heightened investor attention to cybersecurity but delivers high returns during other times, suggesting an element of market underreaction to latent cyber risks.

Taking a different approach, \cite{CelenyMaréchal2023} introduce a method for estimating cyber risk using doc2vec trained on the MITRE ATT\&CK cybersecurity knowledge base and applied it to 10-K statements. Unlike the dictionary-based approaches, their model captures broader contextual information from the entire document, resulting in a more comprehensive and accurate cyber risk score. They show that portfolios sorted by their cyber risk score achieve substantial excess returns—up to 18.72\% annually with a significant risk premium of 6.93\% on a long-short portfolio. Their analysis highlights the critical role of cyber risk in the cross-section of stock returns, and their doc2vec-based approach outperforms traditional dictionary-based methods, such as the one developed by \cite{FlorackisLoucaMichaelyWeber2023}. The study's robustness tests further validate the method, showing that the cyber risk factor remains consistent over time and is not influenced by the exclusion of cybersecurity firms, offering a more accurate measure of latent cyber risk exposure.

Additionally, \cite{LiuMarshXiao2022W} explore how firms' sensitivity to cybercrime impacts the pricing of individual stocks and equity portfolios. Using a news-based cybercrime index and corroborating their findings with Google search trends, the authors demonstrate a significant negative correlation between cybercrime exposure and subsequent stock returns. Their study shows that firms with higher sensitivity to cybercrime tend to underperform in the market, and they further highlight how corporate governance, IT investments, and industry dynamics shape firms' vulnerability to cyber threats. Moreover, they find that high cybercrime beta stocks consistently outperform their low-beta counterparts, particularly following significant cyber incidents. These findings emphasize the importance of incorporating cybercrime exposure into asset pricing models. \cite{JiangKhannaYangZhou2023} use a cyber dictionary from NIST to count the number of cyber-related words in Item 1. A. Combining it with firm characteristics, they perform a logit ridge regression where the dependent variable is the probability for a firm to experience a cyberattack in the future. 

Lastly, \cite{GomesMihetRisbabh2023} examine how cyber risk drives firm-level innovation, especially in firms that develop cybersecurity measures internally. Using the cyber risk score from \cite{FlorackisLoucaMichaelyWeber2023} alongside patent data shows that firms heavily exposed to cyber risk are incentivized to innovate in cybersecurity solutions, which can contribute to long-term growth.

\section{Data and methodology}
\subsection{Market data}

\noindent We download public equity data from Wharton Research Data Services\footnote{\href{https://wrds-www.wharton.upenn.edu/}{https://wrds-www.wharton.upenn.edu/}} (WRDS), and their API. The data originated from the Center for Research in Security Prices\footnote{\href{https://crsp.org/}{https://crsp.org/}} (CRSP) and S\&P Global Market Intelligence's Compustat database\footnote{\href{https://www.marketplace.spglobal.com/en/datasets/compustat-financials-(8)}{https://www.marketplace.spglobal.com/en/datasets/compustat-financials-(8)}}. We report the list of variables in Table \ref{tab:variable_descriptions} and Table \ref{table_fin_stats} report their statistics after cleaning.

\begin{figure}[H] 
    \noindent\makebox[\textwidth]{%
    \includegraphics[scale=0.65]{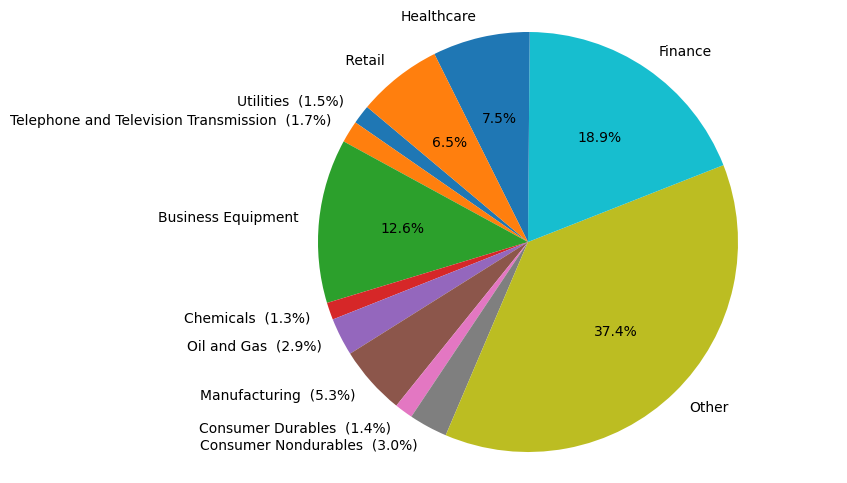}}
    
    \caption[\footnotesize Industry distribution]{\textbf{Industry distribution}}\bigskip
    \footnotesize{Distribution of firms in the 12 Fama-French industries. Standard Industrial Classification (SIC) codes are obtained from CRSP. The conversion table, from SIC to 12 FamaFrench industries, is available on the Kenneth French data repository.}
    \label{camembert}
\end{figure}

We use a pre-existing Python script that retrieves all available data from WRDS about various firms and filters out those that have not filed 10-K forms with the SEC. We extract monthly stock returns and financial ratios for 7,079 firms between January 2007 and December 2023. We depict the industry distribution of these firms using the Fama-French 12 industry distribution in Figure \ref{camembert}.

We also download the one-month Treasury bill rate and returns on the market, book-to-market (HML), size (SMB), momentum (UMD), investment (CMA), and operating profitability (RMW) factors from the Kenneth French data repository\footnote{\href{http://mba.tuck.dartmouth.edu/pages/faculty/ken.french/data\_library.html}{http://mba.tuck.dartmouth.edu/pages/faculty/ken.french/data\_library.html}}.

\begin{center}
[Insert Table 1 here]
\end{center}

\begin{center}
[Insert Figure 1 here]
\end{center}

\subsection{10-K statements}

\noindent 10-K statements are financial filings publicly traded companies submit annually to the U.S. Securities and Exchange Commission (SEC). They contain information such as companies' financial statements, risk factors, and executive compensation. 10-K statements will later be used to build a cybersecurity risk measure. The index files from the SEC's Edgar archives\footnote{\href{https://www.sec.gov/Archives/edgar/full-index/}{https://www.sec.gov/Archives/edgar/full-index/}} are used to download and structure the 10-K. These index files contain information about all the documents filed by all firms for a specific quarter. Each line of the index file corresponds to a 10-K and is structured as follows:

\begin{center}
    CIK|Company Name|Form Type|Date Filed|Filename
\end{center}

\begin{figure}[H] 
    \noindent\makebox[\textwidth]{%
    \includegraphics[scale=0.4]{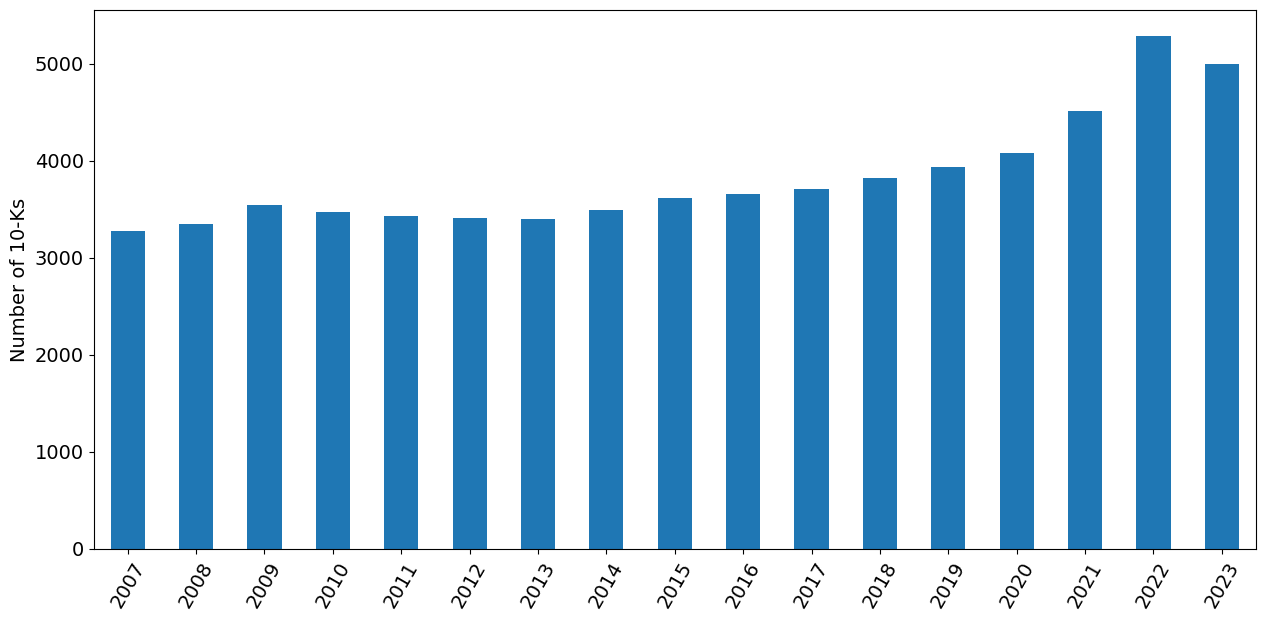}}
    
    \caption[\footnotesize Number of 10-Ks per year]{\textbf{Number of 10-Ks per year}}\bigskip
    \footnotesize{Number of companies in the study sample that have filed a 10-K statement through the years.}
     \label{histo_10k_years}
\end{figure}

\noindent Where Filename is the URL under which an HTML version of the document is available. Their Central Index Key (CIK) is used to identify firms. The CIK consists of a number used by the SEC to identify corporations and individuals who have filed disclosures. We use a Python script that goes through these index files and identifies URLs corresponding to 10-K statements using the Form Type entry. These URLs are matched to one of the 7,079 firms using the CIK entry. 64,988 10-K statements are identified, corresponding to 2.73 statements per firm on average. The evolution of the number of 10-K filled annually is reported in Figure \ref{histo_10k_years}.

\begin{center}
[Insert Figure 2 here]
\end{center}

\subsection{MITRE ATT\&CK\ description}

\begin{figure}[H] 
    \noindent\makebox[\textwidth]{%
    \includegraphics[scale=0.35]{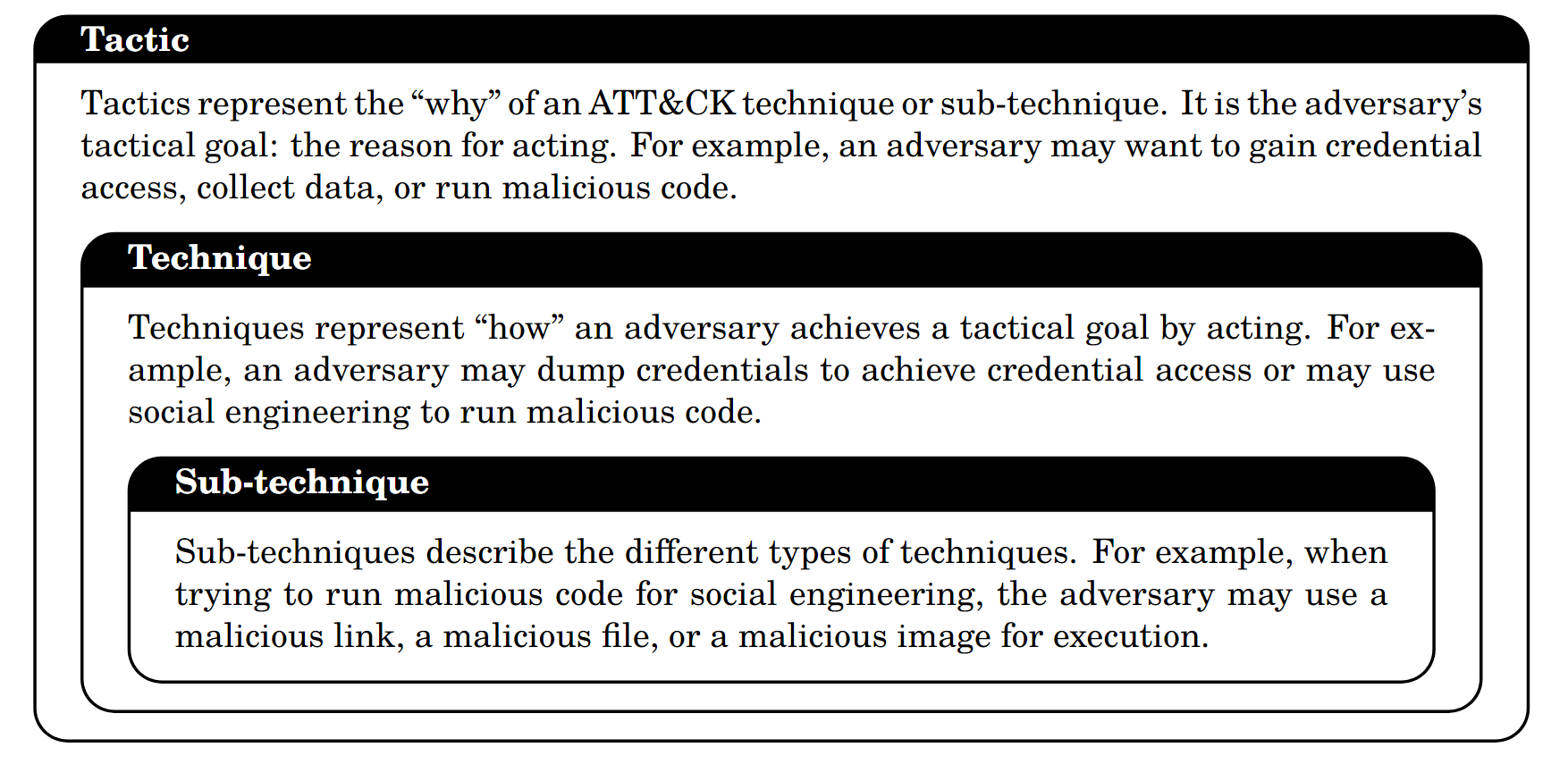}}
    
    \caption[\footnotesize Structure of MITRE ATT\&CK]{\centering \textbf{Structure of MITRE ATT\&CK}\bigskip
    }
    \label{fig:MITRE_structure}
\end{figure}

The MITRE ATT\&CK\footnote{\href{https://attack.mitre.org/}{https://attack.mitre.org/}} cybersecurity knowledge base is used as a reference for cyber attack descriptions. This knowledge base was created in 2013 to document cyber attack tactics, techniques, and procedures. It is structured by tactics, techniques, and sub-techniques as depicted in Figure \ref{fig:MITRE_structure}. There are 14 tactics: reconnaissance,  resource development, initial access, execution, persistence, privilege escalation, defense evasion, credential access, discovery, lateral movement, collection, command and control, exfiltration, and impact. There are 785 sub-techniques across all tactics. Two examples of sub-techniques are given in Table \ref{tab:MITRE_examples}.

\begin{center}
[Insert Figure 3 here]
\end{center}

\begin{center}
[Insert Table 2 here]
\end{center}

\noindent Figure \ref{fig:MITRE_structure} and Table \ref{tab:MITRE_examples} are taken from \cite{CelenyMaréchal2023}. The Data section closely follows their approach, and much of their code has been repurposed to suit our requirements. The additional data primarily originates from 2023.

\subsection{Cyber score}

To compute the cyber scores of interest, We start with the 14 individual MITRE ATT\&CK tactics: Reconnaissance, Resource, Development, Initial Access	Execution, Persistence, Privilege Escalation, Defense Evasion, Credential, Access, Discovery, Lateral Movement, Collection, Command and Control, Exfiltration, Impact. To reduce this dimensionality, We aggregate them with clustering methods (see \ref{chapter_cluster}) that yield the following \quotes{supertactics}: Command and data manipulation, Credential movement, Persistence and evasion, Preparation and reconnaissance. For comparison, we also add the overall score, aggregating all 14 categories into one, corresponding to the score obtained in \cite{CelenyMaréchal2023}. Finally, We add a variation of the overall score that relates better to the risk notion: the cyber sentiment score. 

\subsubsection{Preprocessing}

Everything related to text processing and its use is done exactly as described in \cite{CelenyMaréchal2023}. We download 10-K statements from the SEC Archives as HTML files. Then, we use the library BeautifulSoup to extract usable texts from HTML.\footnote{\href{https://www.crummy.com/software/BeautifulSoup/}{https://www.crummy.com/software/BeautifulSoup/}} We remove punctuation and numbers and set all letters to lowercase. Finally, we apply the Python script of \cite{CelenyMaréchal2023} that uses the \quotes{wordfreq} and NLTK libraries to divide the text into sentences, remove stop-words such as \quotes{the}, \quotes{is}, \quotes{and}, \ldots) and remove the most common words.\footnote{\href{https://pypi.org/project/wordfreq/}{https://pypi.org/project/wordfreq/}}\footnote{\href{https://www.nltk.org/}{https://www.nltk.org/}}

After pre-processing, the average length of the MITRE ATT\&CK sub-technique descriptions is 39.7 words. We use a Python algorithm to merge consecutive sentences from 10-K statements into paragraphs with an average length of close to 40 words after pre-processing. This results in an average of 640 paragraphs per 10-K statement with 46 words per paragraph. The standard deviation is 2.8 words per paragraph and 309 paragraphs per 10-K statement.

\subsubsection{Paragraph Vector algorithm (doc2vec)}

As in \cite{CelenyMaréchal2023}, we use the paragraph to vector model proposed by \cite{LeMikolov2014}, which is an extension of the word2vec model (\citealp{MikolovChenCorradoDean2013}). There are various advantages to working with this NLP approach compared to others, such as the dictionary approach. First, the comprehension of the method is semantical, meaning that it is not limited to a count of word frequencies. The word order impacts the resulting vector, and paragraphs with similar or synonym words will have close vector representations. Second, training the model with specific text that involves a particular vocabulary allows the incorporation of relatively unknown words. Finally, the resulting vectors have a dimension usually much smaller than vectors resulting from the dictionary approach. 

\begin{figure}[H] 
    \noindent\makebox[\textwidth]{%
    \includegraphics[scale=0.4]{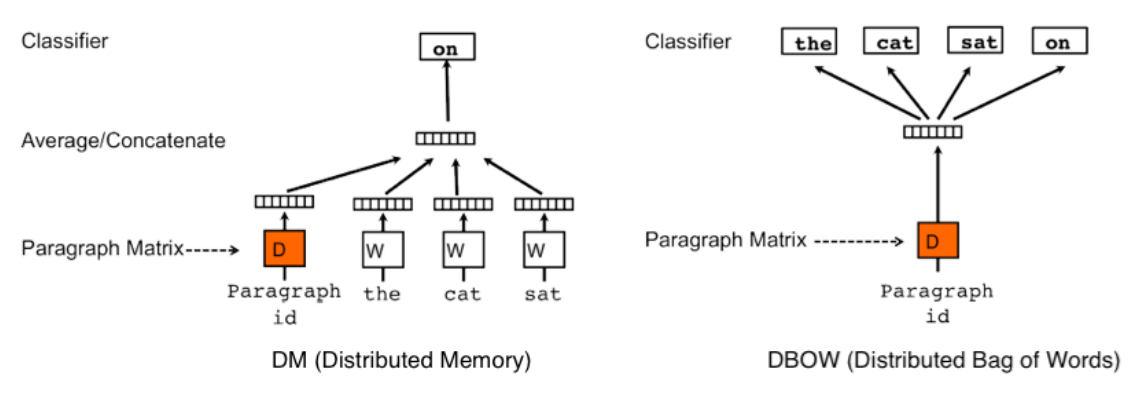}}
    
    \caption[\footnotesize Illustration of doc2vec training]{\textbf{Illustration of doc2vec training}}\bigskip
    \footnotesize{Illustration of the training of the neural network of the two versions of doc2vec, distributed memory model (DM) and distributed bag-of-words model (DBOW). The figure is taken from \cite{LeMikolov2014}.}
    \label{doc2vec_model}
\end{figure}

Two model versions exist: the distributed memory model (DM) and the distributed bag-of-words model (DBOW). In the DM, a neural network is trained as follows. First, a word is removed in a paragraph. Then, inputting the paragraph vector representation and context words (also in vector representation) surrounding the missing word, the neural network is optimized by trying to guess the missing word. In the DBOW, the neural network is trained to predict a series of words sampled from a paragraph using only the vector representation of the paragraph as input. Figure \ref{doc2vec_model} illustrates the training process of the two models.

\begin{center}
[Insert Figure 4 here]
\end{center}

The training data and details, the hyperparameters and their validation, and the final model choice are extensively covered in \cite{CelenyMaréchal2023}. This work uses their saved doc2vec model.\footnote{\url{https://github.com/technometrics-lab/17-Cyber-risk_and_the_cross-section_of_stock_returns}}


\subsubsection{Cosine similarity}

Using the doc2vec method previously described, all paragraphs of interest can be embedded into vectors. A common way to attribute a similarity score to two paragraph vectors is to take the cosine of the angle they form. Other ways exist, but only measuring the angle was proven more effective than considering the vectors' magnitude (see \citealp{AdosoglouLombardoPardalos2021}). This is because the latter is more affected by the random initialization of weights during training in the neural network that outputs the vectors. 

\subsubsection{Cyber tactics clustering}\label{chapter_cluster}

We disentangle the overall cyber score obtained in \cite{CelenyMaréchal2023}. The idea is that the risk coming from different areas of cyber security might not be similarly priced and, therefore, should not be aggregated into a single score but rather be separated into sub-cyber scores. A natural way of splitting the overall score into different categories comes from the written structure of MITRE ATT\&CK, with 14 categories already mentioned in chapter 4.1. However, it is believed that splitting the overall score to such an extent might result in a loss of explanatory power and highly correlated sub-cyber scores. Therefore, aggregating the 14 tactics into a few super tactics might mitigate the negative effect of splitting the overall score.

On the other hand, with the doc2vec method and the similarity score, we can transform every 785 sub-techniques (paragraphs) of MITRE ATT\&CK into vectors and compare their similarity. This process results in a similarity matrix of dimension 785 by 785, onto which clustering methods can be applied. Indeed, the similarity matrix can be understood as the representation of a network where every 785 nodes (paragraphs) are connected by edge values weighted by their similarity. In this context, we present three classical clustering methods.

The first and most simplistic clustering method is K-Means. Note that since the input similarity matrix is based on cosine similarity, it is rather designated as spherical K-Means, where the distance between each point to class into K categories is understood as the angle between the vectors defined by those points rather than the Euclidian distance between those points. Either version of K-Means works as follows: It begins by randomly setting initial cluster centroids, then iteratively assigns each data point (paragraphs) to the nearest centroid and updates the centroids by recalculating their mean positions among their associated data points. The process is repeated until convergence. Note that although the K-Means algorithm always converges, it is relatively dependent on the initial centroid guess. The user must choose the number of clusters K without prior knowledge. The algorithm generally produces rough results but often reveals an initial simple structure in the similarity of the provided data.

The second method is much more powerful as it requires no prior hyperparameters; thus, the number of clusters is an output of the method. The Louvain method, explained in \cite{BlondelGuillaumeLambiotteLefebvre2008}, provides a straightforward way to identify clusters (groups of nodes within a graph that are more densely connected) in a network. To explain the Louvain method, we first need to introduce the notion of modularity. It is defined as a value in the $[-1/2, 1]$ range that measures the density of links within communities compared to between. For a weighted graph, the modularity is defined as:

\[
Q = \frac{1}{2m} \sum_{i=1}^{N} \sum_{j=1}^{N} \left[ S_{ij} - \frac{k_i k_j}{2m} \right] \delta(c_i, c_j),
\]

where $S_{ij}$ represents the edge weight between nodes $i$ and $j$, in this case, this is the similarity matrix. $k_i$ and $k_j$ are the sum of the weights of the edges attached to nodes $i$ and $j$, respectively. $m$ is the sum of all the graph's edge weights. $N$ is the total number of nodes in the graph.  $c_i$ and $c_j$ are the communities to which the nodes $i$ and $j$ belong and $\delta$ is the Kronecker delta function.
The Louvain method works as follows. Initially, each node is assigned to its own community. Then, the method iterates through two phases: the first phase optimizes modularity locally by moving individual nodes between communities to maximize the increase in modularity. The second phase aggregates the nodes in each community in the first phase into single nodes and builds a new network, where the communities found in the first phase are treated as nodes. Phases one and two are repeated until no further improvement in modularity is possible. The final partitioning of nodes into communities is returned as a result.

The third clustering method is spherical K-means on a dimensionally reduced similarity matrix. The spectral clustering method works as follows. First, the degree matrix $D$ is constructed. it consists in a diagonal matrix where each entry $D_{ii}$ represents the sum of similarities for node $i$ and is computed as $D_{ii} = \sum_{j} S_{ij}$. The Laplacian matrix $L$ is defined as $L=D-S$. The spectral clustering algorithm computes the eigenvectors and eigenvalues of the Laplacian matrix $L$. Let $\lambda_1, \lambda_2, ..., \lambda_N$ be the eigenvalues and $v_1, v_2, ..., v_N$ be the corresponding eigenvectors. After obtaining the eigenvectors, we select the $K$ eigenvectors corresponding to the $K$ smallest eigenvalues (excluding the smallest eigenvalue, typically zero). We arrange these eigenvectors as columns in a matrix $V$ of dimension K by N. Finally, we cluster the rows of the matrix $V$ using the k-means clustering method. The power of this approach is that we can choose the number of features necessary to perform a satisfying clustering (reducing from N=785 to K<20, for example, can radically improve the clustering by getting rid of superfluous dimensions).

Finally, scoring is needed to find the best clustering output produced by the wide range of hyperparameters and method choices. We propose a rather simple but efficient approach that requires initial labelalization of each node. Each paragraph (node) is a sub-technique belonging to one of the 14 tactics of MITRE ATT\&CK. Thus, they naturally already belong exclusively to 14 sub-clusters. The discrimination of clustering methods works on the following two requirements.

First, we want the paragraphs belonging to one tactic (sub-cluster) to belong to the same super-tactic, \textit{i.e.}, the same cluster found by the method. Indeed, the paragraphs are initially classed by the creator of MITRE ATT\&CK together because they share common characteristics. It would not be very sensible to spread them across different super-tactics (clusters) once the clustering method is applied. Thus, a measure of sub-cluster heterogeneity among clusters is needed. We use Shannon entropy, defined as follows:
\begin{equation}
    H_{sub_j}=-\sum_{i=1}^{nb. clusters} P(sub_j)_i \log P(sub_j)_i
\end{equation}
Where $P(sub_j)_i$ is the proportion of paragraphs of sub-cluster (tactic) j belonging to cluster (super tactic) $i$. Intuitively, if we are in an ideal case and the paragraphs of a sub-cluster $j$ are entirely contained in cluster 1 we would have $P(sub_j)_1=1$ and $P(sub_j)_i=0$ for $i\neq1 $, thus leading to $H_{sub_j}=0$ being minimal (mind the minus sign in the equation and the logarithm on number lower than 1). If we start to spread the paragraph of the sub-cluster among other clusters, the $P(sub_j)_i$ becomes different from 0 and 1, and $H_{sub_j}$ gradually increases. To reduce the 14 $H_{sub_j}$ to one score of discrimination, we sum them all, thus obtaining the Entropy sum, the sub-cluster heterogeneity measure among clusters. The heterogeneity is high when the Entropy sum is low.

Second, we need a score to counter the following extreme case. All sub-clusters, but one may be classed into one cluster and the last sub-cluster into a second cluster. This would lead to a minimum Entropy sum of 0 but would have no value for our application. We want the sub-cluster to be reasonably spread out among the clusters. To translate this idea into a meaningful score, we create a Balanced score, defined as the standard deviation of the label counts. In other words, the clustering method produced an ordered list of 785 values corresponding to the label of the cluster each paragraph belongs to. For each label, we count the number of occurrences on the list. If the paragraphs are relatively well spread out across the cluster, then taking the standard deviation of all the count of the labels should be low since each cluster would contain approximately the same number of paragraphs. The last case to worry about is that the balanced score could be low, but the paragraph would be randomly spread across the cluster, thus not reflecting the initial structure of MITRE ATT\&CK tactics (sub-clusters). To counter that, it is sufficient to consider the Entropy sum. 

Considering the method that outputs the lowest Entropy sum and the lowest balanced score, we discriminate the different clustering methods' outputs. Note that there is no guideline regarding the optimal trade-off between the two scores, \textit{i.e.} what additional amount is optimal to forfeit to the Entropy sum to lower the Balanced score and inversely. 

Finally, the whole clustering process described here must be seen more as a guideline tool. Indeed, after choosing the best method, we class each paragraph in the cluster where most of its sub-cluster belongs, regardless of the method's output for the misplaced paragraphs. The structure of MITRE ATT\&CK is probably more coherent than the output of any unsupervised clustering method. However, it is still advantageous to consider the new clustering structure output since it is based on the cosine similarity matrix, and it could maximize the likelihood of reducing the correlation between the different sub-cyber scores that will also be based on cosine similarities.

\subsubsection{Setting the cyber score}
At this point, each paragraph of a 10-K can be transformed into a vector, and the same can be done with the 785 paragraphs of MITRE ATT\&CK. Then, each paragraph of the 10-K can be compared to each paragraph of MITRE ATT\&CK. This leads to each paragraph of the 10-K being associated with 785 cosine similarities. \cite{CelenyMaréchal2023} computes the cyber score of a 10-K by associating the maximum out of the 785 cosine similarities to each paragraph and then taking the average of the top 99\% of these maxima. 

Similarly, we define a sub-cyber score by associating to each paragraph the cosine similarities of a subset of paragraphs of MITRE ATT\&CK. For example, each paragraph would be associated with 120 cosine similarities (instead of 785), where 120 would correspond to the 120 paragraphs of MITRE ATT\&CK that belong to the same category (cluster or sub-cluster/ super tactic or tactic). Then, finding the sub-cyber score associated with a super tactic or tactic would be the same as described in the previous paragraph; we take the maximum out of the 120 cosine similarities for each paragraph and then compute the average of the top 99\% of these maxima. 

\subsubsection{Sentiment analysis}
To establish a cyber sentiment score, we opt for a straightforward approach. We define the cyber score as it was done in  \cite{CelenyMaréchal2023}, but instead of taking the maximum, we take 0 if the paragraph does not contain a word from a specific list and the maximum as usual if it does contain a word from the specific list.  

The specific list is defined in \cite{HassanHollandervLentTahoun2019} and is reported in the annex. It contains words relative to \quotes{risk} or \quotes{uncertainty} and was created using the Oxford English Dictionary. 

\subsection{Asset pricing tests}

\subsubsection{Univariate sorts}
Five portfolios are constructed based on a cyber score of interest. Firms are classified each quarter based on their most recent known cyber score from the previous quarter. These firms are then divided into five categories corresponding to the quintiles of their cyber scores. Consequently, the firms in the top 20\% of cyber scores are placed in Portfolio 5 (P5). After that, each firm is weighted within its portfolio according to its market capitalization known from the end of the previous quarter. The cyber-based portfolios are updated quarterly.

A first quantitative test consists of observing whether the average returns of each portfolio change monotonically with the increasing cyber score. The idea is to see if returns are affected by this cyber classification, thus hinting at a potential cyber-related risk structure.  

Next, we assess portfolios' returns, controlling for pricing factors, by using pricing factors recognized in the literature (factors included in the CAPM, in \citealp{FamaFrench1992} (FFC) and in \citealp{FamaFrench2015} (FF5)), we observe if their linear combinations are sufficient to explain the returns of the portfolio or if statistically significant alpha (intercept) appear, meaning that the profitability of the portfolios based on cyber score can not be entirely explained by common pricing factor and new ones are needed.

\subsubsection{Double sorts}

The interest in the double sorting method is the same as in univariate sorting. We want to see if returns are affected by the cyber classification. However, the cyber score may be a proxy \textit{i.e.} something that mimics another firm characteristic, such as the size, the book-to-market ratio, or market beta. To avoid that, we sort the firms according to one of the three characteristics mentioned. These firms are then divided into five categories corresponding to the quintiles of their characteristic. Consequently, the firms in the top 20\% of the characteristic of interest (for example, firms with the highest book-to-market ratio) are placed in category 5 (Q5). Then, for the firm of each category Q1 to Q5, we construct a portfolio based on a cyber score as described previously to obtain 25 portfolios, five for each category. 

\subsubsection{Cross-sectional tests}

\cite{FamaMacBeth1973} proposes the following method.
First, estimate betas using time series regressions with 2-year rolling windows (24 months). This corresponds to the following regression for each asset i with $t \in [T-24, T] $:
\begin{equation}
R_{i,t} = \alpha_{i,t} + \sum_{k} \beta^k_{i,T} F_{k,t} + \epsilon_{i,t}, \quad \forall i
\end{equation}
Each asset return $R_i$ is regressed on non-firm-specific pricing factors $F_k$. Conversely, the cyberscore. Consequently, we obtain time series of betas specific to both asset and factor: $\{ \beta^k_{i} \}_{T=01/2009, ..., 12/2023}$ (ranging from January 2009 to December 2023, in this example). Then, we build twenty portfolios based on the cyber score analogously to the five cyber score-based portfolios described earlier. Knowing the weight $x_{i,p,T}$ of each asset inside each portfolio through time, we compute the factor exposures of the portfolios:
\begin{equation}
    \beta^k_{p,T} = \sum_{i=1}^{20} x_{i,p,T} \cdot \beta^k_{i,T}
\end{equation}
After that, the risk premia (gamma) are computed for each time t with $p=1, \ldots, 20$:
\begin{equation}
    R_{p,t} = \gamma_{t}^0 + \sum_k \gamma_{t}^k \beta_{p,t-1}^k  + \sum_j c_{t}^j   \lambda_{p,t-1}^j  + \epsilon_{p,t}^*, \quad \forall t
\end{equation}
Consequently, to determine each  $\{ \gamma^k_{t} \}_{t=01/2009, ..., 12/2023}$, 20 portfolio returns are used each time in the linear regression. The additional terms are aggregated firm-specific factors. In our case, we only have one such factor (so $j$ is omitted in the following expression), the cyber score:
\begin{equation}
    \lambda_{p,t} = \sum_{i=1}^{20} x_{i,p,t} \cdot \lambda_{i,t}
\end{equation}
The remaining coefficient $\{ c_{t} \}_{t=01/2009, \ldots, 12/2023}$ is determined alongside $\{ \gamma^k_{t} \}_{t=01/2009, \ldots, 12/2023}$ during the linear regression. Finally, a t-test is applied on each time series $\{ \gamma^k_{t},c_{t} \}_{t=01/2009, \ldots, 12/2023}$ to assess the statistical significance of each risk premia. 

\subsection{Time-series tests}

\cite{GibbonsRossShanken1989} introduces a statistical test to assess portfolio pricing efficiency:
\begin{equation}
    R_{i,t} = \alpha_{i} + \sum_{p} \beta_{i,p} R_{p,t} + \epsilon_{i,t}, \quad \forall i,
\end{equation}
where $R_{i,t}$ and $R_{p,t}$ are assets and portfolio returns, respectively. If the portfolios were carefully selected, they could correctly predict the asset returns, thus suppressing the need for alphas (intercepts that contain contributions to their asset returns not taken into account by the explanatory portfolios). GRS provides a statistical test for this null hypothesis: $H_0$: $\alpha_i = 0 \quad \forall i$. \cite{Cochrane2005} generalize this idea by including traded factors $F_k$ instead as explanatory variables and portfolio excess returns as the endogenous one:
\begin{equation}
    R^e_{p,t} = \alpha_{p} + \sum_{k} \beta_{p,k} F_{k,t} + \epsilon_{p,t}
\end{equation}
In that case, the GRS score that tests jointly the zero alphas follows a F-distribution:
\begin{equation}\label{GRS_eq}
\frac{(T - N - K)}{N} \frac{\hat{\alpha}' \hat{\Sigma}^{-1} \hat{\alpha}}{1+ \hat{\mu}' \hat{\Omega}^{-1} \hat{\mu}}  \sim F_{N,T-N-K},
\end{equation}
where $T$ is the number of time periods, $N$ is the number of portfolios,
$K$ is the number of factors, $\hat{\Sigma}$ is the residual covariance matrix, $\hat{\alpha}$ is the vector of alphas, $\hat{\mu}$ is the vector of average factor returns and $\hat{\Omega}$ is the
covariance matrix of factors. Note that both $\hat{\Sigma}$  and $\hat{\Omega}$ must be estimated with the maximum likelihood estimator (biased version). The tests will be performed four times on four series of 20 portfolios constructed according to the cyber score, the size, the book-to-market ratio, and the market beta of involved firms. 

\subsubsection{Bayesian approach}
\cite{BarillasShanken2018} introduces three methods to test pricing factors. The first method is close to the GRS and commonly tests zero alpha for pricing factors and portfolio returns. The second method tests, for a given set of factors, if a subset of those factors is sufficient to price portfolio returns. The third method, on which we focus here, allows us to find which subset of factors, among a large given set of factors,  are the best pricing factors. It is a \quotes{relative} method, meaning that no returns are required for the test. To produce the test, they first introduce the marginal likelihood associated with a given subset of factors:
\begin{equation}\label{marginal_likelihood}
    \text{ML} = \text{ML}_U (f|Mkt) \cdot \text{ML}_R (f^*|Mkt, f) \cdot \text{ML}_R(r|Mkt, f, f^*)
\end{equation}
Where $f$ are the factors of the subset, $f^*$ are the factors excluded from the subset (but in the general set), and $Mkt$ is the market excess returns. Note that the third term can be ignored; it will later be canceled out since it is common to any subset of factors. ML$_U(Y|X)$ and ML$_R(Y|X)$ are based on the equations $Y_{t,n} = \alpha_n + X_{t}  \beta_n+ \epsilon_{t,n}$ and $Y_{t,n} =   X_{t}  \beta_n+ \epsilon_{t,n}$. They can be computed as follows:\footnote{$\overset{(T \times N)}{Y} = \overset{(1 \times N)}{\alpha} +  \overset{(T \times K)}{X} \overset{(K \times N)}{\beta} + { \overset{(T \times N)}{\epsilon}}$, note that $\alpha$ has not the correct dimension here, it is to reflect the fact that $\alpha$ is constant across t for a given n.}
\begin{equation}
    \text{ML}_U(Y|X) = \left| X'X \right|^{-\frac{N}{2}} \left| S \right|^{-\frac{T-K}{2}} Q 
\end{equation}
\begin{equation}
    \text{ML}_R(Y|X) = \left| X'X \right|^{-\frac{N}{2}} \left| S_R \right|^{-\frac{T-K}{2}}  
\end{equation}
where $|S|=|\epsilon'\epsilon |$ and $|S_R|=|\epsilon'\epsilon |$ are the determinants of the N × N cross-product matrices of associated OLS residuals ($R$ stand for restricted since $\alpha_n=0$ is imposed on the second linear equation), $T$ is the number of periods, $K$ the number of factors in the regression (number of columns in X), and N the number of endogenous variable on the RHS of the linear equations (number of columns in Y). For example, $\text{ML}_U (f^*|Mkt, f)$ could be associated with $[f^*_{1,t},f^*_{2,t}]=[\alpha_1,\alpha_2]+[Mkt_t,f_{3,t},f_{4,t}] \beta +[\epsilon^*_{1,t},\epsilon^*_{2,t}]$ with $N=2$, $K=3$ and $\beta$ a $3\times 2$ matrix and the general set containing four factors $[f_1,f_2,f_3,f_4]$ (two included, two excluded marked by $^*$ in this example). The scalar $Q$ is given by:
\begin{equation}
    Q = \left( 1 + \frac{a}{a + k} \left( \frac{W}{T} \right) \right)^{-\frac{T - K}{2}} \left( 1 + \frac{k}{a} \right)^{-\frac{N}{2}}
\end{equation}
\begin{equation}
    a=\frac{1+\hat{\mu}' \hat{\Omega}^{-1} \hat{\mu}}{T}
\end{equation}
\begin{equation}
    k=\frac{\hat{\mu}' \hat{\Omega}^{-1} \hat{\mu}}{N}(1-\text{prior}^2)
\end{equation}
\begin{equation}
    W=T\frac{\hat{\alpha}' \hat{\Sigma}^{-1} \hat{\alpha}}{1+ \hat{\mu}' \hat{\Omega}^{-1} \hat{\mu}},
\end{equation}

%
where $\hat{\Sigma}$ is the residual covariance matrix, $\hat{\alpha}$ is the vector of alphas, $\hat{\mu}$ is the vector of average $X$ factor, and $\hat{\Omega}$ is the covariance matrix of $X$ factors. Note that both $\hat{\Sigma}$ and $\hat{\Omega}$ must be estimated with the maximum likelihood estimator (biased version). Finally, the prior is an arbitrary number. \cite{BarillasShanken2018} use $1.25$, $1.5$, $2$, and $3$ in their empirical test. Intuitively, the prior help to set $k$, the expected increment to the squared Sharpe ratio $Sh(X)^2=\hat{\mu}' \hat{\Omega}^{-1} \hat{\mu}$ from the addition of one more factor. Once the relevant marginal likelihoods are computed, the probability $p_j$ associated with a subset of factors $M_j$ being better pricing factors than other subsets is given by:
\begin{equation}\label{BGRS_proba}
    p_j= \frac{\text{ML}_j \times P(M_j)}{\sum_i \text{ML}_i \times P(M_i)},
\end{equation}
where ML$_j$ is the marginal likelihood associated with the subset $M_j$ and $P(M_j)$ is the prior probability of the subset $M_j$. In general, \cite{BarillasShanken2018} advise all prior probabilities to be constant and equal since there is no particular reason to favor a specific subset of factors. Note that the third term in Eq. \ref{marginal_likelihood} cancels at this last step. 

Following this methodology in \cite{BarillasShanken2018}, $p_j$ can be computed with subsets, including the cyber score as a factor and others without, to compare its pricing ability.

\section{Results}
\subsection{Clustering of MITRE ATT\&CK}

\begin{landscape}
 \begin{figure}[H] 
 \centering
    \noindent\makebox[\textwidth]{%
    \centering
    \includegraphics[width=1.4\textwidth,keepaspectratio]{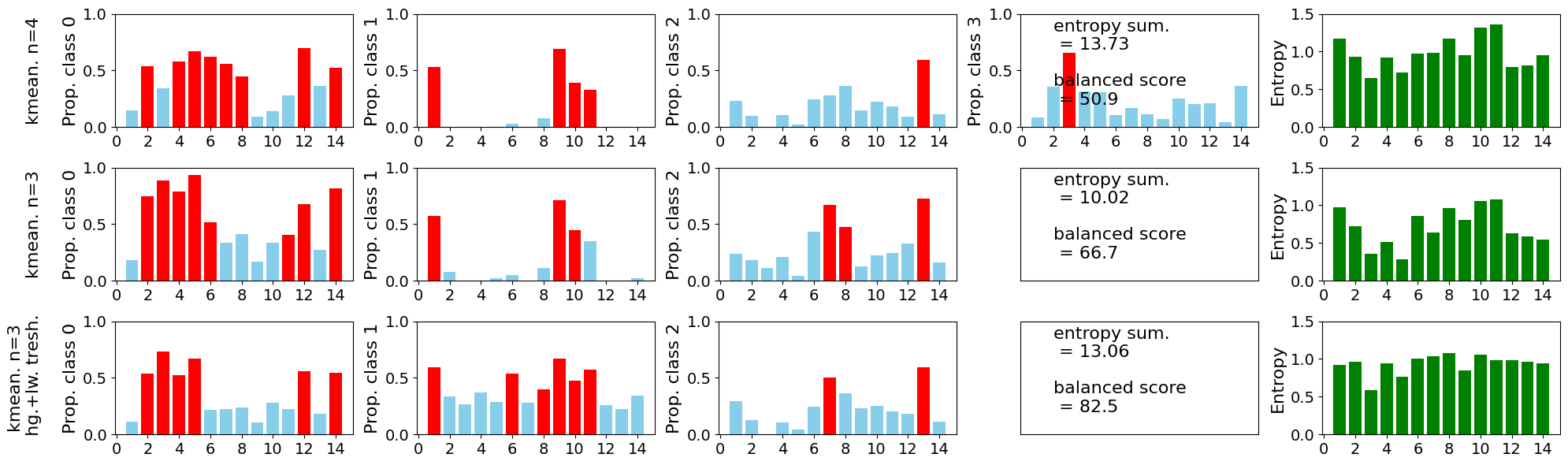}
    }
    \caption[\footnotesize Clustering results part.1]{\textbf{Clustering results part.1}}\label{clustering1}\bigskip
    \footnotesize{
    \begin{flushleft}
    This figure presents the results of each clustering method indicated on the left. The figure in red and blue represents $P(sub_j)_i$, the proportion of paragraphs of sub-cluster (tactic) j belonging to cluster (super tactic) i. The 14 sub-cluster labels are on the x-axis of each figure, and the cluster labels correspond to the columns (class 0 to 3, here). If the proportion is in red, it means it is the highest in the cluster (in other clusters/columns, the same sub-cluster will be in blue). I also report the entropy sum and the balanced score on the figure for each method. Finally, the individual Shannon entropy of each sub-cluster is reported in green in the last column. In the name of the method, I also indicate the hyperparameters of the method. Here, $n$ corresponds to the number of clusters imposed by the k-means method. \quotes{hg. tresh.} and \quotes{lw. tresh.} corresponds to a change applied to the similarity matrix. If the value in the similarity matrix is lower than 0.25, it is changed to 0 (lower threshold), and if the similarity is higher than 0.85, it is changed to 0.5 (higher threshold). In part.2 and part.3 \quotes{egn} corresponds to the $K$ eigenvectors in the spectral clustering. I also made the output clusters of each method match. Hence, the comparison is simpler (otherwise, what the Louvain method called cluster 2 is not necessarily cluster 2 for the k-means method). The following list shows the corresponding number of each tactic : 1: Persistence, 2: Command and Control, 3: Impact, 4: Initial Access, 5: Resource Development, 6: Collection, 7: Exfiltration, 8: Credential Access, 9: Privilege Escalation, 10: Execution, 11: Defense Evasion, 12: Reconnaissance, 13: Lateral Movement, 14: Discovery.
    \end{flushleft}} 
\end{figure}
\end{landscape}

\begin{landscape}
 \begin{figure}[H] 
 \centering
    \noindent\makebox[\textwidth]{%
    \centering
    \includegraphics[width=1.4\textwidth,keepaspectratio]{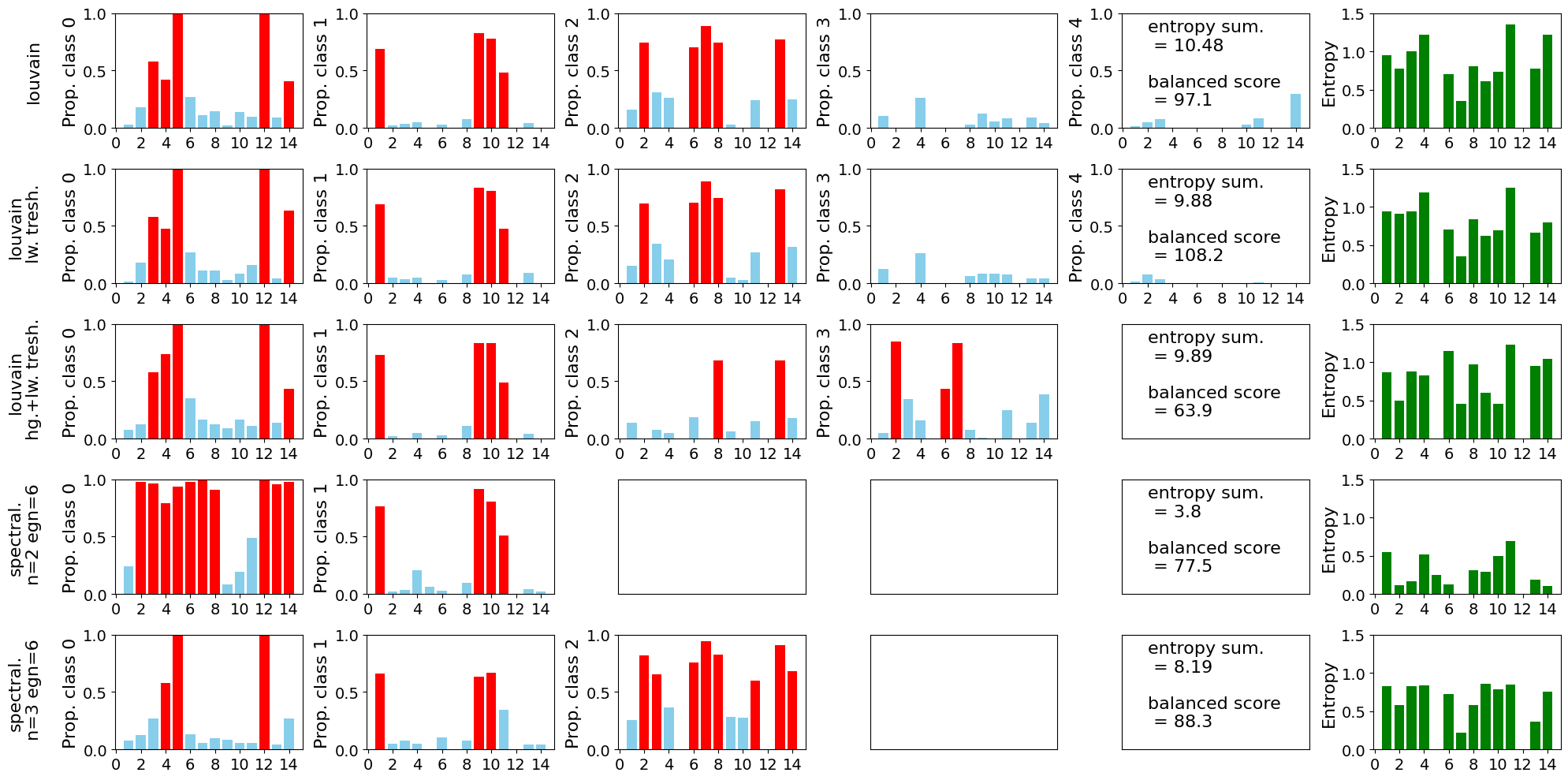}
    }
    \caption[\footnotesize Clustering results part.2]{\textbf{Clustering results part.2}}\label{clustering2}\bigskip
    \footnotesize{
    \begin{flushleft}

    \end{flushleft}} 
\end{figure}
\end{landscape}

\begin{landscape}
 \begin{figure}[H]
 \centering
    \noindent\makebox[\textwidth]{%
    \centering
    \includegraphics[width=1.4\textwidth,keepaspectratio]{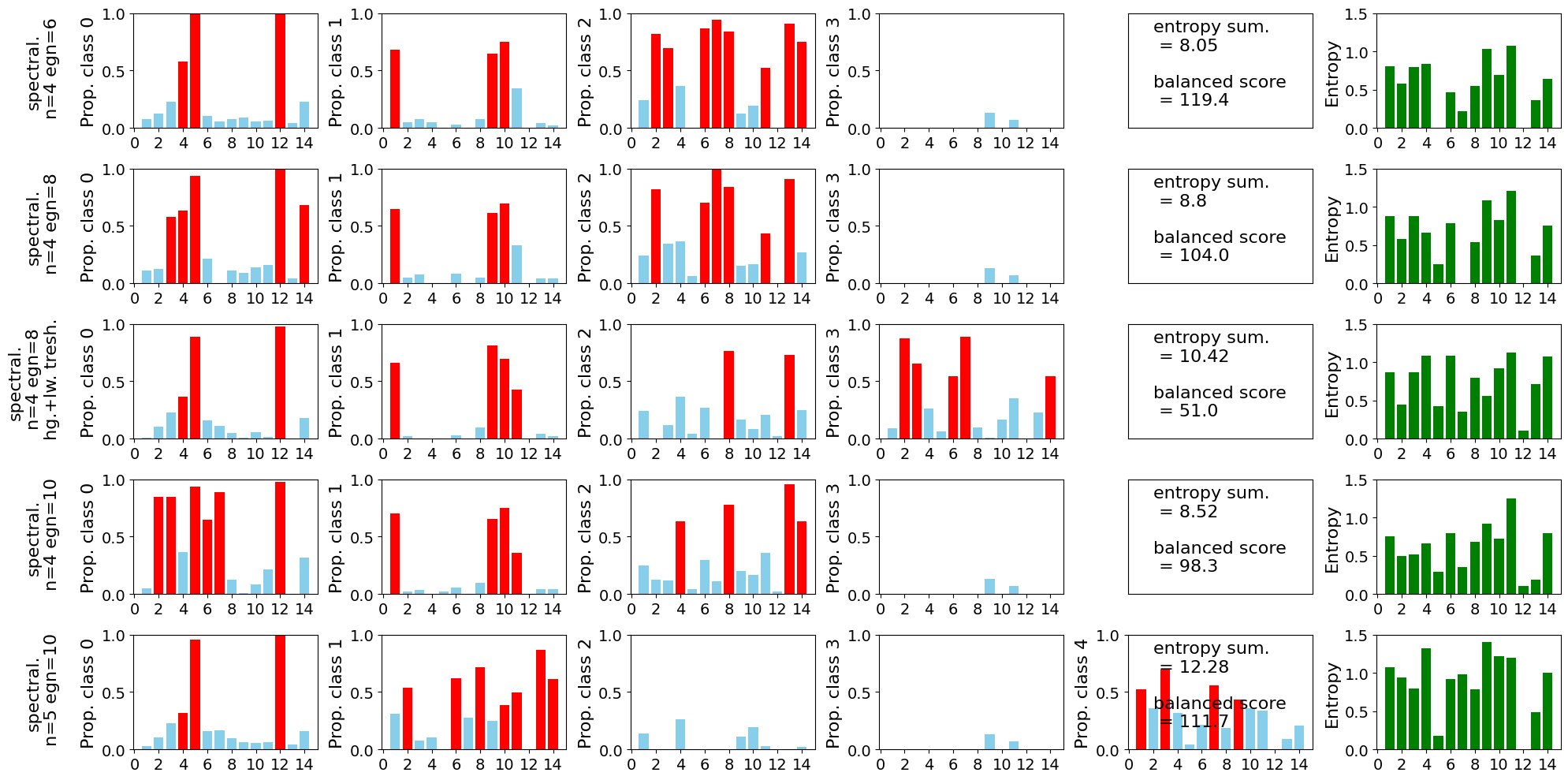}
    }
    \caption[\footnotesize Clustering results part.3]{\textbf{Clustering results part.3}}\label{clustering3}\bigskip
    \footnotesize{
    \begin{flushleft}
    
    \end{flushleft}} 
\end{figure}
\end{landscape}

\begin{figure}[H] 
    \noindent\makebox[\textwidth]{%
    \includegraphics[width=0.9\textwidth,keepaspectratio]{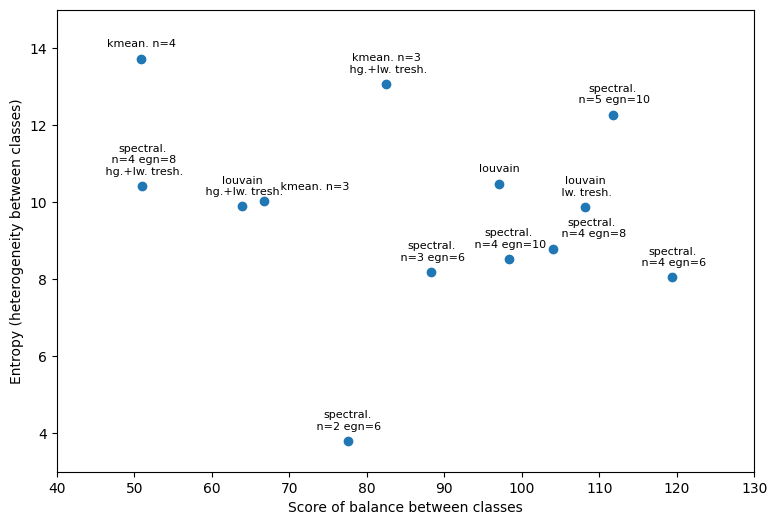}
    }
    \caption[\footnotesize Comparison of clustering scores: Entropy sum and Balanced score]{\textbf{Comparison of clustering scores: Entropy sum and Balanced score}}\label{clustering_scores}\bigskip
    \footnotesize{Each clustering method of Figure \ref{clustering1}, \ref{clustering2}, and \ref{clustering3} is presented here using their respective entropy sum and balanced score. Recall that the aim was to reduce both scores to distinguish the best clustering method. Also, note that there is no guideline regarding what additional amount
is optimal to forfeit to the entropy sum to lower the balanced score and inversely. 
}
\end{figure}

We apply the clustering methods on the cosine similarity matrix created from MITRE ATT\&CK paragraphs vector embeddings. This allows for identifying the relevant sub-cyber score tied to previously mentioned super tactics (command and data manipulation, credential movement, persistence and evasion, and preparation and reconnaissance). We report the results of various attempts with different clustering methods in Figures \ref{clustering1}, \ref{clustering2}, and \ref{clustering3}. The K-means methods provide a coherent but crude initial structure that we report in Figure \ref{clustering1}. Indeed, the paragraphs tend to be well spread across the super tactics (clusters) but at the cost of heterogeneity, with the exclusivity of a tactic in a super tactic being inexistent. This results in a low balanced score at the cost of entropy, as depicted in Figure \ref{clustering_scores}. 

Figure \ref{clustering2} shows the performance of the Louvain method. This method greatly improves the heterogeneity, especially with tactics 5 and 12 (resource development and reconnaissance) exclusive to cluster 1 (the super tactic: preparation and reconnaissance). However, not putting a threshold on the inputted similarity matrix component induces the Louvain method to create two superfluous clusters. Hence, we include those restrictions. Indeed, when comparing two paragraphs of MITRE ATT\&CK, it is not uncommon to encounter sentences with similar structures for different semantic content. Thus, we tone down the similarity of a highly too similar paragraph with a higher threshold. Conversely, we define a lower threshold such that similarities that are too low and, therefore, most likely noise that reflects no similarity are set to zero. 

The last method can be seen as a safeguard for the output of the Louvain method. Applying the spectral clustering method, we retrieve the structure previously encountered with higher heterogeneity than with K-means. If the hyperparameters are correctly tuned, the output is similar to the Louvain method's, particularly for $n=4$ and $egn=6$. We report the results in Figure \ref{clustering3}. Including more dimensions (higher $egn$ value) adds noise and decreases the clustering quality.

Finally, we select the output of the Louvain method as a baseline to group the tactics without splitting them across super tactics. Although Figure \ref{clustering_scores} shows that outputs of other methods may be slightly better, we favor the Louvain method since no additional hyperparameters tuning is required.

\begin{center}
[Insert Figure 5 here]
\end{center}

\begin{center}
[Insert Figure 6 here]
\end{center}

\begin{center}
[Insert Figure 7 here]
\end{center}

\begin{center}
[Insert Figure 8 here]
\end{center}

\noindent This yields the following cluster/super tactics. we name each of them after their content: 

\textbf{Preparation and Reconnaissance}: 
This super tactic encompasses adversaries' tactics to prepare and gather information before launching an attack. \textbf{Impact} involves actions that disrupt, destroy, or manipulate systems and data to achieve the attacker’s objectives. \textbf{Initial Access} includes techniques adversaries use to gain an initial foothold within a network, such as exploiting vulnerabilities or spear phishing. \textbf{Resource Development} entails the acquisition of resources like infrastructure, tools, and credentials necessary to support operations. \textbf{Reconnaissance} involves gathering information about the target environment to identify potential entry points and vulnerabilities. \textbf{Discovery} refers to techniques used to explore and map the target environment, such as network scanning and enumeration.\newline

\textbf{Persistence and Evasion}: 
Once inside a target network, adversaries employ these tactics to maintain their foothold and avoid detection. \textbf{Persistence} ensures the attacker can maintain access even if the system is rebooted or credentials are changed by installing malware or creating rogue accounts. \textbf{Privilege Escalation} involves gaining higher-level permissions to access more sensitive information and critical systems. \textbf{Execution} refers to running malicious code on a victim system, often necessary to carry out the attacker’s objectives. \textbf{Defense Evasion} includes a variety of methods to avoid detection and thwart defensive measures, such as disabling security software, obfuscating code, or using fileless malware.\newline

\textbf{Credential Movement}: 
This group focuses on techniques used to steal and use credentials to move within a network. \textbf{Credential Access} involves obtaining account names, passwords, and other secrets that allow attackers to authenticate themselves as legitimate users. Techniques include keylogging, credential dumping, and brute force attacks. \textbf{Lateral Movement} is moving through a network to find and access additional targets or more valuable data. This can be done using remote services, exploiting trust relationships, or leveraging legitimate credentials to access other systems and resources.\newline

\textbf{Command and Data Manipulation}: 
In this phase, adversaries exert control over compromised systems and manipulate data to achieve their goals. \textbf{Command and Control} involves establishing a communication channel with the compromised environment to issue commands and control malware. This can be achieved through web traffic, DNS, or custom communication protocols with command servers. \textbf{Collection} refers to gathering sensitive information from compromised systems, such as capturing screenshots, logging keystrokes, or accessing stored files. \textbf{Exfiltration} involves transferring the collected data from the target network to an external location controlled by the adversary, often using encrypted channels or covert methods to avoid detection.\newline

\newpage

\subsection{Cyber scores statistical descriptions}

\begin{figure}[H] 
    \noindent\makebox[\textwidth]{%
    \includegraphics[width=0.9\textwidth,keepaspectratio]{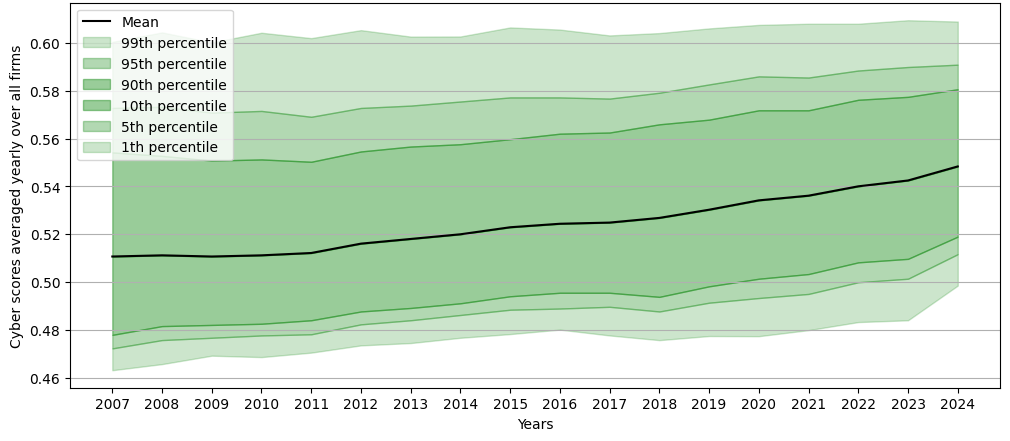}}
    \caption[\footnotesize Evolution of the overall cyber score averaged yearly over all firms ]{\textbf{Evolution of the overall cyber score averaged yearly over all firms}}\label{overall_averaged_yearly}\bigskip
    \footnotesize{The figure shows the evolution of the overall cyber score over all firms yearly. Each year provides a distribution of the cyber score over all firms that can be sorted to provide percentiles of interest and the averaged cyber score for a given year.}
\end{figure}

\begin{figure}[H] 
    \noindent\makebox[\textwidth]{%
    \includegraphics[width=0.9\textwidth,keepaspectratio]{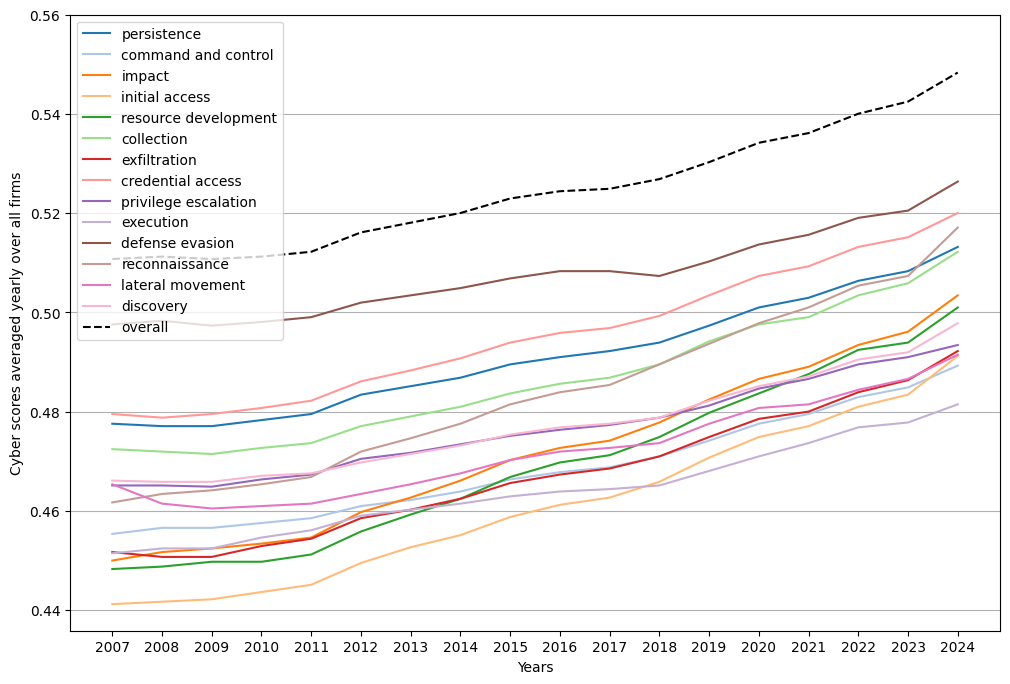}}
    \caption[\footnotesize Evolution of the 14 sub-cyber scores averaged yearly over all firms]{\textbf{Evolution of the sub-cyber scores related to the 14 tactics averaged yearly over all firms}}\label{14_averaged_yearly}\bigskip
    \footnotesize{The figure shows the evolution of the 14 sub-cyber scores averaged over all firms yearly. The overall cyber score is also included to allow comparison.}
\end{figure}

\begin{figure}[H] 
    \noindent\makebox[\textwidth]{%
    \includegraphics[width=0.9\textwidth,keepaspectratio]{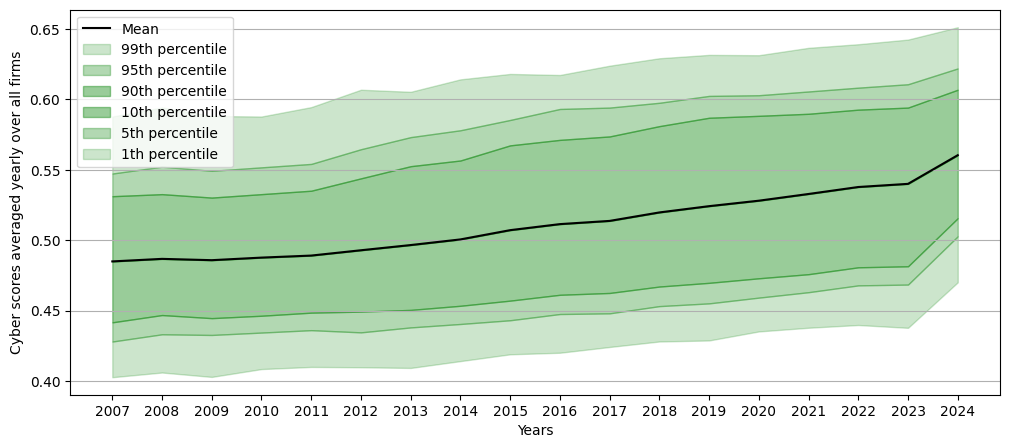}
    }
    \caption[\footnotesize Evolution of the cyber sentiment score averaged yearly over all firms ]{\textbf{Evolution of the cyber sentiment score averaged yearly over all firms}}\label{sentiment_averaged_yearly}\bigskip
    \footnotesize{The figure shows the evolution of the cyber sentiment score over all firms yearly. Each year provides a distribution of the cyber score over all firms that can be sorted to provide percentiles of interest and the averaged cyber score for a given year.}
\end{figure}

\begin{figure}[H] 
    \noindent\makebox[\textwidth]{%
    \includegraphics[width=0.9\textwidth,keepaspectratio]{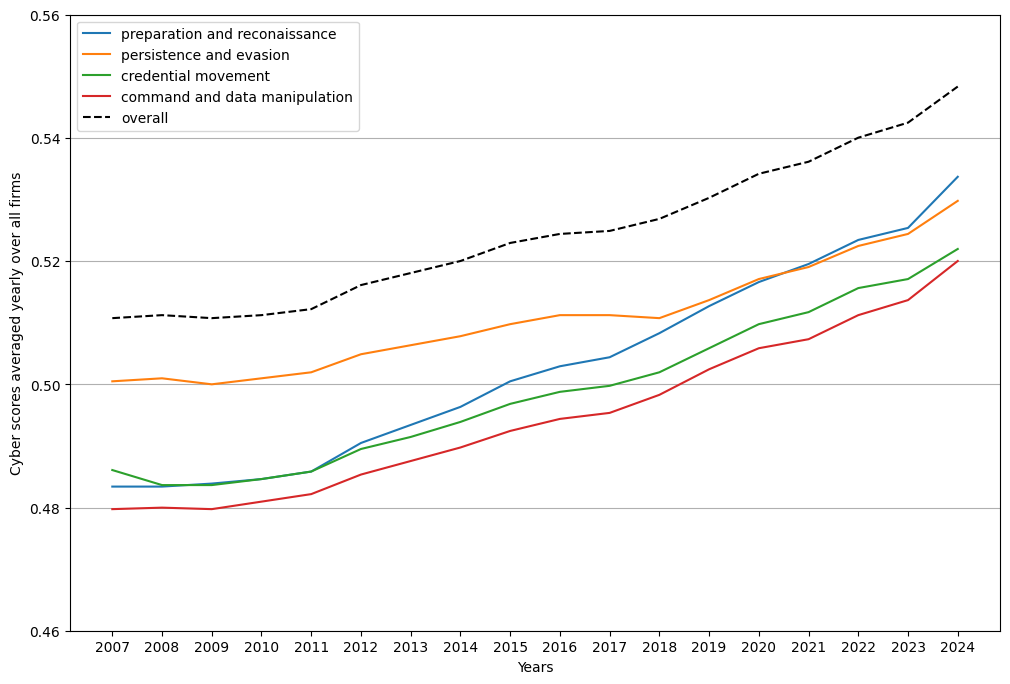}
    }
    \caption[\footnotesize Evolution of the four sub-cyber scores averaged yearly over all firms]{\textbf{Evolution of the sub-cyber scores related to the four super-tactics averaged yearly over all firms}}\label{super_averaged_yearly}\bigskip
    \footnotesize{The figure shows the evolution of the four sub-cyber scores averaged over all firms yearly. The overall cyber score is also included to allow comparison.}
\end{figure}

From the identified super tactics, we construct the cyber scores from the 10-Ks of each firm through the years. Table \ref{cyber_stats} presents the statistics related to each cyber score (the 14 tactics of MITRE ATT\&CK, the four super tactics, the overall score, and the cyber sentiment score). Although their distribution appears similar, several facts must be considered. First, the statistics are for the whole sample, but the distribution is time-varying as Figures \ref{overall_averaged_yearly}, \ref{14_averaged_yearly}, \ref{sentiment_averaged_yearly}, and \ref{super_averaged_yearly} suggest. This means that cyber scores evolving at different rates could be misrepresented. Second, the cosine similarity implies, in theory, a distribution ranging from $-1$ to $1$, whereas the scores are much more narrowly distributed empirically. Thus, the slight variation observed in Table \ref{cyber_stats} is more meaningful than simple noise. 

Two additional aspects must also be reported. First, some tactics lose relevance in the 10-Ks over time, and evidence of the cyber scores reflecting cyber risk has yet to be presented. However, this first feature is encouraging since the cyber scores are evolving differently, showcasing a shift of cyber-related information in the 10-Ks. Second, the cyber sentiment score has a higher 99 percentile than the other score, implying that taking out non-risk-related scores effectively removes points previously belonging to the top one percentile.

\begin{center}
[Insert Table 3 here]
\end{center}

\begin{center}
[Insert Figure 9 here]
\end{center}

\begin{center}
[Insert Figure 10 here]
\end{center}

\begin{center}
[Insert Figure 11 here]
\end{center}

\begin{center}
[Insert Figure 12 here]
\end{center}

\begin{center}
[Insert Figure 13 here]
\end{center}

\begin{figure}[H] 
    \noindent\makebox[\textwidth]{%
    \includegraphics[scale=0.6]{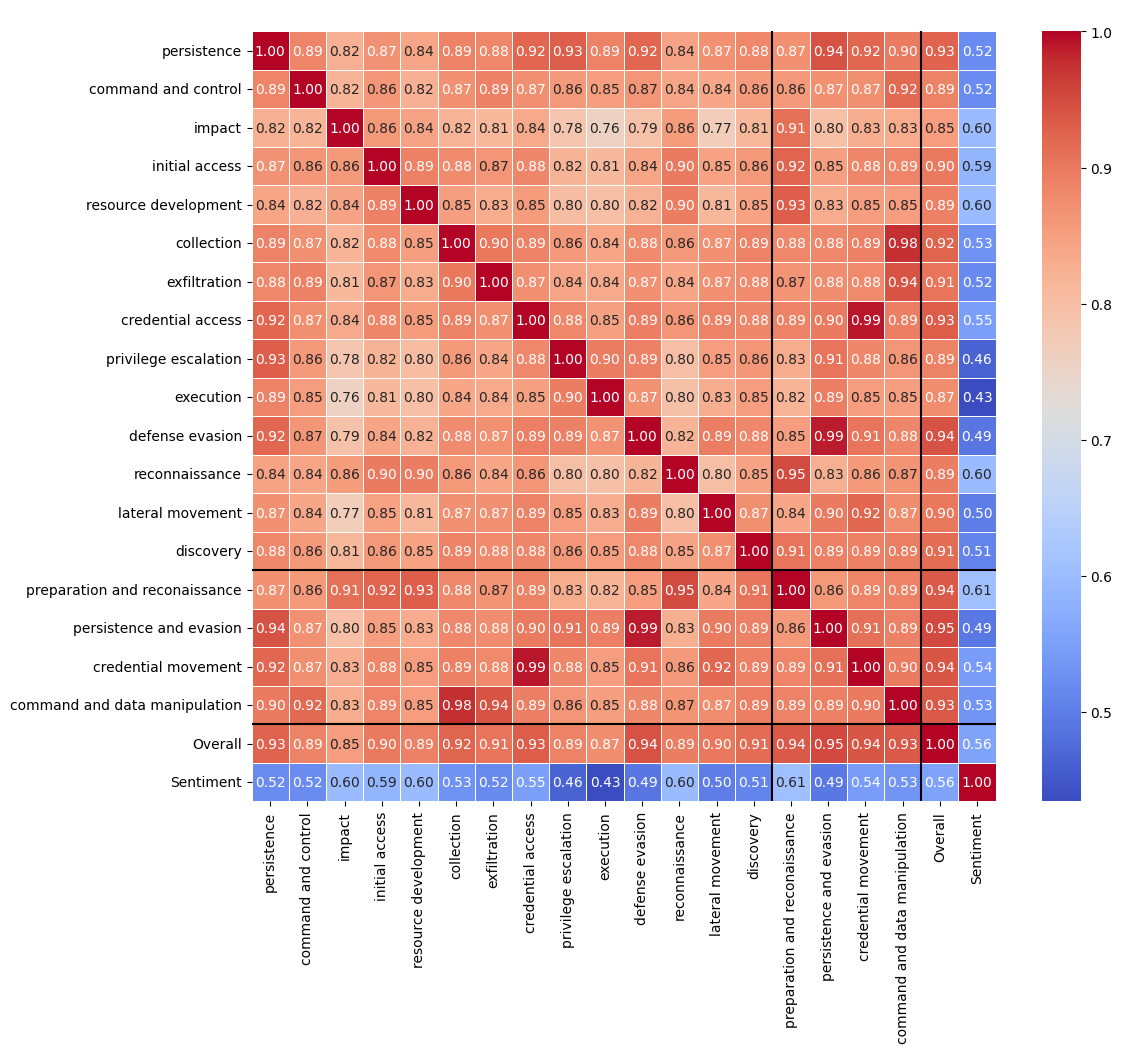}
    }
    \caption[\footnotesize Correlations of all cyber scores]{\textbf{Correlations of all cyber scores}}\label{corr_all_cyber}\bigskip
    \footnotesize{Firm-wise correlations of the sub-cyber scores of the 14 MITRE ATT\&CK tactics, the 4 aggregated sub-cyber scores of the super-tactics, as well as the overall cyber score and the cyber sentiment score are presented here.}
\end{figure}

We present the correlation between cyber scores at the firm level (non-aggregated) in Figure \ref{corr_all_cyber}. Unsurprisingly, the correlations between all scores are high, except for the sentiment score, which differs in its construction.  This is expected since all scores come from the same doc2vec neural network output.

\newpage

\subsection{Cyber scores and financial characteristics}

To ensure that the cyber scores are innovative and not the combination of other existent characteristics of the firm, we present the linear regression of the cyber scores of interest in Tables \ref{tab:determinants_cyber_overall}, \ref{tab:determinants_cyber_sentiment}, \ref{tab:determinants_cyber_command_and_data_manipulation}, \ref{tab:determinants_cyber_credential_movement}, \ref{tab:determinants_cyber_persistence_and_evasion}, and \ref{tab:determinants_cyber_preparation_and_reconaissance}.  Compared to \cite{CelenyMaréchal2023}, we add the following variables: readability, secret, risk length table, volume per capital, and humans per capital. we describe all variables in Table \ref{tab:variable_descriptions}. The first three variables were part of the tested explanatory variables in \cite{FlorackisLoucaMichaelyWeber2023}. Including the risk length table shows the critical improvement made with the cyber score of this paper. Indeed, \cite{FlorackisLoucaMichaelyWeber2023} report t-statistics of $40.80$ and $20.59$ for models 1 and 2, respectively. Those t-statistics are significantly lower, improving our score's independence with non-semantic variables.

We include a new explanatory variable in the models with the following underlying idea: If a limited number of employees are in a firm with highly valuable assets, those assets are more likely to be technological and could be cyber risky. This risk would then be reported in the 10-Ks and thus be reflected in the cyber score. This choice proves relevant as the Tables report t-statistics close to 10 for all cyber scores. The coefficient negative sign supports the view that the lower the human capital ratio, the higher it should be reflected on the cyber score. \footnote{Additionally, we compute the covariance and correlation of each cyber score with the idiosyncratic volatility of firms. The results are presented in Appendix \ref{idiosyncratic_cov_corr}. We thank Prof. Julien Hugonnier for this comment.} 

The statistical significance of other coefficients does not depart too much from the previous studies, with most of the variables being statistically non-significant or with the same sign as in \cite{CelenyMaréchal2023}, especially at the firm level. Note that, despite adding new variables with higher t-statistics, the $R^2$ within is still low. It shows that additional variables cannot fully explain the different cyber scores. Furthermore, different t-statistics are obtained for each cyber score for the same coefficient. This suggests that each score could proxy for an intrinsically different risk for the firm.

\begin{figure}[H] 
    \noindent\makebox[\textwidth]{%
    \includegraphics[scale=0.5]{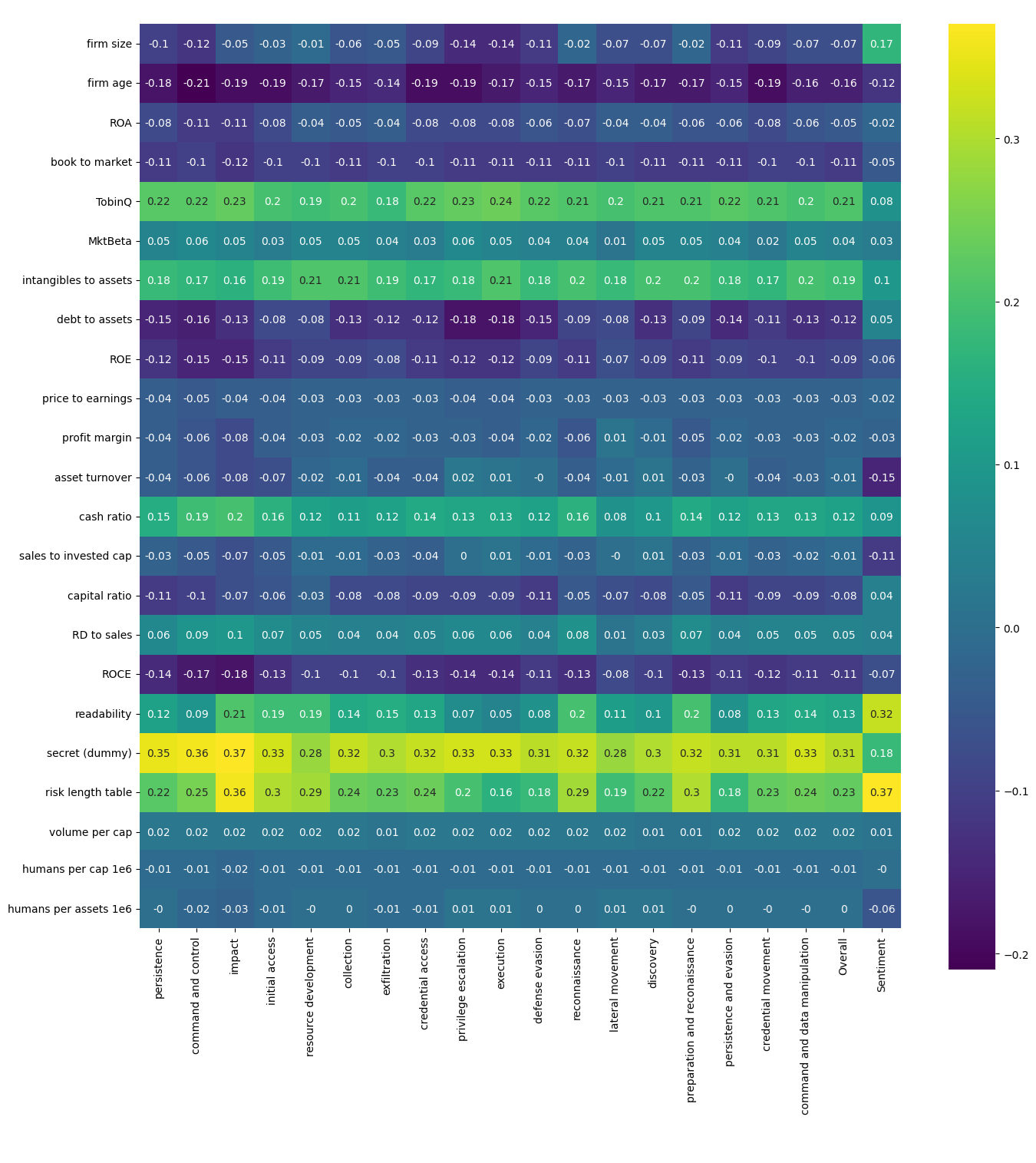}}
    \caption[\footnotesize Correlations of all cyber scores with financial characteristics]{\textbf{Correlations of all cyber scores with financial characteristics}}\bigskip
    \footnotesize{Firm-wise correlations of the sub-cyber scores of the 14 MITRE ATT\&CK tactics, the four aggregated sub-cyber scores of the super-tactics, as well as the overall cyber score and the cyber sentiment score with the financial characteristics of the firms.}
     \label{fig:corr_all_cyber_with_financial_variables}
\end{figure}

Figure \ref{fig:corr_all_cyber_with_financial_variables} displays the correlation of all cyber scores with the mentioned variables. As expected, variables with generally higher t-statistics tend to correlate more to the cyber scores. However, The correlation with \quotes{secret} must be taken cautiously since it is a dummy variable and the cyber score is close to 0.5; the correlation may be spurious or, at best, not informative.

\begin{figure}[H] 
    \noindent\makebox[\textwidth]{%
    \includegraphics[scale=0.5]{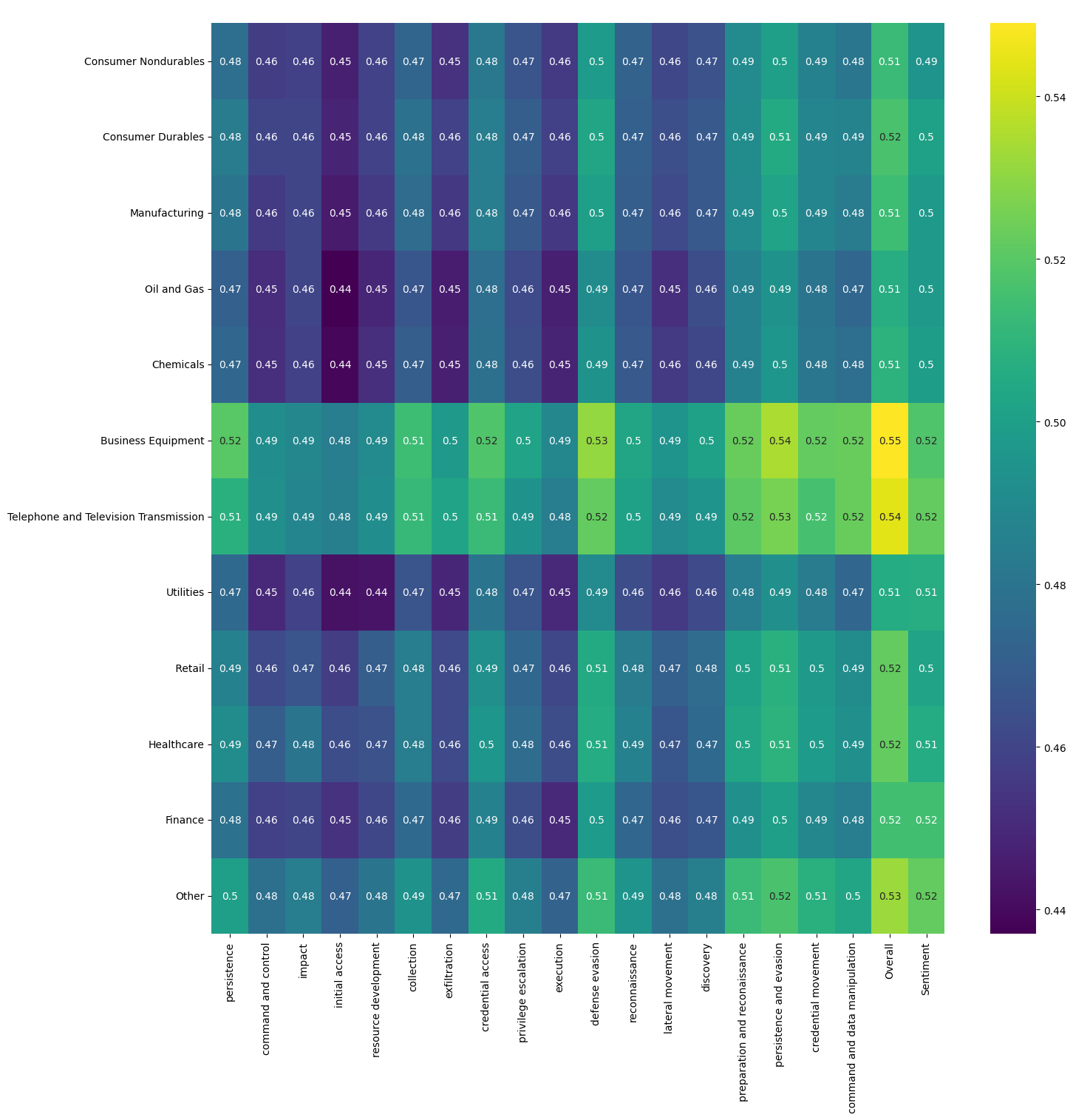}
    }
    \caption[\footnotesize Average cyber score across industries]{\textbf{Average cyber score across industries}}\bigskip
    \footnotesize{The respective average cyber scores of each firm (from 2009-Q1 to 2023-Q4) are computed and averaged across the industry the firms belong to, thus obtaining the different averaged cyber scores aggregate for each industry. Firms are classified into industries using the Fama-French 12 industry classification.}
     \label{fig:average_across_industry}
\end{figure}

Figure \ref{fig:average_across_industry} shows the average cyber scores across industries. The overall score is always higher than other scores when controlling for industry. This was already the case in Table \ref{cyber_stats}. The cyber sentiment score displays much fewer differences across industries when compared to other scores. This suggests that the score does not contain additional information or even destroy some of it at the industry level. As mentioned in \cite{CelenyMaréchal2023}, industries that rely more heavily on technology, like Business Equipment or Telephone and Television Transmission, potentially report their cyber risk and thus have higher cyber scores. One can also observe that the different cyber scores vary differently across the industry, further highlighting the potential changes in the source of cyber risk disclosed in the 10-Ks. 

\begin{center}
[Insert Table 4 here]
\end{center}

\begin{center}
[Insert Table 5 here]
\end{center}

\begin{center}
[Insert Table 6 here]
\end{center}

\begin{center}
[Insert Table 7 here]
\end{center}

\begin{center}
[Insert Table 8 here]
\end{center}

\begin{center}
[Insert Table 9 here]
\end{center}

\begin{center}
[Insert Figure 14 here]
\end{center}

\begin{center}
[Insert Figure 15 here]
\end{center}

\subsection{Univariate sorts}

This section answers the following questions: Does a portfolio sorted according to a given cyber score display a structure in its returns? Are the pricing factors usually found in the literature enough to fully explain (not generate alphas on) the return of the cyber-based portfolios? 

Tables \ref{univariate_overall}, \ref{univariate_command_and_data_manipulation}, \ref{univariate_credential_movement}, \ref{univariate_persistence_and_evasion}, and \ref{univariate_preparation_and_reconaissance} all display increasing average excess returns, along with the associated cyber score, with P5 being the portfolio of firms with the highest cyber score. If a given cyber score is an adequate proxy for the associated cyber risk, it implies that taking additional cyber risk grants additional returns. This idea is further explored in the next section.

In Table \ref{univariate_sentiment}, we report the average returns of portfolios sorted on the cyber sentiment score. Despite all returns being statistically significant at the 1\% level, we cannot observe a monotonic increase in average returns across the scores, and P3 displays the highest average excess returns in P3. This could indicate that the score does not reflect any risk or that investors are unaware of the type of risk it reflects. We provide additional insight in the following section.

Second, Tables \ref{univariate_overall}, \ref{univariate_command_and_data_manipulation}, \ref{univariate_credential_movement}, \ref{univariate_persistence_and_evasion}, and \ref{univariate_preparation_and_reconaissance}, show that the linear regression using the pricing factors respectively contained in the CAPM, FFC, and FF5 models all grant a statistically significant alpha for P5 only (with statistical significance at the 5\% level for CAPM and 1\% level otherwise). Alphas are increasing monotonically across the portfolios after controlling for other sources of risk associated with the pricing factors involved. With this evidence in mind, the alphas could partly reflect the cyber risk of each portfolio. This is, however, not the case for the cyber sentiment score in Table \ref{univariate_sentiment} since there is no overall strong statistical significance or increasing trends when portfolios are sorted across this variable.

\begin{center}
[Insert Table 10 here]
\end{center}

\begin{center}
[Insert Table 11 here]
\end{center}

\begin{center}
[Insert Table 12 here]
\end{center}

\begin{center}
[Insert Table 13 here]
\end{center}

\begin{center}
[Insert Table 14 here]
\end{center}

\begin{center}
[Insert Table 15 here]
\end{center}

\newpage

\subsection{Double sorts}

We aim to determine if organizing quarterly portfolios first based on specific characteristics and then based on the cyber score results in a structure in their average excess returns. The idea is that controlling for additional characteristics rejects the hypothesis that the cyber scores are a proxy for another firm variable. Thus, the cyber score would capture the cyber risk exposure and the associated additional returns. In other words, the increasing cyber score, which should reflect the increasing cyber risk, still displays increasing related returns even if the set of firms to analyze is already organized and structured according to another characteristic unrelated to cyber risk.

Table \ref{double_sort_table} displays the average returns of various double-sorted portfolios. Notably, there is a clear increasing trend in average returns as the overall cyber score quintile increases, which contrasts with the findings of \cite{CelenyMaréchal2023}, where the trend was less pronounced. In this analysis, only the first quintile of the book-to-market ratio does not consistently show an increase. Note that Q3 of market beta and Q5 of the size have a difference problem of 0.01\% at Q3. However, this quantitatively marginal result may be spurious. 

For the cyber sentiment score, as previously observed in Table \ref{univariate_sentiment}, there are no additional returns for an increase in the score, so the cyber sentiment score certainly does not reflect any risk. 

The returns obtained with other cyber scores display an interesting aspect. Command and data manipulation, credential movement, and persistence and evasion strongly suggest a monotonic increasing trend and, therefore, cyber risks premia, except for the lower quintile, where the conclusion might seem slightly less evident. However, this is not the case for preparation and reconnaissance, for which the trend is nonmonotonic almost everywhere. This could suggest two things. It could be that investors do not acknowledge the risk this cyber score reflects. Or, it could be that preparation and reconnaissance do not reflect any risk. 

\begin{center}
[Insert Table 16 here]
\end{center}

\subsection{Cross-sectional tests}

We test whether a cyber score increase drives a return increase in cyber-based portfolios, controlling for other well-known pricing factors using the regression method described in \cite{FamaMacBeth1973}. Table \ref{FamaMcBeth_overall}, \ref{FamaMcBeth_sentiment}, \ref{FamaMcBeth_command_and_data_manipulation}, \ref{FamaMcBeth_credential_movement}, \ref{FamaMcBeth_persistence_and_evasion}, and \ref{FamaMcBeth_preparation_and_reconaissance} displays the results for each cyber score using a different pricing model that include the cyber score. 

The cyber sentiment score probably reflects no meaningful reality regarding the firms; therefore, constructing portfolios based on it reveals no particular structure, as observed in Table \ref{FamaMcBeth_sentiment}. No risk premia is observed for any of the involved pricing factors (including the cyber sentiment score), and no statistically significant alpha exists.

The overall cyber score on Table \ref{FamaMcBeth_overall} displays positive additional returns for an increased cyber score that is statistically significant at the 10\% level when included as the only explanatory variable or with the market factor and at the 5\% level when included with the pricing factors from \cite{FamaFrench1992}. However, a collinearity problem appears for the fifth model using additional factors from \cite{FamaFrench2015}. The cyber score aggregated for all the firms of the cyber-based portfolios constructs a factor that might be colinear to CMA. Therefore, the statistical significance of the cyber score is affected and strongly reduced. When compared to \cite{CelenyMaréchal2023}, they appear to not suffer from collinearity and have a less high adjusted $R^2$. 

The command and data manipulation score on Table \ref{FamaMcBeth_command_and_data_manipulation} grants similar results, also suffering from collinearity. On the contrary, the remaining score in Tables \ref{FamaMcBeth_credential_movement}, \ref{FamaMcBeth_persistence_and_evasion} and \ref{FamaMcBeth_preparation_and_reconaissance} do not display collinearity and their respective cyber score appear statistically significant at the 5\% level. Note that on all Tables (except for the cyber sentiment score), the coefficients of the cyber score appear positive, further pointing toward cyber scores that effectively reflect cyber risks rewarded on the market. A standard deviation of the overall cyber score ($0.03$ in Table \ref{cyber_stats}) generates an additional return of $0.03 \cdot 0.04 = 0.12\%$ compared to \cite{CelenyMaréchal2023} with $0.18\%$.

\begin{center}
[Insert Table 17 here]
\end{center}

\begin{center}
[Insert Table 18 here]
\end{center}

\begin{center}
[Insert Table 19 here]
\end{center}

\begin{center}
[Insert Table 20 here]
\end{center}

\begin{center}
[Insert Table 21 here]
\end{center}

\begin{center}
[Insert Table 22 here]
\end{center}

\subsection{Time series tests}

We use the GRS test to examine whether adding the long-short portfolio $P5-P1$ as a pricing factor enhances the explanatory power of the five-factor model of \cite{FamaFrench2015}. We want to test if we can globally reduce the unexplained part of returns close to zero. We conduct four GRS tests on 20 portfolios sorted quarterly, on firms' cyber score, size, market beta, and book-to-market, respectively. 

Tables \ref{GRS_overall}, \ref{GRS_sentiment}, \ref{GRS_data_and_manipulation}, \ref{GRS_credential_movement}, \ref{GRS_persistence_and_evasion} and \ref{GRS_preparation_and_reconaissance} display the four GRS tests for each cyber score of interest. All tables give similar results, with the probability of alphas being commonly zero increasing as we add the cyber factor P5-P1 and the average $R^2$ increasing. There is an exception when portfolios are sorted on size. Their associated probabilities seem to decrease, but it is important to note that the probability is already high before adding the cyber factor. It is hard to quantify if the alphas are closer to zero when they are already commonly near zero. Also, note that the difference of probabilities in the case of market beta is positive but small compared to the improvement provided by the cyber factor when explaining the return of portfolios sorted on the cyber score or the book to market.

\begin{center}
[Insert Table 23 here]
\end{center}

\begin{center}
[Insert Table 24 here]
\end{center}

\begin{center}
[Insert Table 25 here]
\end{center}

\begin{center}
[Insert Table 26 here]
\end{center}

\begin{center}
[Insert Table 27 here]
\end{center}

\begin{center}
[Insert Table 28 here]
\end{center}

\subsection{Bayesian asset pricing tests}
\begin{figure}[H]
    \centering
    \begin{subfigure}[b]{0.4\textwidth}
        \centering
        \includegraphics[width=\textwidth]{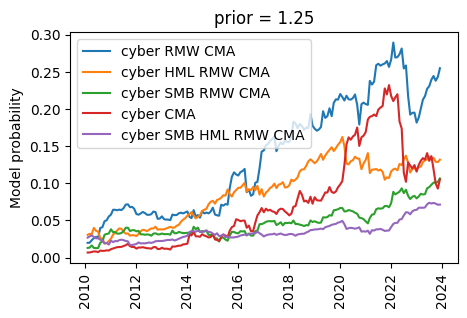}
        
    \end{subfigure}
    \hspace{1cm}
    \begin{subfigure}[b]{0.4\textwidth}
        \centering
        \includegraphics[width=\textwidth]{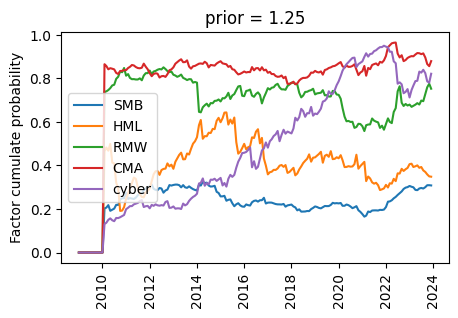}
      
    \end{subfigure}
    \caption[\footnotesize Factor model posterior probabilities using overall  cyber score]{\textbf{\small Factor model posterior probabilities using overall  cyber score}}\label{BGRS_overall}
    \bigskip
    \footnotesize{
    \begin{flushleft}
The first figure depicts the probabilities of being a better set of pricing factors for the shown subset compared to all possible subsets of factors. I present only the top five models, ranked by the probability at the end of the sample, meaning that all other subsets have lower pricing abilities than the ones presented here. HML and SMB refer to the book-to-market and size factors of \cite{FamaFrench1992}. CMA and RMW refer to the investment and operating profitability factors of \cite{FamaFrench2015}. \quotes{cyber} refers to the long-short portfolio built on the cyber score of interest (P5-P1). The prior multiple is 1.25, and the study period is from January 2010 to December 2022. The second figure shows the cumulative probabilities, \textit{i.e.} the sum of probabilities of all the pricing subsets containing the factor on a similar time range.
    \end{flushleft}} 
\end{figure}

\begin{figure}[H]
    \centering
    \begin{subfigure}[b]{0.4\textwidth}
        \centering
        \includegraphics[width=\textwidth]{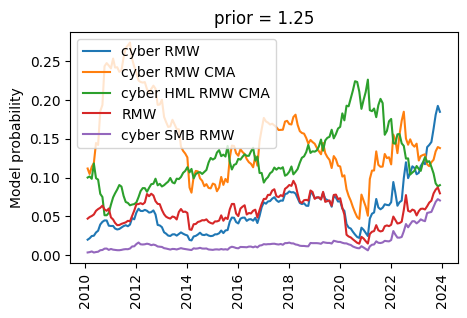}
        
    \end{subfigure}
    \hspace{1cm}
    \begin{subfigure}[b]{0.4\textwidth}
        \centering
        \includegraphics[width=\textwidth]{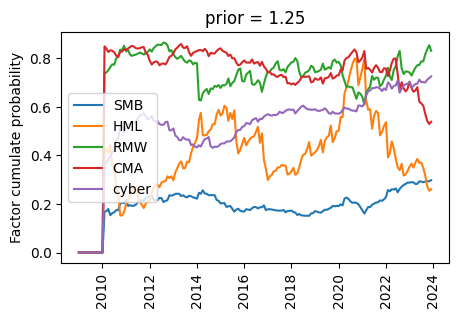}
      
    \end{subfigure}
    \caption[\footnotesize Factor model posterior probabilities using cyber sentiment score]{\textbf{\small Factor model posterior probabilities using cyber sentiment score}}
    \label{BGRS_sentiment}
\end{figure}

\begin{figure}[H]
    \centering
    \begin{subfigure}[b]{0.4\textwidth}
        \centering
        \includegraphics[width=\textwidth]{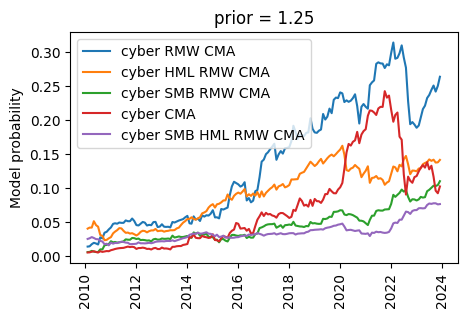}
        
    \end{subfigure}
    \hspace{1cm}
    \begin{subfigure}[b]{0.4\textwidth}
        \centering
        \includegraphics[width=\textwidth]{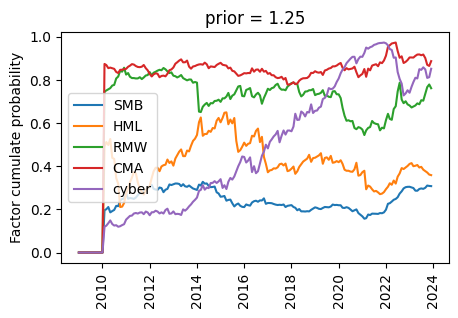}
      
    \end{subfigure}
    \caption[\footnotesize Factor model posterior probabilities using command and data manipulation cyber score]{\textbf{\small Factor model posterior probabilities using command and data manipulation cyber score}}
    \label{BGRS_command_and_data_manipulation}
\end{figure}

\begin{figure}[H]
    \centering
    \begin{subfigure}[b]{0.4\textwidth}
        \centering
        \includegraphics[width=\textwidth]{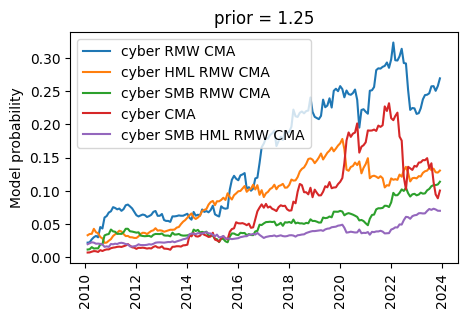}
        
    \end{subfigure}
    \hspace{1cm}
    \begin{subfigure}[b]{0.4\textwidth}
        \centering
        \includegraphics[width=\textwidth]{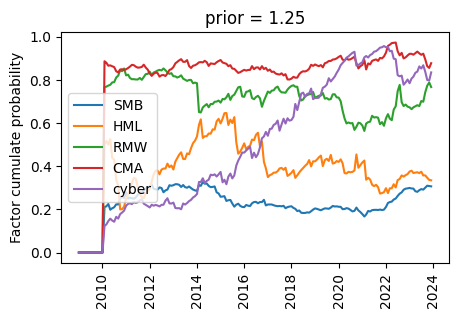}
      
    \end{subfigure}
    \caption[\footnotesize Factor model posterior probabilities using credential movement cyber score]{\textbf{\small Factor model posterior probabilities using credential movement cyber score}}
    \label{BGRS_credential_movement}
\end{figure}

\begin{figure}[H]
    \centering
    \begin{subfigure}[b]{0.4\textwidth}
        \centering
        \includegraphics[width=\textwidth]{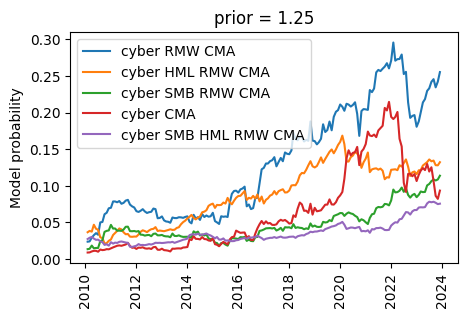}
        
    \end{subfigure}
    \hspace{1cm}
    \begin{subfigure}[b]{0.4\textwidth}
        \centering
        \includegraphics[width=\textwidth]{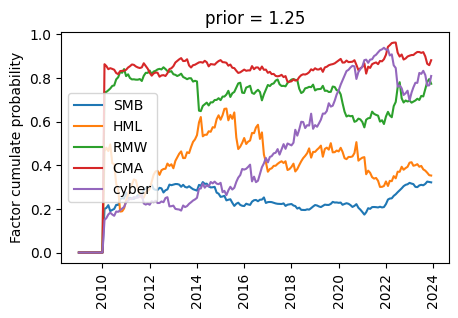}
      
    \end{subfigure}
    \caption[\footnotesize Factor model posterior probabilities using persistence and evasion cyber score]{\textbf{\small Factor model posterior probabilities using persistence and evasion cyber score}}
    \label{BGRS_persistence_and_evasion}
\end{figure}

\begin{figure}[H]
    \centering
    \begin{subfigure}[b]{0.4\textwidth}
        \centering
        \includegraphics[width=\textwidth]{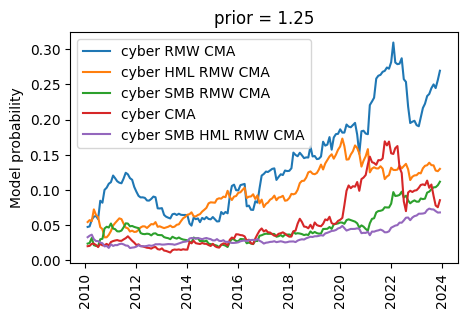}
        
    \end{subfigure}
    \hspace{1cm}
    \begin{subfigure}[b]{0.4\textwidth}
        \centering
        \includegraphics[width=\textwidth]{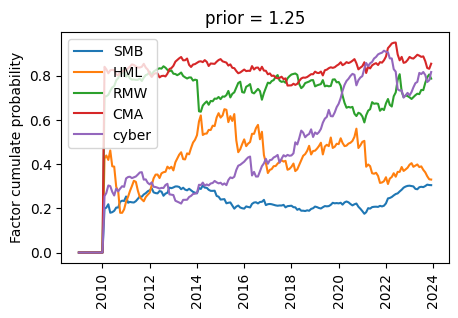}
      
    \end{subfigure}
    \caption[\footnotesize Factor model posterior probabilities using preparation and reconnaissance cyber score]{\textbf{\small Factor model posterior probabilities using preparation and reconnaissance cyber score}}
    \label{BGRS_preparation_and_reconaissance}
\end{figure}

We conduct an additional test to evaluate the cyber factor P5-P1 as a reliable pricing factor. Using the Bayesian GRS (BGRS) test described earlier, we can retrieve the best subset of pricing factors among a large set of pricing factors. Figures \ref{BGRS_overall}, \ref{BGRS_sentiment}, \ref{BGRS_command_and_data_manipulation}, \ref{BGRS_credential_movement}, \ref{BGRS_persistence_and_evasion} and \ref{BGRS_preparation_and_reconaissance} presents the BGRS test for each cyber score of interest. Keeping only the subsets with the higher probabilities at the end of the time range, one can notice that the top five subsets always include the cyber factor (except for the cyber sentiment score case, where one of the top five subsets does not include the cyber factor). When considering the cumulative probabilities associated with each pricing factor, one can see that the cyber factors have always presented a growing trend over the years, meaning that the cyber factors have become more relevant as a pricing factor over time.

\begin{center}
[Insert Figure 16 here]
\end{center}

\begin{center}
[Insert Figure 17 here]
\end{center}

\begin{center}
[Insert Figure 18 here]
\end{center}

\begin{center}
[Insert Figure 19 here]
\end{center}

\begin{center}
[Insert Figure 20 here]
\end{center}

\begin{center}
[Insert Figure 21 here]
\end{center}

\subsection{Additional tests}

In this section, we conduct further tests to expand the range and depth of understanding of the cyber scores we developed. These additional tests aim to identify any potential limitations and verify the scores' behavior in a real case. 

\begin{center}
[Insert Table 29 here]
\end{center}

In Table \ref{P5_and_P20_time_series_diff}, we display the probabilities associated with the differences in mean returns between the overall and other cyber-based portfolios. Although we give evidence that the cyber scores reflect different realities related to the cyber subject in the 10-Ks, when portfolios are constructed from these scores, their returns do not display any statistically significant variation across different scores. In the context of market perception of risk, the results suggest there is no distinction between the various domains of cyber risk, and the market perceives a single aggregated risk related to cybersecurity.

In December 2020, a significant cyber attack on SolarWinds, a major IT management company, was uncovered, marking one of the most extensive and sophisticated cyber espionage operations. The attackers, believed to be state-sponsored, infiltrated SolarWinds' Orion software, which was used by numerous high-profile clients, including Fortune 500 companies and various U.S. government agencies. The attackers gained unprecedented access to sensitive data across multiple networks by embedding malicious code in a routine software update. This breach highlighted vulnerabilities within supply chain security and underscored the broader implications for firms at risk of cyber attacks. The incident serves as a case study for analyzing the financial impact on companies deemed to be cyber-risky. Such analysis using this event was performed in \cite{FlorackisLoucaMichaelyWeber2023} where they analyzed cumulative abnormal returns from their cyber-based portfolios setting the 14 December 2020, the day where the attack was disclosed to the SEC, as the t=0 day of the event.

\begin{center}
[Insert Table 30 here]
\end{center}

\begin{center}
[Insert Figure 22 here]
\end{center}

\begin{figure}[H] 
    \noindent\makebox[\textwidth]{%
    \includegraphics[scale=0.6]{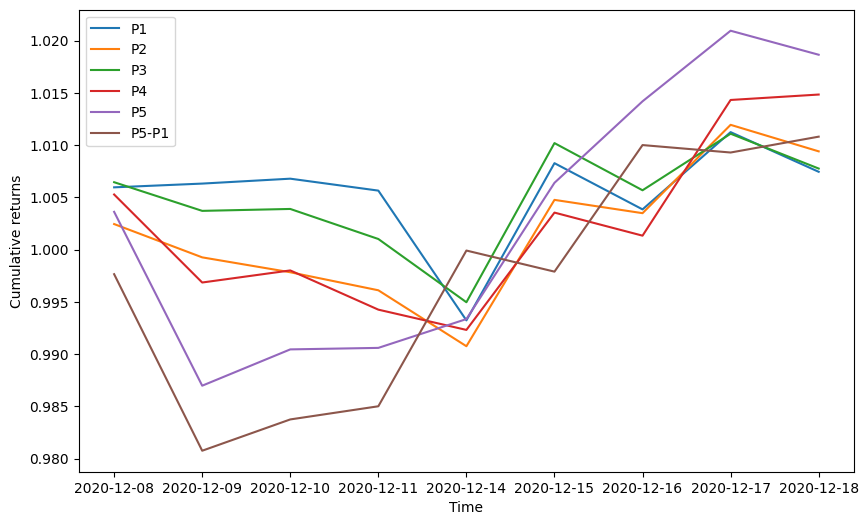}
    }
    \caption[\footnotesize Cumulative returns of cyber-based portfolio around SolarWinds breach]{\textbf{Cumulative returns of cyber-based portfolio around SolarWinds breach}}\bigskip
    \footnotesize{Evolution of the portfolio based on the overall cyber score if 1 dollar was invested the December 7, 2020. Note that the closed trading days do not appear.}
     \label{fig:cumu_overall}
\end{figure}

We conduct a similar analysis on Table \ref{event_overall}. None of the abnormal returns were statistically significant. Figure \ref{fig:cumu_overall} illustrates the cumulative returns of the cyber-based portfolio around this event. The results are unconventional. Returns were higher in the lower-tier cyber-based portfolio in the days leading up to the event. Moreover, when the event occurred, all portfolios declined except for P5. However, two important factors need to be considered. First, each portfolio aggregates over 600 firms. The SolarWinds breach might still be too financially localized to impact such a large number of firms, and the effect could be diluted among unaffected firms (those not associated with SolarWinds or not perceived by the market as affected). Second, none of the variations are statistically significant, and opposing behaviors likely mitigate the event's overall impact. For instance, during the shock, investors might have shifted their investments to other stocks considered safe but still related to cybersecurity, or the event might have increased interest in cybersecurity and boosted investment in P5 firms.

\begin{figure}[H] 
    \noindent\makebox[\textwidth]{%
    \includegraphics[scale=0.7]{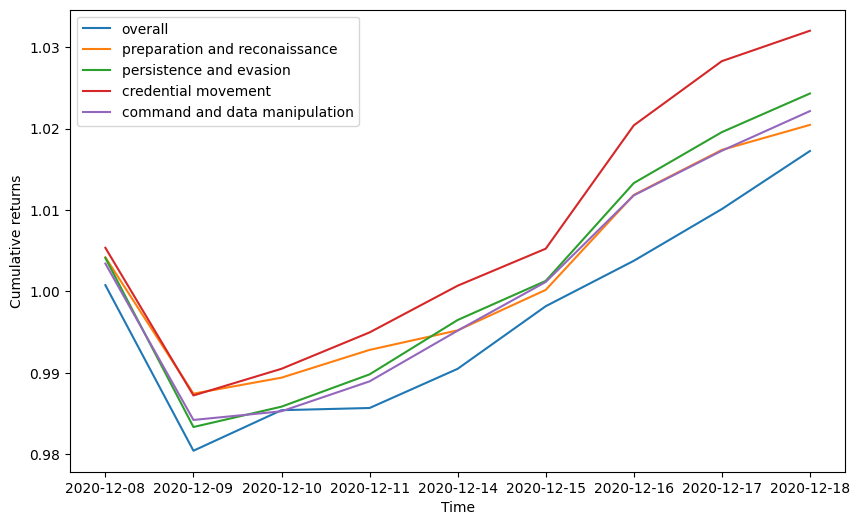}
    }
    \caption[\footnotesize Cumulative returns of cyber-based portfolio (P20) around SolarWinds breach]{\textbf{\small Cumulative returns of cyber-based portfolio (P20) around SolarWinds breach}}\bigskip
    \footnotesize{}
     \label{fig:cumu_P20}
\end{figure}

\begin{figure}[H] 
    \noindent\makebox[\textwidth]{%
    \includegraphics[scale=0.7]{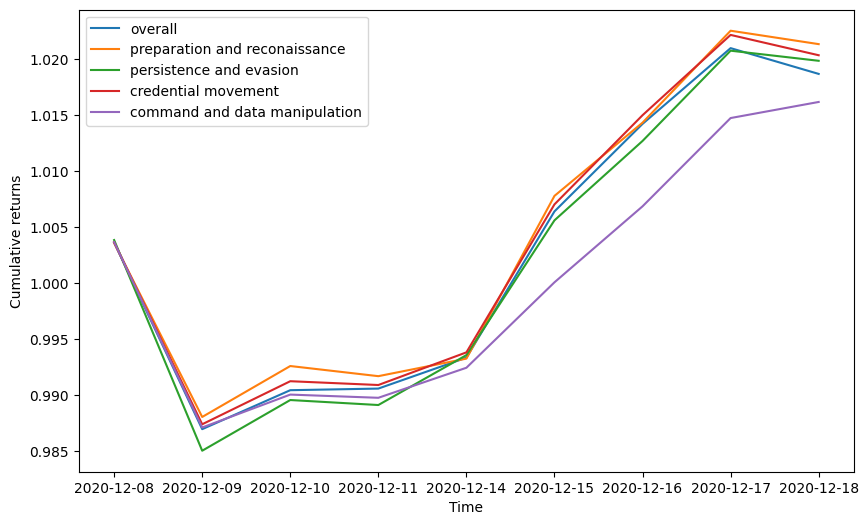}
    }
    \caption[\footnotesize Cumulative returns of cyber-based portfolio (P5) around SolarWinds breach]{\textbf{\small Cumulative returns of cyber-based portfolio (P5) around SolarWinds breach}}\bigskip
    \footnotesize{}
     \label{fig:cumu_P5}
\end{figure}

Finally, we display in Tables \ref{event_P20} and \ref{event_P5} the cumulative abnormal returns of P20 and P5 but constructed with different cyber scores. There is not enough statistical significance to infer anything. Note that the cumulative returns still reach higher returns in the P20 case (figure \ref{fig:cumu_P20}) than in the P5 case (Figure \ref{fig:cumu_P5}), and the portfolios based on the various cyber scores seem to behave similarly which support the hypothesis that the market
perceives a single aggregated risk related to cybersecurity.

These results contrast those of \cite{FlorackisLoucaMichaelyWeber2023}, who find a statistically significant drop in their top cyber-based portfolio returns around the event.

\begin{center}
[Insert Table 31 here]
\end{center}

\begin{center}
[Insert Figure 23 here]
\end{center}

\begin{center}
[Insert Table 32 here]
\end{center}

\begin{center}
[Insert Figure 24 here]
\end{center}

\section{Conclusion}

In this study, we use a doc2vec model to transform paragraphs of the MITRE ATT\&CK database's descriptions of cyber attacks into vectors. Comparing those vectors based on their cosine similarity, we apply clustering methods such as K-means, Louvain, and spectral clustering to infer groups of cyber attacks belonging to four defined types (super-tactics): command and data manipulation, credential movement, persistence and evasion and preparation and reconnaissance. Those clusters were recurrent through different trials using the three methods and different hyper-parameters. They were also chosen to preserve the underlying written structures of MITRE ATT\&CK by using a two-score system that ensures the equal distribution of paragraphs across super-tactics and their exclusivity to these super-tactics. 
Then, we use the doc2vec model to transform paragraphs of annual statements, more precisely 10-Ks, into vectors. Building the cosine similarity between 10-K vectors and vectors belonging to specific super tactics allows me to infer the semantic similarity of the 10-Ks to the four types of cyber attacks. We define those cosine similarities as the cyber score of a 10-K for a given super tactic. We also build an additional cyber sentiment score. This score considers only paragraphs' cyber scores when they contain words related to a \quotes{risk} or \quotes{uncertainty} vocabulary. 

We find that the different cyber scores cannot be unexplained by the linear combination of standard financial variables and non-semantic variables of the firms they belong to (the highest $R^2$ within among all tested cyber scores is 0.43). The independence of those newly found variables supports their innovative nature. All aggregate cyber scores have increased over the years and are higher in industry sectors (from the 12 Fama-French industries classification) involving assisting and workflow-related technology like Telephone and Television Transmission or Business Equipment.

We conduct asset pricing and statistical tests involving portfolios sorted on firms' cyber scores to assess if the cyber scores reflect cyber risks. Since all results for each cyber score are similar to the overall cyber score and the previous study \cite{CelenyMaréchal2023} only used this aggregated cyber score, we report here only the results related to this score. This does not apply to the cyber sentiment score, for which it appeared clear that no risk premium was involved.

Organizing firms into portfolios based on their cyber scores allows for the observation of increasing average excess returns as the portfolio's cyber score increases. The portfolio with the lowest quintile of cyber score, P1, has an average excess return of $0.82\%$. The portfolio with the highest quintile of cyber score, P5, has an average excess return of $1.44\%$ (both statistically significant at the $1\%$ level). Thus, a long-short portfolio P5-P1 destroys performance. Then, controlling for common pricing factors, we find that P5 has an alpha of $0.29\%$ at the $1\%$ level. Conversely, other portfolios, P1 to P4, have increasing alphas but are statistically insignificant. This threshold in significance between P4 and P5 highlights the fact that we cannot tell from a firm's cyber score, at which point it truly highlights cyber security in its 10-Ks. Therefore, lower portfolios contain a variety of firms that may be classified according to noise without meaning. We recommend that future studies using a similar work frame focus solely on P5 instead of P5-P1, as has been done until now. Sorting the firms into a first unrelated category and then according to their cyber score also reveals a similar structure of returns, as previously mentioned, with the top cyber-based portfolios performing better. Thus, the structure is robust, controlling for other firm characteristics.

\cite{FamaMacBeth1973} regressions show a risk premium associated with the cyber score and all disentangled cyber scores. In contrast, the cyber sentiment score does not drive any risk premium. The GRS test of \cite{GibbonsRossShanken1989} shows that the long-short portfolio P5-P1 helps to price various assets when used with the other well-known pricing factors of the five-factor model \cite{FamaFrench2015}. Furthermore, the BGRS tests from \cite{BarillasShanken2018} also highlight that P5-P1 is an important cyber-based pricing factor. According to the test, this importance is rising with time. Interestingly, these observations are valid for all cyber scores, including the cyber sentiment score.
  
Last, we cannot reject the hypothesis that the return of P5-P1, built with different cyber scores, is statistically different. Then, we conduct an event study using the cyber-breach of SolarWind in December 2020. The analysis provided no conclusive results, except that portfolios based on different cyber scores behave similarly. These last two observations could prove that the market does not differentiate between the various types of cyber risk and perceives them as a single aggregate cyber risk. This conclusion is reasonable when we observe the definition of the four types of cyber attacks. In a sense, they are not mutually exclusive since, in a cyber attack, command and data manipulation is the natural next step of credential movement, which is the next step of persistence, and evasion, which is the next step of preparation and reconnaissance.

\clearpage

\bibliographystyle{styles/jfe}
\bibliography{bibliography.bib}

\newpage
\section*{Tables and Figures}
\begin{table}[ht]
  \begin{adjustbox}{width=\textwidth,center}
    \begin{tabular}{lrrrrrrrrr}
    \hline
    Variable & Mean & Std & Min & Max & P1 & P25 & P50 & P75 & P99 \\
    \hline
    firm size & 20.17 & 2.49 & 13.11 & 26.33 & 13.68 & 18.54 & 20.33 & 21.87 & 25.66 \\
    firm age & 2.42 & 1.15 & -2.48 & 4.13 & -1.39 & 1.83 & 2.67 & 3.24 & 4.05 \\
    ROA & -0.15 & 0.54 & -4.30 & 0.49 & -3.04 & -0.13 & 0.02 & 0.06 & 0.37 \\
    book to market & 0.73 & 1.19 & 0.00 & 99.55 & 0.02 & 0.26 & 0.50 & 0.86 & 4.56 \\
    TobinQ & 2.11 & 2.14 & 0.35 & 24.15 & 0.56 & 1.03 & 1.39 & 2.24 & 11.80 \\
    MktBeta & 1.15 & 0.87 & -3.00 & 5.91 & -1.15 & 0.65 & 1.08 & 1.55 & 3.99 \\
    intangibles to assets & 0.15 & 0.21 & 0.00 & 8.10 & 0.00 & 0.00 & 0.05 & 0.24 & 0.78 \\
    debt to assets & 0.57 & 0.30 & 0.03 & 1.81 & 0.05 & 0.34 & 0.55 & 0.78 & 1.48 \\
    ROE & -0.08 & 0.61 & -5.88 & 1.60 & -2.96 & -0.07 & 0.07 & 0.15 & 0.87 \\
    price to earnings & -0.87 & 132.55 & -2001.73 & 455.35 & -568.94 & -3.93 & 12.05 & 22.64 & 295.69 \\
    profit margin & -0.42 & 6.46 & -111.45 & 1.00 & -27.21 & 0.22 & 0.39 & 0.62 & 0.96 \\
    asset turnover & 0.82 & 0.74 & 0.00 & 4.09 & 0.01 & 0.25 & 0.66 & 1.16 & 3.42 \\
    cash ratio & 2.07 & 3.89 & 0.01 & 36.13 & 0.01 & 0.23 & 0.68 & 1.98 & 20.28 \\
    sales to invested cap & 1.39 & 1.50 & 0.00 & 10.42 & 0.01 & 0.44 & 0.94 & 1.77 & 8.13 \\
    capital ratio & 0.31 & 0.32 & -0.10 & 1.97 & 0.00 & 0.03 & 0.24 & 0.47 & 1.51 \\
    RD to sales & 0.75 & 5.08 & 0.00 & 89.34 & 0.00 & 0.00 & 0.00 & 0.06 & 22.67 \\
    ROCE & -0.00 & 0.44 & -3.21 & 1.30 & -1.98 & -0.01 & 0.09 & 0.17 & 0.93 \\
    readability & 16.08 & 1.07 & 7.14 & 19.89 & 13.19 & 15.51 & 16.31 & 16.81 & 18.08 \\
    secret (dummy) & 0.28 & 0.45 & 0.00 & 1.00 & 0.00 & 0.00 & 0.00 & 1.00 & 1.00 \\
    risk length table & 5.03 & 1.46 & 0.00 & 7.69 & 0.00 & 4.84 & 5.35 & 5.81 & 6.84 \\
    volume per cap & 0.28 & 6.52 & -1.91 & 3485.03 & -0.01 & 0.06 & 0.13 & 0.23 & 1.97 \\
    humans per capital $\times 10^6$ & 8.40 & 157.41 & 0.00 & 16689.97 & 0.00 & 0.50 & 1.50 & 4.19 & 71.49 \\
    humans per assets $\times 10^6$ & 4.29 & 14.04 & 0.00 & 879.49 & 0.00 & 0.38 & 1.75 & 4.08 & 42.02 \\
    \hline
    \end{tabular}
 \end{adjustbox}
\caption[\footnotesize Descriptive statistics of the firm characteristics]{\textbf{Descriptive statistics of the firm characteristics}}\label{table_fin_stats}
\bigskip

\footnotesize{
This table provides descriptive statistics for various firm characteristics from 2009 to 2023. Mean, standard deviation (Std), minimum (Min), and maximum (Max) values are reported. Percentiles (P1, P25, P50, P75, P99) are also included. Firm-level characteristics are winsorized at the 1st and 99th percentile (by year). The characteristics are defined in Table \ref{tab:variable_descriptions}}
\end{table}

\begin{table}[h]
\noindent\makebox[\textwidth]{
    \centering
    \begin{tabular}{l|c|l}
    \hline
    \hline
    & & \multicolumn{1}{c}{Description}\\
    \hline
    Tactic & Credential Access & \multirow{3}{70mm}{Adversaries may forge web cookies that can be used to gain access to web applications or Internet services. Web applications and services (hosted in cloud SaaS environments or on-premise servers) often use session cookies to authenticate and authorize user access.}\\[4.9ex]
    Technique & Forge Web Credentials & \\[4.9ex]
    Sub-technique & Web Cookies & \\[4.9ex]
    \hline
    Tactic & Reconnaissance & \multirow{3}{70mm}{Adversaries may gather employee names that can be used during targeting. Employee names can be used to derive email addresses as well as to help guide other reconnaissance efforts and/or craft more believable lures. }\\[4ex]
    Technique &  Gather Victim Identity Information & \\[4ex]
    Sub-technique & Employee Names & \\[4ex]
    \hline
    \hline
    \end{tabular}
}
\caption[\footnotesize MITRE ATT\&CK sub-technique examples]{\centering \textbf{Examples of sub-techniques from MITRE ATT\&CK}\bigskip


}
\label{tab:MITRE_examples}
\end{table}

\newpage

\thispagestyle{empty}

\thispagestyle{empty}

\begin{table}[H]
    \begin{adjustbox}{width=\textwidth,center}
    \begin{tabular}{lrrrrrrrrr}
    \hline
    & Mean & Std & Min & Max & P1 & P25 & P50 & P75 & P99 \\
    \hline
    Persistence & 0.49 & 0.03 & 0.27 & 0.64 & 0.44 & 0.47 & 0.49 & 0.51 & 0.58 \\
    Command and Control & 0.47 & 0.03 & 0.28 & 0.62 & 0.42 & 0.45 & 0.47 & 0.49 & 0.55 \\
    Impact & 0.47 & 0.03 & 0.25 & 0.59 & 0.41 & 0.45 & 0.47 & 0.50 & 0.55 \\
    Initial Access & 0.46 & 0.03 & 0.23 & 0.59 & 0.40 & 0.44 & 0.46 & 0.48 & 0.55 \\
    Resource Development & 0.47 & 0.03 & 0.23 & 0.62 & 0.41 & 0.44 & 0.47 & 0.49 & 0.56 \\
    Collection & 0.49 & 0.03 & 0.29 & 0.64 & 0.43 & 0.46 & 0.48 & 0.51 & 0.57 \\
    Exfiltration & 0.47 & 0.03 & 0.23 & 0.64 & 0.41 & 0.44 & 0.46 & 0.49 & 0.56 \\
    Credential Access & 0.50 & 0.03 & 0.29 & 0.64 & 0.43 & 0.47 & 0.49 & 0.52 & 0.58 \\
    Privilege Escalation & 0.48 & 0.03 & 0.27 & 0.64 & 0.43 & 0.46 & 0.47 & 0.49 & 0.56 \\
    Execution & 0.46 & 0.03 & 0.29 & 0.61 & 0.42 & 0.44 & 0.46 & 0.48 & 0.55 \\
    Defense Evasion & 0.51 & 0.03 & 0.29 & 0.65 & 0.46 & 0.49 & 0.50 & 0.52 & 0.59 \\
    Reconnaissance & 0.48 & 0.03 & 0.32 & 0.61 & 0.42 & 0.46 & 0.48 & 0.51 & 0.57 \\
    Lateral Movement & 0.47 & 0.03 & 0.26 & 0.64 & 0.43 & 0.45 & 0.47 & 0.49 & 0.56 \\
    Discovery & 0.48 & 0.03 & 0.31 & 0.63 & 0.43 & 0.46 & 0.47 & 0.49 & 0.56 \\
    Preparation and Reconnaissance & 0.50 & 0.03 & 0.33 & 0.64 & 0.44 & 0.48 & 0.50 & 0.53 & 0.58 \\
    Persistence and Evasion & 0.51 & 0.03 & 0.29 & 0.65 & 0.46 & 0.49 & 0.51 & 0.53 & 0.59 \\
    Credential Movement & 0.50 & 0.03 & 0.29 & 0.65 & 0.44 & 0.48 & 0.50 & 0.52 & 0.59 \\
    Command and Data Manipulation & 0.50 & 0.03 & 0.29 & 0.64 & 0.44 & 0.47 & 0.49 & 0.52 & 0.58 \\
    Overall & 0.53 & 0.03 & 0.33 & 0.65 & 0.47 & 0.50 & 0.52 & 0.54 & 0.61 \\
    Sentiment & 0.51 & 0.05 & 0.00 & 0.72 & 0.42 & 0.48 & 0.51 & 0.54 & 0.63 \\
    \hline
    \end{tabular}
    \end{adjustbox}
    \caption[\footnotesize Descriptive statistics of cyber scores]{\textbf{Descriptive statistics of cyber scores}}\label{cyber_stats}\bigskip
    \footnotesize{
     This table provides descriptive statistics for the 14 MITRE ATT\&CK tactics cyber score, the four aggregated sub-cyber scores of the super-tactics, the overall cyber score, and the cyber sentiment score. The statistics are computed from all firms from 2009 to 2023.}
\end{table}

\newpage
\thispagestyle{empty}
\begin{table}[H]
    \vspace{-4em}
    \begin{adjustbox}{width=0.4\textwidth,center}
    \begin{tabular}{lcc}
        \hline
        \hline
        \multicolumn{3}{c}{Dependent variable: Firm-level indicator of cyber score}\\
        \hline
        & Model 1 & Model 2\\
        \hline
        Constant & \textbf{50.514}$^{***}$ & \textbf{53.229}$^{***}$ \\
         & [46.66] & [65.46]\\
        Firm Size (ln) & \textbf{0.008} & \textbf{0.039}\\
        & [0.16] & [1.37] \\
        Firm Age (ln)& \textbf{-0.346}$^{***}$ & \textbf{-0.492}$^{***}$\\
        & [-2.93] & [-8.80] \\
        ROA & \textbf{0.027} & \textbf{0.014} \\
        & [0.25] & [0.08] \\
        Book to Market & \textbf{-0.028}$^{***}$ & \textbf{-0.138}$^{***}$\\
        & [-2.66] & [-5.04] \\
        Market Beta & \textbf{-0.057}$^{*}$ & \textbf{-0.115}$^{***}$\\
        & [-1.95] & [-2.83] \\
        Intangibles/Assets & \textbf{-0.335}$^{*}$ & \textbf{1.133}$^{***}$\\
        & [-1.75] & [5.51] \\
        Debt/Assets & \textbf{-0.486}$^{**}$ & \textbf{1.088}$^{***}$ \\
        & [-2.13] & [2.59]\\
        ROE & \textbf{-0.009} & \textbf{0.011} \\
        & [-0.19] & [0.12] \\
        Price/Earnings & \textbf{0.0003}$^{**}$ & \textbf{0.00001} \\
        & [2.10] & [0.04] \\
        Profit Margin & \textbf{0.001} & \textbf{0.023}$^{***}$\\
        & [0.17] & [3.08] \\
        Asset Turnover & \textbf{-0.056} & \textbf{-0.438}$^{***}$\\
        & [-0.66] & [-3.52] \\
        Cash Ratio & \textbf{-0.0003} & \textbf{0.005} \\
        & [-0.04] & [0.31] \\
        Sales/Invested Capital & \textbf{0.013} & \textbf{0.134}$^{**}$\\
        & [0.39] & [2.30]\\
        Capital Ratio & \textbf{0.048} & \textbf{-2.200}$^{***}$\\
        & [0.25] & [-6.95] \\
        R\&D/Sales & \textbf{-0.005} & \textbf{-0.004}\\
        & [-0.89] & [-0.42]\\
        ROCE & \textbf{0.040} & \textbf{0.306}$^{*}$ \\
        & [0.43] & [1.92] \\
        Readability & \textbf{0.090}$^{***}$ & \textbf{-0.164}$^{***}$\\
        & [2.81] & [-2.93] \\
        Secret & \textbf{0.205}$^{*}$ & \textbf{0.711}$^{***}$\\
        & [1.80] & [6.98] \\
        Risk Length Table & \textbf{0.139}$^{***}$ & \textbf{0.254}$^{***}$\\
        & [4.67] & [7.36] \\
        Volume per Cap. & \textbf{0.001} & \textbf{0.007}$^{***}$\\
        & [0.27] & [3.79] \\
        Humans per Cap. & \textbf{-0.0004}$^{***}$ & \textbf{-0.000}\\
        & [-11.02] & [-1.18] \\
        \hline
        Year fixed effect & Yes & Yes\\
        Industry fixed effect & No & Yes\\
        Firm fixed effect & Yes & No\\
        Observations & 25531 & 25531\\
        $R^2$ within & 0.3193& 0.2672\\
        \hline
        \hline
    \end{tabular}
    \end{adjustbox}
\caption[\footnotesize Determinants of firm-level overall cyber score]{\textbf{Determinants of firm-level overall cyber score}}\bigskip
\footnotesize{This table reports the results of cyber score regressions on firm characteristics. Year-,
industry-, and firm-fixed effects are controlled. T-statistics are reported in brackets. The variables are standardized, and the standard errors are clustered at the firm level. $*$, $**$, and $***$ indicate significance at the 10\%, 5\% and 1\%
levels, respectively. All characteristics are defined in Table \ref{tab:variable_descriptions}.}
\label{tab:determinants_cyber_overall}
\end{table}

\thispagestyle{empty}
\begin{table}[H]
    \vspace{-4em}
    \begin{adjustbox}{width=0.4\textwidth,center}
    \begin{tabular}{lcc}
        \hline
        \hline
        \multicolumn{3}{c}{Dependent variable: Firm-level indicator of cyber score}\\
        \hline
        & Model 1 & Model 2\\
        \hline
        Constant & \textbf{43.921}$^{***}$ & \textbf{36.239}$^{***}$ \\
         & [18.44] & [26.16]\\
        Firm Size (ln) & \textbf{0.1046} & \textbf{0.2436}$^{***}$\\
        & [1.12] & [5.99] \\
        Firm Age (ln)& \textbf{-0.6886}$^{***}$ & \textbf{-0.4258}$^{***}$\\
        & [-2.91] & [-5.15] \\
        ROA & \textbf{-0.3729}$^{**}$ & \textbf{-0.4243}* \\
        & [-2.07] & [-1.89] \\
        Book to Market & \textbf{-0.0068} & \textbf{-0.0861}**\\
        & [-0.22] & [-2.07] \\
        Market Beta & \textbf{-0.0759} & \textbf{0.0189}\\
        & [-1.22] & [0.32] \\
        Intangibles/Assets & \textbf{0.4} & \textbf{1.5571}$^{***}$\\
        & [1.02] & [5.33] \\
        Debt/Assets & \textbf{0.1785} & \textbf{3.1032}$^{***}$ \\
        & [0.37] & [5.77]\\
        ROE & \textbf{-0.082} & \textbf{-0.0041} \\
        & [-0.78] & [-0.03] \\
        Price/Earnings & \textbf{-0.0002} & \textbf{-0.0003} \\
        & [-0.89] & [-0.83] \\
        Profit Margin & \textbf{-0.0059} & \textbf{0.0095}\\
        & [-0.56] & [0.86] \\
        Asset Turnover & \textbf{-0.0891} & \textbf{-0.1494} \\
        & [-0.51] & [-0.8] \\
        Cash Ratio & \textbf{0.0121} & \textbf{0.0217} \\
        & [0.72] & [1.02] \\
        Sales/Invested Capital & \textbf{-0.0733} & \textbf{-0.0916} \\
        & [-1.14] & [-1.08]\\
        Capital Ratio & \textbf{-0.12} & \textbf{-3.4501}$^{***}$\\
        & [-0.31] & [-8.31] \\
        R\&D/Sales & \textbf{-0.0225} & \textbf{-0.021}\\
        & [-1.48] & [-1.48]\\
        ROCE & \textbf{0.2894} & \textbf{0.4123}* \\
        & [1.47] & [1.74] \\
        Readability & \textbf{0.1335} & \textbf{0.2959}$^{***}$\\
        & [1.1] & [2.85] \\
        Secret & \textbf{0.4}$^{*}$ & \textbf{0.5921}$^{***}$\\
        & [1.88] & [4.31] \\
        Risk Length Table & \textbf{0.568}$^{***}$ & \textbf{0.7207}$^{***}$\\
        & [7.21] & [10.8] \\
        Volume per Cap. & \textbf{-0.0044} & \textbf{-0.0029} \\
        & [-0.9] & [-0.99] \\
        Humans per Cap. & \textbf{-0.0014}$^{***}$ & \textbf{0.0005}$^{***}$ \\
        & [-8.1] & [6.33] \\
        \hline
        Year fixed effect & Yes & Yes\\
        Industry fixed effect & No & Yes\\
        Firm fixed effect & Yes & No\\
        Observations & 25531 & 25531\\
        $R^2$ within & 0.2221 & 0.2088\\
        \hline
        \hline
    \end{tabular}
    \end{adjustbox}
\caption[\footnotesize Determinants of firm-level cyber sentiment score]{\textbf{Determinants of firm-level cyber sentiment score}}\bigskip
\footnotesize{This table reports the results of cyber score regressions on firm characteristics. Year-, industry-, and firm-fixed effects are controlled. T-statistics are reported in brackets. The variables are standardized, and the standard errors are clustered at the firm level. $*$, $**$, and $***$ indicate significance at the 10\%, 5\% and 1\% levels, respectively. All characteristics are defined in Table \ref{tab:variable_descriptions}.}
\label{tab:determinants_cyber_sentiment}
\end{table}

\thispagestyle{empty}
\begin{table}[H]
    \vspace{-4em}
    \begin{adjustbox}{width=0.4\textwidth,center}
    \begin{tabular}{lcc}
        \hline
        \hline
        \multicolumn{3}{c}{Dependent variable: Firm-level indicator of cyber score}\\
        \hline
        & Model 1 & Model 2\\
        \hline
        Constant & \textbf{48.709}$^{***}$ & \textbf{49.389}$^{***}$ \\
         & [42.27] & [53.32]\\
        Firm Size (ln) & \textbf{-0.014} & \textbf{0.0394}\\
        & [-0.27] & [1.32] \\
        Firm Age (ln)& \textbf{-0.5965}$^{***}$ & \textbf{-0.4996}$^{***}$\\
        & [-4.72] & [-8.04] \\
        ROA & \textbf{0.0168} & \textbf{0.0293} \\
        & [0.14] & [0.15] \\
        Book to Market & \textbf{-0.0224}$^{**}$ & \textbf{-0.1448}$^{***}$\\
        & [-1.97] & [-5.48] \\
        Market Beta & \textbf{-0.0446} & \textbf{-0.1203}$^{***}$\\
        & [-1.45] & [-2.69] \\
        Intangibles/Assets & \textbf{-0.3246} & \textbf{1.3698}$^{***}$\\
        & [-1.53] & [6.14] \\
        Debt/Assets & \textbf{-0.626}$^{**}$ & \textbf{0.8444}$^{*}$ \\
        & [-2.4] & [1.83]\\
        ROE & \textbf{-0.0561} & \textbf{-0.0365} \\
        & [-1.06] & [-0.39] \\
        Price/Earnings & \textbf{0.0} & \textbf{-0.0003} \\
        & [0.09] & [-0.96] \\
        Profit Margin & \textbf{-0.0017} & \textbf{0.0259}$^{***}$\\
        & [-0.29] & [3.23] \\
        Asset Turnover & \textbf{0.047} & \textbf{-0.5544}$^{***}$\\
        & [0.54] & [-4.03] \\
        Cash Ratio & \textbf{0.0028} & \textbf{0.0129} \\
        & [0.3] & [0.81] \\
        Sales/Invested Capital & \textbf{-0.0259} & \textbf{0.1951}$^{***}$\\
        & [-0.7] & [3.06]\\
        Capital Ratio & \textbf{0.1883} & \textbf{-2.2148}$^{***}$\\
        & [0.88] & [-6.28] \\
        R\&D/Sales & \textbf{-0.0063} & \textbf{0.0003}\\
        & [-0.79] & [0.03]\\
        ROCE & \textbf{0.097} & \textbf{0.3042}$^{*}$ \\
        & [0.93] & [1.73] \\
        Readability & \textbf{0.0869}$^{**}$ & \textbf{-0.1231}$^{**}$\\
        & [2.47] & [-2.02] \\
        Secret & \textbf{0.293}$^{**}$ & \textbf{0.8431}$^{***}$\\
        & [2.47] & [7.77] \\
        Risk Length Table & \textbf{0.118}$^{***}$ & \textbf{0.2876}$^{***}$\\
        & [3.6] & [6.68] \\
        Volume per Cap. & \textbf{-0.0034} & \textbf{0.006}$^{***}$ \\
        & [-1.05] & [2.66] \\
        Humans per Cap. & \textbf{-0.0003}$^{***}$ & \textbf{-0.0001} \\
        & [-9.34] & [-1.47] \\
        \hline
        Year fixed effect & Yes & Yes\\
        Industry fixed effect & No & Yes\\
        Firm fixed effect & Yes & No\\
        Observations & 25531 & 25531\\
        $R^2$ within & 0.3224 & 0.2677\\
        \hline
        \hline
    \end{tabular}
    \end{adjustbox}
\caption[\footnotesize Determinants of firm-level command and data manipulation cyber score]{\textbf{Determinants of firm-level command and data manipulation cyber score}}\bigskip
\footnotesize{This table reports the results of cyber score regressions on firm characteristics. Year-, industry-, and firm-fixed effects are controlled. T-statistics are reported in brackets. The variables are standardized, and the standard errors are clustered at the firm level. $*$, $**$, and $***$ indicate significance at the 10\%, 5\% and 1\% levels, respectively. All characteristics are defined in Table \ref{tab:variable_descriptions}.}
\label{tab:determinants_cyber_command_and_data_manipulation}
\end{table}

\thispagestyle{empty}
\begin{table}[H]
     \vspace{-4em}
    \begin{adjustbox}{width=0.4\textwidth,center}
    \begin{tabular}{lcc}
        \hline
        \hline
        \multicolumn{3}{c}{Dependent variable: Firm-level indicator of cyber score}\\
        \hline
        & Model 1 & Model 2\\
        \hline
        Constant & \textbf{49.779}$^{***}$ & \textbf{50.925}$^{***}$ \\
         & [41.1] & [57.92]\\
        Firm Size (ln) & \textbf{-0.0834} & \textbf{-0.005}\\
        & [-1.5] & [-0.16] \\
        Firm Age (ln)& \textbf{-0.6247}$^{***}$ & \textbf{-0.5823}$^{***}$\\
        & [-4.97] & [-9.58] \\
        ROA & \textbf{0.0446} & \textbf{0.0669} \\
        & [0.4] & [0.34] \\
        Book to Market & \textbf{-0.0085} & \textbf{-0.1247}$^{***}$\\
        & [-0.63] & [-3.88] \\
        Market Beta & \textbf{-0.0529} & \textbf{-0.1494}$^{***}$\\
        & [-1.64] & [-3.4] \\
        Intangibles/Assets & \textbf{-0.2478} & \textbf{1.0274}$^{***}$\\
        & [-1.2] & [4.62] \\
        Debt/Assets & \textbf{-0.588}$^{**}$ & \textbf{1.0181}$^{**}$ \\
        & [-2.33] & [2.29]\\
        ROE & \textbf{0.008} & \textbf{0.0121} \\
        & [0.15] & [0.13] \\
        Price/Earnings & \textbf{0.0003}$^{*}$ & \textbf{-0.0} \\
        & [1.92] & [-0.04] \\
        Profit Margin & \textbf{-0.0015} & \textbf{0.022}$^{***}$\\
        & [-0.26] & [2.97] \\
        Asset Turnover & \textbf{-0.164}$^{*}$ & \textbf{-0.5367}$^{***}$\\
        & [-1.83] & [-3.96] \\
        Cash Ratio & \textbf{-0.0005} & \textbf{-0.0024} \\
        & [-0.06] & [-0.15] \\
        Sales/Invested Capital & \textbf{0.0404} & \textbf{0.1481}$^{**}$\\
        & [1.17] & [2.42]\\
        Capital Ratio & \textbf{0.2001} & \textbf{-2.1857}$^{***}$\\
        & [0.98] & [-6.42] \\
        R\&D/Sales & \textbf{-0.0093} & \textbf{-0.011}\\
        & [-1.19] & [-1.18]\\
        ROCE & \textbf{0.0148} & \textbf{0.2837} \\
        & [0.15] & [1.61] \\
        Readability & \textbf{0.1344}$^{***}$ & \textbf{-0.1081}$^{*}$\\
        & [3.7] & [-1.78] \\
        Secret & \textbf{0.2014}$^{*}$ & \textbf{0.7763}$^{***}$\\
        & [1.65] & [7.15] \\
        Risk Length Table & \textbf{0.144}$^{***}$ & \textbf{0.2837}$^{***}$\\
        & [4.1] & [7.02] \\
        Volume per Cap. & \textbf{0.0006} & \textbf{0.008}$^{***}$ \\
        & [0.12] & [3.59] \\
        Humans per Cap. & \textbf{-0.0005}$^{***}$ & \textbf{-0.0} \\
        & [-11.29] & [-0.64] \\
        \hline
        Year fixed effect & Yes & Yes\\
        Industry fixed effect & No & Yes\\
        Firm fixed effect & Yes & No\\
        Observations & 25531 & 25531\\
        $R^2$ within & 0.3099 & 0.2564\\
        \hline
        \hline
    \end{tabular}
    \end{adjustbox}
\caption[\footnotesize Determinants of firm-level credential movement cyber score]{\textbf{Determinants of firm-level credential movement cyber score}}\bigskip
\footnotesize{This table reports the results of cyber score regressions on firm characteristics. Year-, industry-, and firm-fixed effects are controlled. T-statistics are reported in brackets. The variables are standardized, and the standard errors are clustered at the firm level. $*$, $**$, and $***$ indicate significance at the 10\%, 5\% and 1\% levels, respectively. All characteristics are defined in Table \ref{tab:variable_descriptions}.}
\label{tab:determinants_cyber_credential_movement}
\end{table}

\thispagestyle{empty}
\begin{table}[H]
    \vspace{-4em}
    \begin{adjustbox}{width=0.4\textwidth,center}
    \begin{tabular}{lcc}
        \hline
        \hline
        \multicolumn{3}{c}{Dependent variable: Firm-level indicator of cyber score}\\
        \hline
        & Model 1 & Model 2\\
        \hline
        Constant & \textbf{49.357}$^{***}$ & \textbf{52.308}$^{***}$ \\
         & [47.79] & [65.49]\\
        Firm Size (ln) & \textbf{-0.0327} & \textbf{0.0116}\\
        & [-0.71] & [0.43] \\
        Firm Age (ln)& \textbf{-0.1343} & \textbf{-0.4462}$^{***}$\\
        & [-1.2] & [-8.4] \\
        ROA & \textbf{0.0298} & \textbf{0.1} \\
        & [0.29] & [0.54] \\
        Book to Market & \textbf{-0.0117} & \textbf{-0.1262}$^{***}$\\
        & [-0.96] & [-5.07] \\
        Market Beta & \textbf{-0.0542}$^{*}$ & \textbf{-0.127}$^{***}$\\
        & [-1.96] & [-3.29] \\
        Intangibles/Assets & \textbf{-0.2192} & \textbf{0.9123}$^{***}$\\
        & [-1.23] & [4.78] \\
        Debt/Assets & \textbf{-0.4056}$^{*}$ & \textbf{1.0515}$^{***}$ \\
        & [-1.85] & [2.6]\\
        ROE & \textbf{0.0071} & \textbf{-0.0028} \\
        & [0.16] & [-0.03] \\
        Price/Earnings & \textbf{0.0001} & \textbf{-0.0001} \\
        & [1.2] & [-0.29] \\
        Profit Margin & \textbf{0.0004} & \textbf{0.0225}$^{***}$\\
        & [0.09] & [3.0] \\
        Asset Turnover & \textbf{-0.0273} & \textbf{-0.4263}$^{***}$\\
        & [-0.36] & [-3.64] \\
        Cash Ratio & \textbf{-0.0018} & \textbf{0.0015} \\
        & [-0.2] & [0.11] \\
        Sales/Invested Capital & \textbf{0.0107} & \textbf{0.1105}$^{**}$\\
        & [0.34] & [2.03]\\
        Capital Ratio & \textbf{-0.0628} & \textbf{-2.1721}$^{***}$\\
        & [-0.36] & [-7.0] \\
        R\&D/Sales & \textbf{-0.004} & \textbf{-0.0051}\\
        & [-0.65] & [-0.54]\\
        ROCE & \textbf{-0.0611} & \textbf{0.1708} \\
        & [-0.72] & [1.12] \\
        Readability & \textbf{0.1256}$^{***}$ & \textbf{-0.1214}$^{**}$\\
        & [3.34] & [-2.15] \\
        Secret & \textbf{0.059} & \textbf{0.6653}$^{***}$\\
        & [0.55] & [6.89] \\
        Risk Length Table & \textbf{0.0929}$^{***}$ & \textbf{0.1794}$^{***}$\\
        & [3.29] & [5.55] \\
        Volume per Cap. & \textbf{-0.0007} & \textbf{0.0056}$^{**}$ \\
        & [-0.21] & [2.29] \\
        Humans per Cap. & \textbf{-0.0002}$^{***}$ & \textbf{-0.0} \\
        & [-7.49] & [-0.82] \\
        \hline
        Year fixed effect & Yes & Yes\\
        Industry fixed effect & No & Yes\\
        Firm fixed effect & Yes & No\\
        Observations & 25531 & 25531\\
        $R^2$ within & 0.2193 & 0.1566\\
        \hline
        \hline
    \end{tabular}
    \end{adjustbox}
\caption[\footnotesize Determinants of firm-level persistence and evasion cyber score]{\textbf{Determinants of firm-level persistence and evasion cyber score}}\bigskip
\footnotesize{This table reports the results of cyber score regressions on firm characteristics. Year-, industry-, and firm-fixed effects are controlled. T-statistics are reported in brackets. The variables are standardized, and the standard errors are clustered at the firm level. $*$, $**$, and $***$ indicate significance at the 10\%, 5\% and 1\% levels, respectively. All characteristics are defined in Table \ref{tab:variable_descriptions}.}
\label{tab:determinants_cyber_persistence_and_evasion}
\end{table}

\thispagestyle{empty}
\begin{table}[H]
    \vspace{-4em}
    \begin{adjustbox}{width=0.4\textwidth,center}
    \begin{tabular}{lcc}
        \hline
        \hline
        \multicolumn{3}{c}{Dependent variable: Firm-level indicator of cyber score}\\
        \hline
        & Model 1 & Model 2\\
        \hline
        Constant & \textbf{48.897}$^{***}$ & \textbf{49.466}$^{***}$ \\
         & [42.14] & [56.36]\\
        Firm Size (ln) & \textbf{0.0069} & \textbf{0.0873}$^{***}$\\
        & [0.13] & [2.99] \\
        Firm Age (ln)& \textbf{-0.7352}$^{***}$ & \textbf{-0.523}$^{***}$\\
        & [-5.92] & [-9.11] \\
        ROA & \textbf{-0.0807} & \textbf{-0.1396} \\
        & [-0.76] & [-0.78] \\
        Book to Market & \textbf{-0.0347}$^{***}$ & \textbf{-0.1499}$^{***}$\\
        & [-2.75] & [-5.58] \\
        Market Beta & \textbf{-0.0577}$^{*}$ & \textbf{-0.0962}$^{**}$\\
        & [-1.92] & [-2.34] \\
        Intangibles/Assets & \textbf{-0.1576} & \textbf{1.5511}$^{***}$\\
        & [-0.73] & [7.31] \\
        Debt/Assets & \textbf{-0.4222}$^{*}$ & \textbf{1.6297}$^{***}$ \\
        & [-1.67] & [3.81]\\
        ROE & \textbf{0.0376} & \textbf{0.0192} \\
        & [0.7] & [0.21] \\
        Price/Earnings & \textbf{0.0002} & \textbf{0.0} \\
        & [1.6] & [0.05] \\
        Profit Margin & \textbf{-0.0009} & \textbf{0.0191}$^{***}$\\
        & [-0.21] & [2.63] \\
        Asset Turnover & \textbf{-0.033} & \textbf{-0.4303}$^{***}$\\
        & [-0.36] & [-3.38] \\
        Cash Ratio & \textbf{0.0055} & \textbf{0.015} \\
        & [0.56] & [0.96] \\
        Sales/Invested Capital & \textbf{0.0072} & \textbf{0.146}$^{**}$\\
        & [0.21] & [2.46]\\
        Capital Ratio & \textbf{0.0664} & \textbf{-2.5668}$^{***}$\\
        & [0.33] & [-7.98] \\
        R\&D/Sales & \textbf{-0.0094} & \textbf{-0.0015}\\
        & [-1.59] & [-0.15]\\
        ROCE & \textbf{0.0523} & \textbf{0.3945}$^{**}$ \\
        & [0.52] & [2.55] \\
        Readability & \textbf{0.0651} & \textbf{-0.1943}$^{***}$\\
        & [1.64] & [-3.14] \\
        Secret & \textbf{0.2292}$^{*}$ & \textbf{0.6711}$^{***}$\\
        & [1.95] & [6.58] \\
        Risk Length Table & \textbf{0.1945}$^{***}$ & \textbf{0.3505}$^{***}$\\
        & [5.87] & [9.16] \\
        Volume per Cap. & \textbf{-0.0021} & \textbf{0.0036} \\
        & [-0.53] & [1.45] \\
        Humans per Cap. & \textbf{-0.0005}$^{***}$ & \textbf{0.0} \\
        & [-10.84] & [0.44] \\
        \hline
        Year fixed effect & Yes & Yes\\
        Industry fixed effect & No & Yes\\
        Firm fixed effect & Yes & No\\
        Observations & 25531 & 25531\\
        $R^2$ within & 0.4331 & 0.3920\\
        \hline
        \hline
    \end{tabular}
    \end{adjustbox}
\caption[\footnotesize Determinants of firm-level preparation and reconnaissance cyber score]{\textbf{Determinants of firm-level preparation and reconnaissance cyber score}}\bigskip
\footnotesize{This table reports the results of cyber score regressions on firm characteristics. Year-, industry-, and firm-fixed effects are controlled. T-statistics are reported in brackets. The variables are standardized, and the standard errors are clustered at the firm level. $*$, $**$, and $***$ indicate significance at the 10\%, 5\% and 1\% levels, respectively. All characteristics are defined in Table \ref{tab:variable_descriptions}.}
\label{tab:determinants_cyber_preparation_and_reconaissance}
\end{table}

\thispagestyle{empty}

\thispagestyle{empty}

\newpage

\newcommand{\tablescale}{0.74}

\begin{table}[H]
\centering

\renewcommand{\arraystretch}{1.2}
\resizebox{\tablescale\textwidth}{!}{%
\begin{tabular}{l *{6}{c}}
\toprule
 & P1 & P2 & P3 & P4 & P5 & P5-P1 \\
\midrule
\multicolumn{7}{l}{\textbf{A. Portfolios sorted by cyber score}} \\
\textbf{avg. excess ret.} & \textbf{0.82***} & \textbf{0.93***} & \textbf{1.04***} & \textbf{1.22***} & \textbf{1.44***} & \textbf{0.62**} \\
 & [3.27] & [3.46] & [3.65] & [4.65] & [4.54] & [2.05] \\
\textbf{CAPM alpha} & \textbf{-0.18} & \textbf{-0.12} & \textbf{-0.08} & \textbf{0.14} & \textbf{0.36**} & \textbf{0.54} \\
 & [-0.85] & [-0.93] & [-0.84] & [1.47] & [2.14] & [1.49] \\
\textbf{FFC alpha} & \textbf{-0.09} & \textbf{-0.05} & \textbf{0.0} & \textbf{0.15*} & \textbf{0.27***} & \textbf{0.36**} \\
 & [-0.88] & [-0.57] & [0.04] & [1.71] & [3.04] & [2.2] \\
\textbf{FF5 alpha} & \textbf{-0.14} & \textbf{-0.1} & \textbf{0.0} & \textbf{0.13} & \textbf{0.29***} & \textbf{0.44***} \\
 & [-1.57] & [-1.18] & [0.01] & [1.47] & [3.16] & [2.88] \\
\midrule
\multicolumn{7}{l}{\textbf{B. Characteristics}} \\
 Nb. firms& 628.48 & 629.1 & 629.01 & 629.1 & 629.67 & - \\

Avg. cyber score & 0.49 & 0.51 & 0.52 & 0.53 & 0.57 & - \\
Sharp Ratio & 0.61 & 0.69 & 0.72 & 0.88 & 1.02 & 0.68 \\
\bottomrule
\end{tabular}%
}

\caption[\footnotesize Average monthly excess returns and alphas for the overall cyber score]{\textbf{Average monthly excess returns and alphas (in percent) using the overall cyber score}}
    \label{univariate_overall}
    \bigskip
    \footnotesize{
    \begin{flushleft}
     FFC refers to the four-factor model of \cite{Carhart1997}, and FF5 refers to the five-factor model of \cite{FamaFrench2015}. Panel B shows the average number of firms in each portfolio and the average cyber risk of the portfolios. T-statistics are reported in brackets. $*$, $**$, and $***$ indicate significance at the
10\%, 5\%, and 1\% levels, respectively. The time ranges from January 2009 to December 2023.\end{flushleft}} 
\end{table}

\begin{table}[H]
\centering

\renewcommand{\arraystretch}{1.2}
\resizebox{\tablescale\textwidth}{!}{%
\begin{tabular}{l *{6}{c}}
\toprule
 & P1 & P2 & P3 & P4 & P5 & P5-P1 \\
\midrule
\multicolumn{7}{l}{\textbf{A. Portfolios sorted by cyber score}} \\
\textbf{avg. excess ret.} & \textbf{0.99***} & \textbf{1.08***} & \textbf{1.24***} & \textbf{1.15***} & \textbf{1.14***} & \textbf{0.14} \\
 & [3.92] & [4.61] & [4.78] & [3.94] & [3.88] & [1.21] \\
\textbf{CAPM alpha} & \textbf{-0.03} & \textbf{0.04} & \textbf{0.19*} & \textbf{0.02} & \textbf{0.05} & \textbf{0.08} \\
 & [-0.27] & [0.31] & [1.91] & [0.31] & [0.59] & [0.58] \\
\textbf{FFC alpha} & \textbf{0.0} & \textbf{0.05} & \textbf{0.17*} & \textbf{0.04} & \textbf{0.05} & \textbf{0.05} \\
 & [0.02] & [0.48] & [1.78] & [0.59] & [0.66] & [0.45] \\
\textbf{FF5 alpha} & \textbf{-0.03} & \textbf{-0.02} & \textbf{0.12} & \textbf{0.07} & \textbf{0.1} & \textbf{0.12} \\
 & [-0.41] & [-0.18] & [1.31] & [1.15] & [1.17] & [1.15] \\
\midrule
\multicolumn{7}{l}{\textbf{B. Characteristics}} \\
 Nb. firms& 628.48 & 629.1 & 629.01 & 629.1 & 629.67 & - \\
Avg. cyber score & 0.46 & 0.49 & 0.51 & 0.53 & 0.57 & - \\
Sharp Ratio & 0.75 & 0.8 & 0.92 & 0.81 & 0.82 & 0.3 \\
\bottomrule
\end{tabular}%
}

\caption[\footnotesize Average monthly excess returns and alphas for the cyber sentiment score]{\textbf{Average monthly excess returns and alphas (in percent) using the cyber sentiment score}}\label{univariate_sentiment}
    
\end{table}

\begin{table}[H]
\centering

\renewcommand{\arraystretch}{1.2}
\resizebox{\tablescale\textwidth}{!}{%
\begin{tabular}{l *{6}{c}}
\toprule
 & P1 & P2 & P3 & P4 & P5 & P5-P1 \\
\midrule
\multicolumn{7}{l}{\textbf{A. Portfolios sorted by cyber score}} \\
\textbf{avg. excess ret.} & \textbf{0.81***} & \textbf{0.91***} & \textbf{1.12***} & \textbf{1.17***} & \textbf{1.46***} & \textbf{0.65*} \\
 & [3.18] & [3.35] & [4.31] & [4.55] & [4.33] & [1.94] \\
\textbf{CAPM alpha} & \textbf{-0.22} & \textbf{-0.13} & \textbf{0.06} & \textbf{0.05} & \textbf{0.39**} & \textbf{0.6} \\
 & [-0.94] & [-1.07] & [0.75] & [0.61] & [2.01] & [1.49] \\
\textbf{FFC alpha} & \textbf{-0.11} & \textbf{-0.07} & \textbf{0.1**} & \textbf{0.04} & \textbf{0.3***} & \textbf{0.42**} \\
 & [-1.0] & [-0.77] & [2.18] & [0.5] & [2.69] & [2.12] \\
\textbf{FF5 alpha} & \textbf{-0.16} & \textbf{-0.12} & \textbf{0.11**} & \textbf{0.03} & \textbf{0.33***} & \textbf{0.49***} \\
 & [-1.66] & [-1.35] & [2.01] & [0.37] & [2.76] & [2.61] \\
\midrule
\multicolumn{7}{l}{\textbf{B. Characteristics}} \\
 Nb. firms& 628.48 & 629.1 & 629.01 & 629.1 & 629.67 & - \\
Avg. cyber score & 0.46 & 0.48 & 0.49 & 0.5 & 0.54 & - \\
Sharp Ratio & 0.59 & 0.68 & 0.82 & 0.82 & 1.04 & 0.71 \\
\bottomrule
\end{tabular}%
}

\caption[\footnotesize Average monthly excess returns and alphas for the command and data manipulation cyber score]{\textbf{Average monthly excess returns and alphas (in percent) using the command and data manipulation cyber score}}
\label{univariate_command_and_data_manipulation}
\end{table}

\begin{table}[H]
\centering

\renewcommand{\arraystretch}{1.2}
\resizebox{\tablescale\textwidth}{!}{%
\begin{tabular}{l *{6}{c}}
\toprule
 & P1 & P2 & P3 & P4 & P5 & P5-P1 \\
\midrule
\multicolumn{7}{l}{\textbf{A. Portfolios sorted by cyber score}} \\
\textbf{avg. excess ret.} & \textbf{0.88***} & \textbf{0.9***} & \textbf{1.04***} & \textbf{1.13***} & \textbf{1.49***} & \textbf{0.61**} \\
 & [3.63] & [3.51] & [3.5] & [4.34] & [4.66] & [2.06] \\
\textbf{CAPM alpha} & \textbf{-0.12} & \textbf{-0.14} & \textbf{-0.1} & \textbf{0.07} & \textbf{0.4**} & \textbf{0.52} \\
 & [-0.6] & [-1.01] & [-1.16] & [1.11] & [2.31] & [1.48] \\
\textbf{FFC alpha} & \textbf{-0.04} & \textbf{-0.07} & \textbf{-0.02} & \textbf{0.09} & \textbf{0.3***} & \textbf{0.34**} \\
 & [-0.41] & [-0.76] & [-0.29] & [1.4] & [3.09] & [2.07] \\
\textbf{FF5 alpha} & \textbf{-0.1} & \textbf{-0.11} & \textbf{-0.02} & \textbf{0.08} & \textbf{0.33***} & \textbf{0.43***} \\
 & [-1.15] & [-1.25] & [-0.32] & [1.23] & [3.19] & [2.74] \\
\midrule
\multicolumn{7}{l}{\textbf{B. Characteristics}} \\
 Nb. firms& 628.48 & 629.1 & 629.01 & 629.1 & 629.67 & - \\
Avg. cyber score & 0.46 & 0.48 & 0.49 & 0.51 & 0.54 & - \\
Sharp Ratio & 0.66 & 0.67 & 0.72 & 0.83 & 1.04 & 0.69 \\
\bottomrule
\end{tabular}%
}

\caption[\footnotesize Average monthly excess returns and alphas for the credential movement cyber score]{\textbf{Average monthly excess returns and alphas (in percent) using the credential movement cyber score}}\label{univariate_credential_movement}
\end{table}

\begin{table}[H]
\centering

\renewcommand{\arraystretch}{1.2}
\resizebox{\tablescale\textwidth}{!}{%
\begin{tabular}{l *{6}{c}}
\toprule
 & P1 & P2 & P3 & P4 & P5 & P5-P1 \\
\midrule
\multicolumn{7}{l}{\textbf{A. Portfolios sorted by cyber score}} \\
\textbf{avg. excess ret.} & \textbf{0.84***} & \textbf{0.95***} & \textbf{1.04***} & \textbf{1.11***} & \textbf{1.49***} & \textbf{0.64**} \\
 & [3.35] & [3.61] & [3.75] & [4.44] & [4.54] & [2.01] \\
\textbf{CAPM alpha} & \textbf{-0.18} & \textbf{-0.08} & \textbf{-0.03} & \textbf{0.05} & \textbf{0.39**} & \textbf{0.58} \\
 & [-0.86] & [-0.54] & [-0.39] & [0.66] & [2.16] & [1.53] \\
\textbf{FFC alpha} & \textbf{-0.1} & \textbf{0.01} & \textbf{0.03} & \textbf{0.08} & \textbf{0.29***} & \textbf{0.39**} \\
 & [-0.89] & [0.13] & [0.38] & [1.47] & [2.9] & [2.23] \\
\textbf{FF5 alpha} & \textbf{-0.15} & \textbf{-0.05} & \textbf{0.03} & \textbf{0.05} & \textbf{0.33***} & \textbf{0.48***} \\
 & [-1.66] & [-0.78] & [0.5] & [0.78] & [3.09] & [2.97] \\
\midrule
\multicolumn{7}{l}{\textbf{B. Characteristics}} \\
 Nb. firms& 628.48 & 629.1 & 629.01 & 629.1 & 629.67 & - \\
Avg. cyber score & 0.48 & 0.49 & 0.5 & 0.52 & 0.55 & - \\
Sharp Ratio & 0.61 & 0.71 & 0.76 & 0.82 & 1.03 & 0.67 \\
\bottomrule
\end{tabular}%
}

\caption[\footnotesize Average monthly excess returns and alphas for the persistence and evasion cyber score]{\textbf{Average monthly excess returns and alphas (in percent) using the persistence and evasion cyber score}}\label{univariate_persistence_and_evasion}
\end{table}

\begin{table}[H]
\centering

\renewcommand{\arraystretch}{1.2}
\resizebox{\tablescale\textwidth}{!}{%
\begin{tabular}{l *{6}{c}}
\toprule
 & P1 & P2 & P3 & P4 & P5 & P5-P1 \\
\midrule
\multicolumn{7}{l}{\textbf{A. Portfolios sorted by cyber score}} \\
\textbf{avg. excess ret.} & \textbf{0.86***} & \textbf{0.85***} & \textbf{1.15***} & \textbf{1.11***} & \textbf{1.43***} & \textbf{0.57**} \\
 & [3.54] & [3.13] & [4.22] & [3.94] & [4.69] & [1.97] \\
\textbf{CAPM alpha} & \textbf{-0.09} & \textbf{-0.23} & \textbf{0.02} & \textbf{0.02} & \textbf{0.37**} & \textbf{0.46} \\
 & [-0.44] & [-1.97] & [0.2] & [0.29] & [2.39] & [1.34] \\
\textbf{FFC alpha} & \textbf{-0.02} & \textbf{-0.16} & \textbf{0.07} & \textbf{0.04} & \textbf{0.28***} & \textbf{0.3*} \\
 & [-0.14] & [-2.45] & [1.23] & [0.56] & [3.07] & [1.75] \\
\textbf{FF5 alpha} & \textbf{-0.09} & \textbf{-0.19} & \textbf{0.11*} & \textbf{0.01} & \textbf{0.3***} & \textbf{0.38**} \\
 & [-0.95] & [-3.08] & [1.76] & [0.19] & [3.17] & [2.42] \\
\midrule
\multicolumn{7}{l}{\textbf{B. Characteristics}} \\
 Nb. firms& 628.48 & 629.1 & 629.01 & 629.1 & 629.67 & - \\
Avg. cyber score & 0.47 & 0.48 & 0.5 & 0.51 & 0.54 & - \\
Sharp Ratio & 0.67 & 0.61 & 0.8 & 0.8 & 1.03 & 0.69 \\
\bottomrule
\end{tabular}%
}

\caption[\footnotesize Average monthly excess returns and alphas for the preparation and reconnaissance cyber score]{\textbf{Average monthly excess returns and alphas (in percent) using the preparation and reconnaissance cyber score}}\label{univariate_preparation_and_reconaissance}
\end{table}

\newpage

\definecolor{mycolor}{RGB}{120,120,120}

\begin{table}[H]
\centering
\resizebox{0.9\textwidth}{!}{%
\begin{tabular}{lccccc|lccccc|lccccc}
\toprule
\textbf{} & \textbf{Q1} & \textbf{Q2} & \textbf{Q3} & \textbf{Q4} & \textbf{Q5} & \textbf{} & \textbf{Q1} & \textbf{Q2} & \textbf{Q3} & \textbf{Q4} & \textbf{Q5} & \textbf{} & \textbf{Q1} & \textbf{Q2} & \textbf{Q3} & \textbf{Q4} & \textbf{Q5} \\ \midrule
\multicolumn{18}{c}{Double sorted portfolios with overall cyber score} \\  \midrule
Beta Q1& 1.01 & 1.01 & 1.13 & 1.37 & 1.42 & BM Q1*& 1.09 & 0.98 & 1.14 & 1.18 & 1.24 & Size Q1& 0.86 & 0.96 & 1.06 & 1.22 & 1.32\\
Beta Q2& 0.89 & 1.00 & 1.07 & 1.25 & 1.33 & BM Q2& 0.86 & 0.92 & 1.01 & 1.25 & 1.37 & Size Q2& 0.87 & 0.96 & 1.05 & 1.25 & 1.32 \\
Beta Q3& 0.89 & 1.02 & 1.01 & 1.22 & 1.28 & BM Q3& 0.85 & 1.01 & 1.04 & 1.23 & 1.33 & Size Q3& 0.87 & 0.95 & 1.05 & 1.25 & 1.32 \\
Beta Q4& 0.84 & 0.90 & 1.04 & 1.23 & 1.25 & BM Q4& 0.83 & 0.99 & 1.03 & 1.22 & 1.35 & Size Q4& 0.86 & 0.95 & 1.08 & 1.26 & 1.33 \\
Beta Q5& 0.83 & 0.92 & 1.02 & 1.21 & 1.30 & BM Q5& 0.87 & 0.95 & 1.07 & 1.22 & 1.33 & Size Q5& 1.09 & 1.15 & 1.20 & 1.19 & 1.39 \\ \midrule
\multicolumn{18}{c}{Double sorted portfolios with cyber sentiment score} \\  \midrule
Beta Q1 & 1.16 & 1.23 & 1.27 & 1.25 & 1.09 & BM Q1*& 1.13 & 1.15 & 1.04 & 1.12 & 1.12 & Size Q1*& 0.97 & 1.20 & 1.10 & 1.15 & 1.07 \\
Beta Q2*& 0.95 & 1.00 & 1.36 & 1.18 & 1.07 & BM Q2*& 0.93 & 1.01 & 1.30 & 1.13 & 1.09 & Size Q2*& 0.96 & 1.12 & 1.21 & 1.13 & 1.07 \\
Beta Q3*& 1.06 & 1.00 & 1.07 & 1.11 & 1.10 & BM Q3*& 0.96 & 1.13 & 1.22 & 1.15 & 1.08 & Size Q3*& 0.96 & 1.11 & 1.19 & 1.13 & 1.07 \\
Beta Q4*& 0.95 & 1.08 & 1.10 & 1.09 & 1.03 & BM Q4*& 0.95 & 1.09 & 1.21 & 1.14 & 1.08 & Size Q4*& 0.95 & 1.03 & 1.24 & 1.11 & 1.08 \\
Beta Q5*& 0.96 & 1.00 & 1.20 & 1.13 & 1.04 & BM Q5*& 0.95 & 1.09 & 1.20 & 1.15 & 1.05 & Size Q5*& 1.23 & 1.16 & 1.20 & 1.17 & 1.29 \\ \midrule
\multicolumn{18}{c}{Double sorted portfolios with command and data manipulation cyber score} \\  \midrule
Beta Q1 & 1.04 & 1.06 & 1.18 & 1.27 & 1.41 & BM Q1*& 1.07 & 0.94 & 1.22 & 1.15 & 1.24 & Size Q1 & 0.90 & 0.89 & 1.16 & 1.13 & 1.33 \\
Beta Q2 & 0.89 & 0.93 & 1.14 & 1.19 & 1.36 & BM Q2 & 0.83 & 0.89 & 1.11 & 1.16 & 1.39 & Size Q2*& 0.89 & 0.91 & 1.16 & 1.12 & 1.33 \\
Beta Q3*& 0.91 & 0.90 & 1.16 & 1.09 & 1.34 & BM Q3 & 0.89 & 0.90 & 1.17 & 1.16 & 1.34 & Size Q3 & 0.87 & 0.91 & 1.16 & 1.14 & 1.34 \\
Beta Q4 & 0.85 & 0.85 & 1.17 & 1.14 & 1.25 & BM Q4 & 0.85 & 0.93 & 1.15 & 1.20 & 1.32 & Size Q4 & 0.85 & 0.96 & 1.11 & 1.21 & 1.34 \\
Beta Q5 & 0.87 & 0.85 & 1.11 & 1.12 & 1.30 & BM Q5 & 0.86 & 0.96 & 1.13 & 1.20 & 1.31 & Size Q5 & 1.11 & 1.19 & 1.16 & 1.16 & 1.41 \\ \midrule
\multicolumn{18}{c}{Double sorted portfolios with credential movement cyber score} \\  \midrule
Beta Q1*& 1.05 & 0.96 & 1.16 & 1.22 & 1.49 & BM Q1*& 1.03 & 1.03 & 1.18 & 1.09 & 1.25 & Size Q1 & 0.90 & 0.93 & 1.06 & 1.10 & 1.39 \\
Beta Q2*& 0.93 & 0.94 & 1.08 & 1.03 & 1.42 & BM Q2 & 0.86 & 0.90 & 1.07 & 1.16 & 1.41 & Size Q2& 0.90 & 0.92 & 1.09 & 1.09 & 1.38 \\
Beta Q3 & 0.90 & 0.99 & 1.03 & 1.08 & 1.37 & BM Q3 & 0.93 & 0.92 & 1.05 & 1.09 & 1.41 & Size Q3 & 0.90 & 0.92 & 1.08 & 1.08 & 1.40 \\
Beta Q4 & 0.85 & 0.88 & 1.09 & 1.09 & 1.31 & BM Q4 & 0.90 & 0.90 & 1.07 & 1.09 & 1.41 & Size Q4 & 0.89 & 0.93 & 1.03 & 1.10 & 1.42 \\
Beta Q5 & 0.90 & 0.87 & 1.04 & 1.11 & 1.35 & BM Q5 & 0.88 & 0.95 & 1.06 & 1.06 & 1.41 & Size Q5 & 1.13 & 1.13 & 1.17 & 1.22 & 1.39 \\ \midrule
\multicolumn{18}{c}{Double sorted portfolios with persistence and evasion cyber score} \\  \midrule
Beta Q1*& 1.02 & 1.07 & 1.12 & 1.25 & 1.46 & BM Q1*& 1.06 & 1.01 & 1.18 & 1.19 & 1.19 & Size Q1 & 0.88 & 0.99 & 1.08 & 1.12 & 1.36 \\
Beta Q2*& 0.88 & 1.06 & 1.05 & 1.13 & 1.40 & BM Q2*& 0.85 & 1.03 & 0.98 & 1.19 & 1.41 & Size Q2 & 0.89 & 1.00 & 1.08 & 1.15 & 1.34 \\
Beta Q3 & 0.90 & 1.05 & 1.06 & 1.09 & 1.33 & BM Q3 & 0.90 & 1.00 & 1.03 & 1.15 & 1.38 & Size Q3 & 0.88 & 0.99 & 1.05 & 1.14 & 1.38 \\
Beta Q4 & 0.86 & 0.97 & 1.05 & 1.11 & 1.29 & BM Q4 & 0.87 & 1.00 & 1.03 & 1.17 & 1.38 & Size Q4 & 0.86 & 1.02 & 1.05 & 1.14 & 1.39 \\
Beta Q5 & 0.87 & 0.91 & 0.99 & 1.18 & 1.32 & BM Q5 & 0.86 & 1.00 & 1.05 & 1.16 & 1.38 & Size Q5 & 1.17 & 1.10 & 1.20 & 1.16 & 1.39 \\ \midrule
\multicolumn{18}{c}{Double sorted portfolios with preparation and reconnaissance cyber score} \\  \midrule
Beta Q1*& 1.10 & 0.91 & 1.28 & 1.21 & 1.41 & BM Q1*& 1.08 & 0.99 & 1.12 & 1.13 & 1.24 & Size Q1*& 0.91 & 0.84 & 1.13 & 1.15 & 1.34 \\
Beta Q2*& 0.93 & 0.87 & 1.20 & 1.13 & 1.35 & BM Q2*& 0.85 & 0.78 & 1.18 & 1.19 & 1.37 & Size Q2*& 0.92 & 0.83 & 1.14 & 1.20 & 1.31 \\
Beta Q3 & 0.92 & 0.90 & 1.13 & 1.11 & 1.29 & BM Q3*& 0.90 & 0.93 & 1.15 & 1.09 & 1.34 & Size Q3*& 0.90 & 0.87 & 1.17 & 1.13 & 1.32 \\
Beta Q4*& 0.90 & 0.77 & 1.20 & 1.07 & 1.27 & BM Q4 & 0.86 & 0.89 & 1.15 & 1.12 & 1.34 & Size Q4*& 0.86 & 0.86 & 1.20 & 1.12 & 1.34 \\
Beta Q5*& 0.89 & 0.78 & 1.13 & 1.12 & 1.29 & BM Q5 & 0.85 & 0.90 & 1.13 & 1.14 & 1.34 & Size Q5 & 1.13 & 1.15 & 1.19 & 1.19 & 1.38 \\ \bottomrule
\end{tabular}%
}
\caption[\footnotesize Average returns of the double sorted portfolios]{\textbf{Average returns of the double sorted portfolios}}
    \label{double_sort_table}
    \bigskip
    \footnotesize{
    \begin{flushleft}
     \textbf{Q1} to \textbf{Q5} represent quintiles. The sorting of firms is done according to market beta (Beta), book-to-market ratios (BM), or firm size (Size) and then on the relevant cyber score. The average returns are given in percent. $*$ indicates that the returns are not increasing monotonically with the quintile of the cyber score (with an incertitude of -0.03\%). 
     \end{flushleft}} 
\end{table}

\newpage

\newpage

\begin{table}[H]
\centering

\renewcommand{\arraystretch}{1.2}
\resizebox{0.45\textwidth}{!}{%
\begin{tabular}{lccccc}
\toprule
 & M.1 & M.2& M.3 & M.4& M.5 \\
\midrule
Market & \textbf{0.011}*** & \textbf{ } & \textbf{0.009}** & \textbf{0.013}*** & \textbf{0.009**} \\
 & [2.886] &  & [2.526] & [3.507] & [2.429] \\
Cyber &  & \textbf{0.054}* & \textbf{0.051}* & \textbf{0.051}** & \textbf{0.04} \\
 &  & [1.925] & [1.807] & [2.097] & [1.547] \\
HML &  &  & & \textbf{0.003} & \textbf{0.003}   \\
 &  &  & &  [1.176] & [0.964]   \\
SMB &  &  & & \textbf{-0.001} & \textbf{0.001}   \\
 &  &  & & [-0.223] & [0.636]   \\
UMD &  &  &  & \textbf{0.002} &  \\
 &  &  &  & [0.766] &  \\
 CMA &  &  &  &  & \textbf{-0.001} \\
 &  &  &  &  & [-0.627] \\
RMW &  &  &  &  & \textbf{0.002} \\
 &  &  &  &  & [0.776] \\
Constant & \textbf{0.001} & \textbf{-0.017} & \textbf{-0.024} & \textbf{-0.029}** & \textbf{-0.019} \\
 & [0.148] & [-1.083] & [-1.586] & [-2.223] & [-1.357] \\
\midrule
$\overline{R^2_{\text{adj}}}$ & 0.067 & 0.158 & 0.22 & 0.296 & 0.309 \\
MAPE & 0.013 & 0.012 & 0.012 & 0.01 & 0.009 \\
\bottomrule
\end{tabular}
}
\caption[\footnotesize Fama-McBeth for overall cyber score]{\textbf{Fama-MacBeth for overall cyber score}}\label{FamaMcBeth_overall}
\bigskip
    \footnotesize{
    \begin{flushleft}
    This table reports the results of Fama-MacBeth regressions of 20 value-weighted portfolios sorted on their cyber score. These portfolios are regressed each month on portfolio value-weighted betas with the market, HML, SMB, MOM, RMW, and CMA. \quotes{Cyber} is the value-weighted cyber score of each portfolio. HML and SMB refer to the book-to-market and size factors from Fama and French (1992). UMD refers to the momentum factor from Carhart (1997). CMA and RMW refer to the investment and operating profitability factors from Fama and French (2015). $\overline{R^2_{\text{adj}}}$ is the average adjusted R-squared, and MAPE is the mean average pricing error (mean average of the absolute value of the residuals).  T-statistics are reported in brackets. *, **, and *** indicate significance at the 10\%, 5\% and 1\% levels, respectively. The period is from January 2009 to December 2023.
    \end{flushleft}} 
\end{table}

\begin{table}[H]
\centering

\renewcommand{\arraystretch}{1.2}
\resizebox{0.45\textwidth}{!}{%
\begin{tabular}{lccccc}
\toprule
 & M.1 & M.2& M.3 & M.4& M.5 \\
\midrule
Market & \textbf{0.005} & \textbf{ } & \textbf{0.004} & \textbf{0.005} & \textbf{0.005} \\
 & [1.335] &  & [0.951] & [1.308] & [1.27] \\
Cyber &  & \textbf{0.003} & \textbf{0.0} & \textbf{0.003} & \textbf{-0.008} \\
 &  & [0.277] & [0.003] & [0.228] & [-0.501] \\
HML &  &  & & \textbf{0.001} & \textbf{-0.001}   \\
 &  &  & & [0.381] & [-0.235]   \\
SMB &  &  & &\textbf{0.001} & \textbf{0.001}   \\
 &  &  & & [0.304] & [0.383]   \\
UMD &  &  &  & \textbf{0.005} &  \\
 &  &  &  & [1.508] &  \\
CMA &  &  &  &  & \textbf{-0.001} \\
 &  &  &  &  & [-0.488] \\
 RMW &  &  &  &  & \textbf{0.001} \\
 &  &  &  &  & [0.486] \\
Constant & \textbf{0.006} & \textbf{0.01} & \textbf{0.008} & \textbf{0.005} & \textbf{0.011} \\
 & [1.643] & [1.607] & [1.209] & [0.754] & [1.439] \\
\midrule
$\overline{R^2_{\text{adj}}}$ & 0.077 & 0.042 & 0.108 & 0.202 & 0.237 \\
MAPE & 0.013 & 0.013 & 0.012 & 0.011 & 0.01 \\
\bottomrule
\end{tabular}
}
\caption[\footnotesize Fama-McBeth for cyber sentiment score]{\textbf{Fama-McBeth for cyber sentiment score}} \label{FamaMcBeth_sentiment}
\end{table}

\begin{table}[H]
\centering

\renewcommand{\arraystretch}{1.2}
\resizebox{0.5\textwidth}{!}{%
\begin{tabular}{lccccc}
\toprule
 & M.1 & M.2& M.3 & M.4& M.5 \\
\midrule
Market & \textbf{0.009}** & \textbf{ } & \textbf{0.007}* & \textbf{0.009}** & \textbf{0.01**} \\
 & [2.328] &  & [1.716] & [2.493] & [2.574] \\
Cyber &  & \textbf{0.044}* & \textbf{0.051}* & \textbf{0.056}** & \textbf{0.038} \\
 &  & [1.673] & [1.914] & [2.335] & [1.401] \\
HML &  &  & & \textbf{0.003} & \textbf{0.003}   \\
 &  &  & & [1.122] & [0.942]   \\
SMB &  &  & & \textbf{0.002} & \textbf{0.002}   \\
 &  &  & & [0.884] & [0.692]   \\
UMD &  &  &  & \textbf{0.0} &  \\
 &  &  &  & [0.005] &  \\
CMA &  &  &  &  & \textbf{-0.001} \\
 &  &  &  &  & [-0.35] \\
  RMW &  &  &  &  & \textbf{0.001} \\
 &  &  &  &  & [0.561] \\
Constant & \textbf{0.003} & \textbf{-0.01} & \textbf{-0.02} & \textbf{-0.026}** & \textbf{-0.016} \\
 & [0.661] & [-0.733] & [-1.439] & [-2.032] & [-1.141] \\
\midrule
$\overline{R^2_{\text{adj}}}$ & 0.053 & 0.154 & 0.206 & 0.315 & 0.322 \\
MAPE & 0.014 & 0.013 & 0.012 & 0.01 & 0.01 \\
\bottomrule
\end{tabular}
}
\caption[\footnotesize Fama-McBeth for command and data manipulation cyber score]{\textbf{Fama-McBeth for command and data manipulation cyber score}}\label{FamaMcBeth_command_and_data_manipulation}
\end{table}

\begin{table}[H]
\centering

\renewcommand{\arraystretch}{1.2}
\resizebox{0.5\textwidth}{!}{%
\begin{tabular}{lccccc}
\toprule
 & M.1 & M.2& M.3 & M.4& M.5 \\
\midrule
Market & \textbf{0.007}** & \textbf{ } & \textbf{0.004} & \textbf{0.009}** & \textbf{0.007*} \\
 & [2.106] &  & [1.036] & [2.454] & [1.894] \\
Cyber &  & \textbf{0.051}* & \textbf{0.056}** & \textbf{0.065}** & \textbf{0.056**} \\
 &  & [1.906] & [2.101] & [2.419] & [2.062] \\
HML &  &  & & \textbf{0.005}* & \textbf{0.003}   \\
 &  &  & & [1.67] & [1.162]   \\
SMB &  &  & & \textbf{-0.001} & \textbf{0.001}   \\
 &  &  & & [-0.379] & [0.46]  \\
UMD &  &  & & \textbf{-0.001} &   \\
 &  &  & & [-0.243] &    \\
CMA &  &  &  &  & \textbf{-0.002} \\
 &  &  &  &  & [-1.145] \\
  RMW &  &  &  &  & \textbf{0.001} \\
 &  &  &  &  & [0.458] \\
Constant & \textbf{0.004} & \textbf{-0.014} & \textbf{-0.02} & \textbf{-0.03}** & \textbf{-0.023} \\
 & [1.04] & [-0.982] & [-1.487] & [-2.228] & [-1.649] \\
\midrule
$\overline{R^2_{\text{adj}}}$ & 0.057 & 0.157 & 0.219 & 0.308 & 0.313 \\
MAPE & 0.013 & 0.013 & 0.012 & 0.01 & 0.009 \\
\bottomrule
\end{tabular}
}
\caption[\footnotesize Fama-McBeth for credential movement cyber score]{\textbf{Fama-McBeth for credential movement cyber score}}\label{FamaMcBeth_credential_movement}
\end{table}

\begin{table}[H]
\centering

\renewcommand{\arraystretch}{1.2}
\resizebox{0.5\textwidth}{!}{%
\begin{tabular}{lccccc}
\toprule
 & M.1 & M.2& M.3 & M.4& M.5 \\
\midrule
Market & \textbf{0.004} & \textbf{ } & \textbf{0.002} & \textbf{0.004} & \textbf{0.003} \\
 & [1.201] &  & [0.472] & [1.222] & [0.801] \\
Cyber &  & \textbf{0.056}* & \textbf{0.061}** & \textbf{0.073}** & \textbf{0.071**} \\
 &  & [1.835] & [2.062] & [2.451] & [2.193] \\
HML &  &  & & \textbf{0.003} & \textbf{0.004}   \\
 &  &  & & [1.255] & [1.357]  \\
SMB &  &  & & \textbf{0.001} & \textbf{0.001}   \\
 &  &  & & [0.442] & [0.474]   \\
UMD &  &  & & \textbf{-0.002} &    \\
 &  &  & & [-0.477] &    \\
CMA &  &  &  &  & \textbf{-0.0} \\
 &  &  &  &  & [-0.192] \\
  RMW &  &  &  &  & \textbf{0.002} \\
 &  &  &  &  & [1.370] \\
Constant & \textbf{0.007} & \textbf{-0.017} & \textbf{-0.021} & \textbf{-0.03}* & \textbf{-0.028} \\
 & [1.644] & [-1.04] & [-1.349] & [-1.944] & [-1.641] \\
\midrule
$\overline{R^2_{\text{adj}}}$ & 0.058 & 0.167 & 0.215 & 0.3 & 0.304 \\
MAPE & 0.013 & 0.012 & 0.012 & 0.01 & 0.009 \\
\bottomrule
\end{tabular}
}
\caption[\footnotesize Fama-McBeth for persistence and evasion cyber score]{\textbf{Fama-McBeth for persistence and evasion cyber score}}\label{FamaMcBeth_persistence_and_evasion}
\end{table}

\begin{table}[H]
\centering

\renewcommand{\arraystretch}{1.2}
\resizebox{0.5\textwidth}{!}{%
\begin{tabular}{lccccc}
\toprule
 & M.1 & M.2& M.3 & M.4& M.5 \\
\midrule
Market & \textbf{0.009}* & \textbf{ } & \textbf{0.006} & \textbf{0.011}** & \textbf{0.01**} \\
 & [1.951] &  & [1.531] & [2.541] & [2.055] \\
Cyber &  & \textbf{0.047}* & \textbf{0.046}* & \textbf{0.049}** & \textbf{0.052**} \\
 &  & [1.848] & [1.888] & [2.262] & [2.299] \\
HML &  &  & & \textbf{0.003} & \textbf{0.004}   \\
 &  &  & & [1.053] & [1.52]   \\
SMB &  &  & & \textbf{0.001} & \textbf{0.001}   \\
 &  &  & & [0.233] & [0.418]   \\
UMD &  &  & & \textbf{0.001}   &  \\
 &  &  & &  [0.168] &    \\
CMA &  &  &  &  & \textbf{-0.001} \\
 &  &  &  &  & [-0.466] \\
  RMW &  &  &  &  & \textbf{0.002} \\
 &  &  &  &  & [0.862] \\
Constant & \textbf{0.003} & \textbf{-0.012} & \textbf{-0.018} & \textbf{-0.024}** & \textbf{-0.025**} \\
 & [0.674] & [-0.908] & [-1.345] & [-2.047] & [-2.008] \\
\midrule
$\overline{R^2_{\text{adj}}}$ & 0.072 & 0.137 & 0.201 & 0.274 & 0.288 \\
MAPE & 0.014 & 0.013 & 0.012 & 0.01 & 0.01 \\
\bottomrule
\end{tabular}
}
\caption[\footnotesize Fama-McBeth for preparation and reconaissance cyber score]{\textbf{Fama-McBeth for preparation and reconaissance cyber score}}\label{FamaMcBeth_preparation_and_reconaissance}
\end{table}

\newpage

\newpage
\begin{table}[H]
\centering

\renewcommand{\arraystretch}{1.2}
\resizebox{0.6\textwidth}{!}{%
\begin{tabular}{lcccccc}
\toprule
& GRS & p-value & $\overline{R2}$ & GRS & p-value & $\overline{R2}$\\
\midrule
& \multicolumn{3}{c}{Sorted on cyber score}
& \multicolumn{3}{c}{Sorted on size}\\
\cmidrule(lr){2-4}
\cmidrule(lr){5-7}
FF5 & 1.451 & 0.107 & 0.868 & 0.737 & 0.783 & 0.876\\
FF5 + CyberFactor & 1.088 & 0.367 & 0.888 & 0.833 & 0.671 & 0.877\\
\\
& \multicolumn{3}{c}{Sorted on market beta}
& \multicolumn{3}{c}{Sorted on book-to-market}\\
\cmidrule(lr){2-4}
\cmidrule(lr){5-7}
FF5 & 1.541 & 0.075 & 0.793 & 1.240 & 0.229 & 0.891\\
FF5 + CyberFactor & 1.495 & 0.090 & 0.806 & 1.021 & 0.441 & 0.895\\
\bottomrule
\end{tabular}
}
\caption[\footnotesize GRS test for overall cyber score]{\textbf{\small GRS test for overall cyber score}}\label{GRS_overall}
\bigskip
    \footnotesize{
    \begin{flushleft}
    This table reports the results of time series regressions of 20 value-weighted portfolios (sorted on the cyber score, the size of firms, the market beta or the book-to-market ratio) on the five-factor model of\cite{FamaFrench2015} (FF5) and the \quotes{CyberFactor}, \textit{i.e.} the factor built as the long-short of extreme quintile portfolios sorted on the relevant cyber score (P5-P1). The p-value is the probability that the alphas of the 20 regressions are jointly zero. A probability lower than 10\% means that the hypothesis that alphas are jointly zero can be rejected at the 10\% level. The study period is from January 2009 to December 2023.
    \end{flushleft}} 
\end{table}

\begin{table}[H]
\centering

\renewcommand{\arraystretch}{1.2}
\resizebox{0.6\textwidth}{!}{%
\begin{tabular}{lcccccc}
\toprule
& GRS & p-value & $\overline{R2}$ & GRS & p-value & $\overline{R2}$\\
\midrule
& \multicolumn{3}{c}{Sorted on cyber score}
& \multicolumn{3}{c}{Sorted on size}\\
\cmidrule(lr){2-4}
\cmidrule(lr){5-7}
FF5 & 1.254 & 0.218 & 0.854 & 0.737 & 0.783 & 0.876\\
FF5 + CyberFactor & 1.198 & 0.263 & 0.864 & 0.748 & 0.771 & 0.877\\
\\
& \multicolumn{3}{c}{Sorted on market beta}
& \multicolumn{3}{c}{Sorted on book-to-market}\\
\cmidrule(lr){2-4}
\cmidrule(lr){5-7}
FF5 & 1.541 & 0.075 & 0.793 & 1.240 & 0.229 & 0.891\\
FF5 + CyberFactor & 1.488 & 0.093 & 0.796 & 1.188 & 0.271 & 0.892\\
\bottomrule
\end{tabular}
}
\caption[\footnotesize GRS test for cyber sentiment score]{\textbf{\small GRS test for cyber sentiment score}}\label{GRS_sentiment}
\end{table}

\begin{table}[H]
\centering

\renewcommand{\arraystretch}{1.2}
\resizebox{0.6\textwidth}{!}{%
\begin{tabular}{lcccccc}
\toprule
& GRS & p-value & $\overline{R2}$ & GRS & p-value & $\overline{R2}$\\
\midrule
& \multicolumn{3}{c}{Sorted on cyber score}
& \multicolumn{3}{c}{Sorted on size}\\
\cmidrule(lr){2-4}
\cmidrule(lr){5-7}
FF5 & 1.558 & 0.070 & 0.855 & 0.737 & 0.783 & 0.876\\
FF5 + CyberFactor & 1.119 & 0.335 & 0.877 & 0.844 & 0.657 & 0.877\\
\\
& \multicolumn{3}{c}{Sorted on market beta}
& \multicolumn{3}{c}{Sorted on book-to-market}\\
\cmidrule(lr){2-4}
\cmidrule(lr){5-7}
FF5 & 1.541 & 0.075 & 0.793 & 1.240 & 0.229 & 0.891\\
FF5 + CyberFactor & 1.472 & 0.099 & 0.807 & 1.026 & 0.436 & 0.895\\
\bottomrule
\end{tabular}
}
\caption[\footnotesize GRS test for command and data manipulation cyber score]{\textbf{\small GRS test for command and data manipulation cyber score}}\label{GRS_data_and_manipulation}
\end{table}

\begin{table}[H]
\centering

\renewcommand{\arraystretch}{1.2}
\resizebox{0.6\textwidth}{!}{%
\begin{tabular}{lcccccc}
\toprule
& GRS & p-value & $\overline{R2}$ & GRS & p-value & $\overline{R2}$\\
\midrule
& \multicolumn{3}{c}{Sorted on cyber score}
& \multicolumn{3}{c}{Sorted on size}\\
\cmidrule(lr){2-4}
\cmidrule(lr){5-7}
FF5 & 1.539 & 0.076 & 0.864 & 0.737 & 0.783 & 0.876\\
FF5 + CyberFactor & 1.192 & 0.268 & 0.884 & 0.770 & 0.746 & 0.877\\
\\
& \multicolumn{3}{c}{Sorted on market beta}
& \multicolumn{3}{c}{Sorted on book-to-market}\\
\cmidrule(lr){2-4}
\cmidrule(lr){5-7}
FF5 & 1.541 & 0.075 & 0.793 & 1.240 & 0.229 & 0.891\\
FF5 + CyberFactor & 1.441 & 0.111 & 0.804 & 0.997 & 0.469 & 0.895\\
\bottomrule
\end{tabular}
}
\caption[\footnotesize GRS test for credential movement]{\textbf{\small GRS test for credential movement}}\label{GRS_credential_movement}
\end{table}

\begin{table}[H]
\centering

\renewcommand{\arraystretch}{1.2}
\resizebox{0.6\textwidth}{!}{%
\begin{tabular}{lcccccc}
\toprule
& GRS & p-value & $\overline{R2}$ & GRS & p-value & $\overline{R2}$\\
\midrule
& \multicolumn{3}{c}{Sorted on cyber score}
& \multicolumn{3}{c}{Sorted on size}\\
\cmidrule(lr){2-4}
\cmidrule(lr){5-7}
FF5 & 1.465 & 0.101 & 0.868 & 0.737 & 0.783 & 0.876\\
FF5 + CyberFactor & 1.128 & 0.326 & 0.890 & 0.884 & 0.608 & 0.878\\
\\
& \multicolumn{3}{c}{Sorted on market beta}
& \multicolumn{3}{c}{Sorted on book-to-market}\\
\cmidrule(lr){2-4}
\cmidrule(lr){5-7}
FF5 & 1.541 & 0.075 & 0.793 & 1.240 & 0.229 & 0.891\\
FF5 + CyberFactor & 1.500 & 0.088 & 0.806 & 1.021 & 0.441 & 0.896\\
\bottomrule
\end{tabular}
}
\caption[\footnotesize GRS test for persistence and evasion cyber score]{\textbf{\small GRS test for persistence and evasion cyber score}}\label{GRS_persistence_and_evasion}
\end{table}

\begin{table}[H]
\centering

\renewcommand{\arraystretch}{1.2}
\resizebox{0.6\textwidth}{!}{%
\begin{tabular}{lcccccc}
\toprule
& GRS & p-value & $\overline{R2}$ & GRS & p-value & $\overline{R2}$\\
\midrule
& \multicolumn{3}{c}{Sorted on cyber score}
& \multicolumn{3}{c}{Sorted on size}\\
\cmidrule(lr){2-4}
\cmidrule(lr){5-7}
FF5 & 1.516 & 0.083 & 0.861 & 0.737 & 0.783 & 0.876\\
FF5 + CyberFactor & 1.216 & 0.248 & 0.879 & 0.807 & 0.703 & 0.877\\
\\
& \multicolumn{3}{c}{Sorted on market beta}
& \multicolumn{3}{c}{Sorted on book-to-market}\\
\cmidrule(lr){2-4}
\cmidrule(lr){5-7}
FF5 & 1.541 & 0.075 & 0.793 & 1.240 & 0.229 & 0.891\\
FF5 + CyberFactor  & 1.546 &  0.076 &  0.807  & 0.990 & 0.477 & 0.896\\
\bottomrule
\end{tabular}
}
\caption[\footnotesize GRS test for preparation and reconaissance cyber score]{\textbf{\small GRS test for preparation and reconaissance cyber score}}\label{GRS_preparation_and_reconaissance}
\end{table}

\newpage

\newpage

\newpage

\begin{table}[h!]
    \centering
    \begin{tabular}{lcc|cc}
        \toprule
        \textbf{Cyber score }& \multicolumn{2}{c}{P5 (top 20\%)} & \multicolumn{2}{c}{P20 (top 5\%)} \\
        \cmidrule(r){2-3} \cmidrule(r){4-5}
          & t-stat. & p-value & t-stat. & p-value \\
        \midrule
        persistence & -0.0372 & 0.9703 & -0.0768 & 0.9388 \\
        command and control & 0.0605 & 0.9518 & 0.7315 & 0.4649 \\
        impact & 0.0692 & 0.9449 & 0.0011 & 0.9992 \\
        initial access & -0.0751 & 0.9402 & 0.1292 & 0.8972 \\
        resource development & 0.3001 & 0.7643 & 0.2834 & 0.7770 \\
        collection & -0.0099 & 0.9921 & 0.0890 & 0.9291 \\
        exfiltration & 0.0041 & 0.9967 & 0.4411 & 0.6593 \\
        credential access & -0.0167 & 0.9867 & 0.3022 & 0.7626 \\
        privilege escalation & -0.0461 & 0.9632 & 0.1641 & 0.8697 \\
        execution & 0.2384 & 0.8117 & 0.1040 & 0.9172 \\
        defense evasion & 0.0474 & 0.9622 & -0.1037 & 0.9174 \\
        reconnaissance & 0.1565 & 0.8757 & 0.2418 & 0.8091 \\
        lateral movement & -0.0665 & 0.9470 & -0.2061 & 0.8368 \\
        discovery & 0.0419 & 0.9666 & 0.1053 & 0.9162 \\
        \hline
        preparation and reconnaissance & 0.1050 & 0.9165 & 0.2395 & 0.8108 \\
        persistence and evasion & -0.1185 & 0.9058 & -0.0888 & 0.9293 \\
        credential movement & 0.0242 & 0.9807 & -0.0176 & 0.9860 \\
        command and data manipulation & 0.0894 & 0.9288 & 0.0601 & 0.9521 \\
        \hline
        sentiment & 0.6125 & 0.5405 & 1.1653 & 0.2446 \\
        \bottomrule
    \end{tabular}
    \caption[\footnotesize Cyber based portfolios returns differences]{\textbf{\small Cyber based portfolio returns differences}} \label{P5_and_P20_time_series_diff}
\bigskip
    \footnotesize{
    \begin{flushleft}
    The table displays the outcomes of Welch's t-test, the statistical method used to evaluate the significance of mean differences with the possibility of different variances, applied to each cyber score time series in comparison to the overall cyber score time series. These time series are the monthly returns of cyber-based portfolios: P5 (constructed with 5 quantiles, taking the top 20\%) and P20 (constructed with 20 quantiles, taking the top 5\%). 
    \end{flushleft}} 
\end{table}

\begin{table}[h!]
    \centering
    \begin{tabular}{lrrrrrr}
        \hline
        & P1 & P2 & P3 & P4 & P5 & P5-P1 \\
        \hline
        CAR[-1,1] & -0.146 & 0.001 & -0.021 & -0.103 & 0.206 & 0.352 \\
        t-statistic & -0.311 & 0.002 & -0.058 & -0.342 & 0.616 & 0.450 \\
        CAR[-1,3] & -0.197 & -0.040 & -0.178 & -0.051 & 0.194 & 0.390 \\
        t-statistic & -0.540 & -0.115 & -0.626 & -0.220 & 0.748 & 0.644 \\
        \hline
    \end{tabular}
    \caption[\footnotesize Cumulative abnormal returns of cyber-based portfolios]{\textbf{\small Cumulative abnormal returns of cyber-based portfolios}} \label{event_overall}
\bigskip
    \footnotesize{
    \begin{flushleft}
    To estimate the cumulative abnormal returns (CAR), I use the market model around December 14, 2020, as t=0. Note that it was a Monday. Therefore, $t=-1$ corresponds to Friday, December 11. The beta of the market model is set up thanks to the returns of the prior year. The abnormal returns are given in percent. The portfolios are based on the overall cyber score.
    \end{flushleft}} 
\end{table}

\newpage

\begin{table}[h!]
\centering
\begin{tabular}{lcc}
\toprule
 & CAR[-1,1] & CAR[-1,3] \\
\midrule
overall & 0.078 & 0.055 \\
 & [0.151] & [0.138] \\
preparation and reconnaissance & 0.036 & 0.146 \\
 & [0.082] & [0.429] \\
persistence and evasion & 0.163 & 0.228 \\
 & [0.301] & [0.543] \\
credential movement & 0.141 & 0.312 \\
 & [0.255] & [0.724] \\
command and data manipulation & 0.207 & 0.227 \\
 & [0.427] & [0.603] \\
\bottomrule
\end{tabular}
\caption[\footnotesize Cumulative abnormal returns of cyber-based P20]{\textbf{\small Cumulative abnormal returns of cyber-based P20}} \label{event_P20}
\bigskip
    \footnotesize{
    \begin{flushleft}
    To estimate the cumulative abnormal returns (CAR), I use the market model around December 14, 2020, as $t=0$. The beta of the market model is set up thanks to the returns of the prior year. The abnormal returns are given in percent. The t-statistics associated with the abnormal returns are given in the parenthesis.
    \end{flushleft}} 
\end{table}

\newpage
\begin{table}[h!]
\centering
\begin{tabular}{lcc}
\toprule
 & CAR[-1,1] & CAR[-1,3] \\
\midrule
overall & 0.206 & 0.194 \\
 & [0.616] & [0.748] \\

preparation and reconnaissance & 0.182 & 0.181 \\
 & [0.639] & [0.818] \\

persistence and evasion & 0.192 & 0.189 \\
& [0.533] & [0.675] \\

credential movement & 0.198 & 0.200 \\
 & [0.565] & [0.735] \\

command and data manipulation & 0.009 & 0.081 \\
 & [0.029] & [0.336] \\

\bottomrule
\end{tabular}
\caption[\footnotesize Cumulative abnormal returns of cyber-based P5]{\textbf{\small Cumulative abnormal returns of cyber-based P5}} \label{event_P5}
\bigskip
    \footnotesize{
    \begin{flushleft}
    To estimate the cumulative abnormal returns (CAR), I use the market model around December 14, 2020, as $t=0$. The beta of the market model is set up thanks to the returns of the prior year. The abnormal returns are given in percent. The t-statistics associated with the abnormal returns are given in the parenthesis.
    \end{flushleft}} 
\end{table}

\newpage

\clearpage
\section*{Appendix}

\setcounter{table}{0}
\renewcommand{\thetable}{A\arabic{table}}
\setcounter{figure}{0}
\renewcommand{\thefigure}{A\arabic{figure}}

\begin{table}[h]
\begin{adjustbox}{width=0.95\textwidth,center}
    \begin{tabular}{lcc}
     Variable & Description & Source \\
    \hline
    Firm size (ln) & ln(total assets [at])  & Compustat\\
    Firm Age (ln) & ln(years) since the firm first appeared in Compustat & Compustat\\
    Book to market ratio & Common equity [ceq] / market equity [prc*shrout] & Compustat and CRSP\\
    Tobin's Q & (Total assets - common equity + market equity) / total assets & Compustat and CRSP \\
    ROA & Net income [ni] / total assets & Compustat\\
    Market Beta & 5-year rolling market beta [beta] & Compustat\\
    Intangible/Assets & Intangible assets [intan] / total assets & Compustat\\
    Debt/assets & Total Debt / Total Assets [debt\_assets] & WRDS Financial Ratios \\
    ROE & Net Income / Book Equity [roe] & WRDS Financial Ratios \\
    Price/Earnings & Stock Price / Earnings [pe\_exi] & WRDS Financial Ratios\\
    Profit Margin & Gross Profit / Sales [gpm] & WRDS Financial Ratios \\
    Asset Turnover & Sales / Total Assets [at\_turn] & WRDS Financial Ratios\\
    Cash Ratio & (Cash + Short-term Investments) / Current Liabilities [cash\_ratio] & WRDS Financial Ratios\\
    Sales/Invested Capital & Sales per dollar of Invested Capital [sale\_invcap] &  WRDS Financial Ratios\\
    Capitalization Ratio &  Long-term Debt / (Long-term Debt + Equity) [capital\_ratio] & WRDS Financial Ratios \\
    R\&D/Sales & R\&D expenses / Sales [RD\_SALE] & WRDS Financial Ratios \\
    ROCE & Earnings Before Interest and Taxes / average Capital Employed [roce] & WRDS Financial Ratios\\
    Readability ($\ln$) & Number of characters in the 10-K & EDGAR - SEC\\
    Risk section length ($\ln$) & Number of sentences in Item 1A of the 10-K &  EDGAR - SEC \\
    Secrets & As defined in \cite{FlorackisLoucaMichaelyWeber2023}&  EDGAR - SEC \\
    Volume per capital & Monthly trading volume / Market capitalization & CRSP\\
    Humans per capital & Monthly number of employees / Market capitalization & Compustat and CRSP\\
    \end{tabular}
    \end{adjustbox}
    \caption[\footnotesize Variable definitions]{\textbf{\small Variable definitions}}\bigskip
    \footnotesize{This table reports the variable names used throughout the paper, their description, and their source. Square brackets indicate variable name definitions in CRSP and Compustat.}
    
\label{tab:variable_descriptions}
\end{table}

\newpage

\textbf{Risk/Uncertainty dictionary : } {\small risk, jeopardize, riskiness, risks, unsettled, treacherous, uncertainty, unpredictability, oscillating, variable, dilemma, perilous, chance, skepticism, tentativeness, possibility, hesitancy, unreliability, pending, riskier, wariness, uncertainties, unresolved, vagueness, uncertain, unsure, dodgy, doubt, irregular, equivocation, prospect, jeopardy, indecisive, bet, suspicion, chancy, variability, risking, menace, exposed, peril, qualm, likelihood, hesitating, vacillating, threat, risked, gnarly, probability, unreliable, disquiet, unknown, unsafe, ambivalence, varying, hazy, imperil, unclear, apprehension, vacillation, unpredictable, unforeseeable, incalculable, speculative, halting, untrustworthy, fear, wager, equivocating, reservation, torn, diffident, hesitant, precarious, fickleness, gamble, undetermined, misgiving, risky, insecurity, changeability, instability, debatable, undependable, doubtful, undecided, incertitude, hazard, dicey, fitful, tricky, indecision, parlous, sticky, wavering, unconfident, dangerous, iffy, defenseless, tentative, faltering, unsureness, hazardous, endanger, fluctuant, queries, quandary, niggle, danger, insecure, diffidence, fluctuating, changeable, precariousness, unstable, riskiest, doubtfulness, vague, hairy, erratic, ambivalent, query, dubious} (\citealp{HassanHollandervLentTahoun2019})

\textcolor{white}{skip}

\textcolor{white}{skip}

\begin{table}[ht]
\centering
\begin{tabular}{l|c|c}
\hline
\textbf{Cyber score} & \textbf{Covariance $\cdot 10^3$}  & \textbf{Correlation} \\
\hline
persistence & 0.2777 & 0.1038 \\
command and control & 0.3146 & 0.1281 \\
impact & 0.2960 & 0.1079 \\
initial access & 0.2237 & 0.0766 \\
resource development & 0.1841 & 0.0629 \\
collection & 0.2035 & 0.0723 \\
exfiltration & 0.1473 & 0.0499 \\
credential access & 0.2515 & 0.0916 \\
privilege escalation & 0.3088 & 0.1286 \\
execution & 0.2704 & 0.1121 \\
defense evasion & 0.1809 & 0.0761 \\
reconnaissance & 0.2145 & 0.0768 \\
lateral movement & 0.1272 & 0.0488 \\
discovery & 0.1640 & 0.0681 \\
\hline
preparation and reconnaissance & 0.1958 & 0.0710 \\
persistence and evasion & 0.1914 & 0.0795 \\
credential movement & 0.2350 & 0.0867 \\
command and data manipulation & 0.2114 & 0.0756 \\
\hline
overall & 0.1799 & 0.0699 \\
sentiment & -0.0408 & -0.0095 \\
\hline
\end{tabular}
\caption[\footnotesize Cyber score correlation and covariance with idiosyncratic volatility]{\textbf{\small Cyber scores correlation and covariance with idiosyncratic volatility}} \label{idiosyncratic_cov_corr}
\bigskip
    \footnotesize{
    \begin{flushleft}
    Correlation and covariance of the different cyber scores with the idiosyncratic volatility of the firms they are associated with. The idiosyncratic volatility at a given time is computed as the root squared of $var(\epsilon_i)=var(r_i)-cov(r_i,r_m)^2/var(r_m)$ taken over the last five years, where $r_i$ and $r_m$ are the excess return of the firm and the market. The covariance is multiplied by $10^3$ to improve readability.
    \end{flushleft}} 
\end{table}

\end{document}